\documentclass[fleqn,usenatbib]{mnras}
\usepackage{newtxtext,newtxmath}
\usepackage[utf8]{inputenc}
\usepackage[T1]{fontenc}

\usepackage{fontawesome5}
\usepackage{graphicx}
\usepackage{amsfonts}
\usepackage{booktabs}
\usepackage{siunitx}
\usepackage{mathtools}
\usepackage{multirow}
\usepackage{array}
\usepackage[normalem]{ulem}
\usepackage{orcidlink}
\usepackage{bm}
\usepackage{cleveref}
\usepackage{xcolor}
\usepackage{caption}
\usepackage{float}
\usepackage{acro}
\usepackage{tabularray}
\UseTblrLibrary{booktabs}
\usepackage{subcaption}
\usepackage{placeins}
\usepackage[justification=RaggedRight]{caption}
\usepackage{transparent}
\usepackage[multiple]{footmisc}
\usepackage[dont-mess-around]{fnpct}
\usepackage{import}
\usepackage{xspace}

\xspaceaddexceptions{+}
\def\msun{M_\odot}
\DeclareSIUnit{\year}{yr}

\newcommand{\msol}{$M_\odot$\xspace}
\newcommand{\ett}{ET$\Delta$\xspace}
\newcommand{\nsat}{\textit{n}_{\text{sat}}}

\newcommand{\mtov}{M_{\text{TOV}}}
\newcommand{\sat}{\mathrm{sat}}
\newcommand{\sym}{\mathrm{sym}}

\usepackage{acro}

\DeclareAcronym{AD}{
  short = AD,
  long  = automatic differentiation
}

\DeclareAcronym{BH}{
  short = BH,
  long  = black hole
}

\DeclareAcronym{BBH}{
  short = BBH,
  long  = binary black hole
}

\DeclareAcronym{BNS}{
  short = BNS,
  long  = binary neutron star
}

\DeclareAcronym{CBC}{
  short        = CBC,
  long         = compact binary coalescence,
  long-plural  = compact binary coalescences
}

\DeclareAcronym{CCSN}{
  short = CCSN,
  long  = core-collapse supernova
}

\DeclareAcronym{CE}{
  short = CE,
  long  = Cosmic Explorer
}

\DeclareAcronym{chiEFT}{
  short = $\chi$EFT,
  long  = chiral effective field theory
}

\DeclareAcronym{CPU}{
  short = CPU,
  long  = central processing unit
}

\DeclareAcronym{CSE}{
  short = CSE,
  long  = speed-of-sound extension scheme
}

\DeclareAcronym{EFT}{
  short = EFT,
  long  = effective field theory
}

\DeclareAcronym{EM}{
  short = EM,
  long  = electromagnetic
}

\DeclareAcronym{EOS}{
  short = EOS,
  long  = equation of state
}

\DeclareAcronym{ET}{
  short = ET,
  long  = Einstein Telescope
}

\DeclareAcronym{FIM}{
   short = FIM,
   long = Fisher information matrix
}

\DeclareAcronym{FOV}{
  short = FOV,
  long = field of view
}

\DeclareAcronym{GPU}{
  short = GPU,
  long  = graphical processing unit
}

\DeclareAcronym{GR}{
  short = GR,
  long  = general relativity
}

\DeclareAcronym{GRB}{
  short = GRB,
  long  = gamma-ray burst
}

\DeclareAcronym{GW}{
  short = GW,
  long  = gravitational wave
}

\DeclareAcronym{HIC}{
  short = HIC,
  long  = heavy-ion collision
}

\DeclareAcronym{JIT}{
  short = JIT,
  long  = just-in-time
}

\DeclareAcronym{JS}{
  short = JS,
  long  = Jensen--Shannon
}

\DeclareAcronym{KDE}{
  short = KDE,
  long  = kernel density estimate
}

\DeclareAcronym{KN}{
  short        = KN,
  long         = kilonova,
  short-plural-form = KNe,
  long-plural  = kilonovae
}

\DeclareAcronym{LCDM}{
    short = $\Lambda$CDM,
    long = $\Lambda$ cold dark matter
}

\DeclareAcronym{LVK}{
  short = LVK,
  long  = LIGO--Virgo--KAGRA Scientific Collaboration
}

\DeclareAcronym{MCMC}{
  short = MCMC,
  long  = Markov chain Monte Carlo
}

\DeclareAcronym{ML}{
  short = ML,
  long  = machine learning
}

\DeclareAcronym{MM}{
  short = MM,
  long  = metamodel
}

\DeclareAcronym{NF}{
  short = NF,
  long  = normalizing flow
}

\DeclareAcronym{NICER}{
  short = NICER,
  long  = Neutron star Interior Composition ExploreR
}

\DeclareAcronym{NS}{
  short = NS,
  long  = neutron star
}

\DeclareAcronym{NSBH}{
  short = NSBH,
  long  = neutron star--black hole
}

\DeclareAcronym{NEP}{
  short = NEP,
  long  = nuclear empirical parameter
}

\DeclareAcronym{NR}{
  short = NR,
  long  = numerical relativity
}

\DeclareAcronym{PE}{
  short = PE,
  long  = parameter estimation
}

\DeclareAcronym{PSD}{
  short = PSD,
  long  = power spectral density
}

\DeclareAcronym{RMF}{
  short = RMF,
  long  = relativistic mean-field theory
}

\DeclareAcronym{SKA}{
  short = SKA,
  long  = Square Kilometre Array
}

\DeclareAcronym{SN}{
  short = SN,
  long  = supernova
}

\DeclareAcronym{SNR}{
  short = SNR,
  long  = signal-to-noise ratio
}

\DeclareAcronym{TOV}{
  short = TOV,
  long  = Tolman--Oppenheimer--Volkoff
}

\DeclareAcronym{TPU}{
  short = TPU,
  long  = tensor processing unit
}

\DeclareAcronym{ZAMS}{
  short = ZAMS,
  long  = zero-age main sequence
}

\title[BNSs in the next-generation era]{Binary neutron stars in the next-generation era: multimessenger detection prospects and constraints on the equation of state, mass distribution, and cosmology}

\author[H.~Koehn et al.]
{Hauke Koehn\orcidlink{0009-0001-5350-7468},$^{1}$\thanks{E-mail: hauke.koehn@uni-potsdam.de} Thibeau Wouters\orcidlink{0009-0006-2797-3808},$^{2,3}$ Gilad Sadeh\orcidlink{0009-0003-0141-6171},$^{4}$ Peter T. H. Pang\orcidlink{0000-0001-7041-3239},$^{3,2}$ Mattia Bulla\orcidlink{0000-0002-8255-5127},$^{5,6,7}$\newauthor
Chris Van Den Broeck\orcidlink{0000-0001-6800-4006},$^{2,3}$ Michael W. Coughlin\orcidlink{0000-0002-8262-2924},$^{8}$ and Tim Dietrich\orcidlink{0000-0003-2374-307X},$^{1,4}$
\\
$^{1}$Institute of Physics and Astronomy, University of Potsdam, Karl-Liebknecht-Str. 24/25, 14476, Potsdam, Germany\\
$^{2}$Institute for Gravitational and Subatomic Physics (GRASP), Utrecht University, Princetonplein 1, 3584 CC Utrecht, the Netherlands\\
$^{3}$Nikhef, Science Park 105, 1098 XG Amsterdam, the Netherlands\\
$^{4}$Max Planck Institute for Gravitational Physics (Albert Einstein Institute), Am M\"{u}hlenberg 1, Potsdam 14476, Germany\\
$^{5}$Department of Physics and Earth Science, University of Ferrara, via Saragat 1, I-44122 Ferrara, Italy\\
$^{6}$INFN, Sezione di Ferrara, via Saragat 1, I-44122 Ferrara, Italy\\
$^{7}$INAF, Osservatorio Astronomico d’Abruzzo, via Mentore Maggini snc, 64100 Teramo, Italy\\
$^{8}$School of Physics and Astronomy, University of Minnesota, Minneapolis, Minnesota 55455, USA
}

\date{Accepted 2026 October 12. Received 2026 October 12; in original form 2026 August 4}
\pubyear{\the\year{}}

\begin{document}
\label{firstpage}
\pagerange{\pageref{firstpage}--\pageref{lastpage}}
\maketitle

\begin{abstract}
Next-generation gravitational-wave (GW) observatories will provide crucial insights into the nature of neutron star (NS) matter and the cosmological expansion history.
We estimate the number of multimessenger detections of binary neutron stars (BNS) with the Einstein Telescope (ET) and Cosmic Explorer (CE) and project the resulting constraints on the equation of state (EOS), BNS mass distribution, and cosmology using joint hierarchical Bayesian inference.
Our results rely on specific assumptions for the merger rate, EOS, and mass distribution; we consider two different mass functions, a narrow one centred around 1.4~$\msun$ and a wide one ranging from 1.1 to 2~$\msun$, and a merger rate with a local value of 107~Gpc$^{-3}$~yr$^{-1}$.
Under this setup, our mock follow-up algorithm results in between $\sim40$ and $\sim100$ short-term electromagnetic counterparts per year for ET, depending on the detector layout and mass distribution.
In a joint network with CE, this number ranges from $\sim 100$ to $\sim500$.
Additionally, we project tens to hundreds of late-time afterglow detections from gamma-ray bursts or kilonovae (KNe).
Based on the generated multimessenger events, we perform an injection campaign to hierarchically recover the EOS, mass distribution, and cosmology in a fully Bayesian framework.
Focusing on KN events from ET alone, the multimessenger data from our largest analysed catalogue with 78 events can constrain the canonical NS radius $R_{1.4}$ to $\lesssim 0.2$~km and the Hubble constant $H_0$ to $\lesssim 1$~km~s$^{-1}$~Mpc$^{-1}$, at least in absence of systematic modelling offsets.
By comparing inference results that are solely based on GW data and those incorporating light-curve information, we find that while the latter has negligible impact on the EOS constraints, it can benefit the inference of cosmological parameters.
\end{abstract}

\begin{keywords}
neutron star mergers -- gravitational waves -- equation of state -- cosmological parameters -- methods: data analysis.
\end{keywords}
\acresetall

\section{Introduction}
\label{sec:Intro}

Multimessenger signals from \acp{BNS} provide the opportunity to constrain fundamental physical parameters such as the \ac{EOS} for dense nuclear matter or the cosmological expansion rate.
While contemporary \ac{GW} detectors and \ac{EM} telescopes successfully identified GW170817~\citep{LIGOScientific:2017vwq, LIGOScientific:2017ync} together with the associated \ac{GRB} GRB170817~\citep{LIGOScientific:2017zic, Goldstein:2017mmi, Savchenko:2017ffs}, the \ac{KN} AT2017gfo~\citep{Coulter:2017wya, Tanvir:2017pws, Valenti:2017ngx, Lipunov:2017dwd, DES:2017kbs, Arcavi:2017vbi}, and the GRB afterglow~\citep{Troja:2017nqp, Margutti:2017cjl, Haggard:2017qne, Hallinan:2017woc, Alexander:2017aly}, a significant increase in multimessenger \ac{BNS} detections is expected to occur only with next-generation \ac{GW} detectors~\citep{Punturo:2010zz, Reitze:2019iox, Evans:2021gyd, ET:2019dnz, Branchesi:2023mws}.

However, it remains uncertain how many bright \ac{BNS} mergers can be anticipated.
This is due to several factors, most notably the largely unknown merger rate and modelling systematics in the \ac{EM} emission.
The current estimate for the local \ac{BNS} merger rate from the fifth \ac{GW} catalogue is 5.1--154.7~Gpc$^{-3}$yr$^{-1}$~\citep{LIGOScientific:2026ctl}, meaning detection prospects for \ac{BNS} \ac{GW} signals span almost two orders of magnitude.
Furthermore, how many \ac{BNS} \ac{GW} signals will be accompanied by a successfully identified \ac{EM} counterpart depends on the brightness of the \ac{KN}, \ac{GRB}, and afterglow emission.
The emission of the quasi-thermal \ac{KN} stems from the decay heat produced by the $r$-process nuclei synthesized in the mildly relativistic ejecta from the merger, but the total amount of ejecta is highly sensitive to the \ac{NS} masses, spins, and \ac{EOS}~\citep[e.g.][]{Radice:2018pdn, Kiuchi:2019lls, Kruger:2020gig, Nedora:2020qtd, Neuweiler:2025klw}.
Therefore, the uncertainty in the \ac{BNS} mass distribution and \ac{EOS} directly impacts estimates for the amount of observable \acp{KN}.
Moreover, current models of the \ac{KN} emission itself suffer from several unknown microphysical systematics~\citep{Sarin:2024tja, Brethauer:2024zxg}, further complicating estimates of the peak flux and, hence, the total \ac{KN} count.
Likewise, the process behind the launch of the ultra-relativistic component as well as the microphysical processes leading to the GRB prompt and afterglow emission remain poorly understood~\citep{Wang:2015vpa, Aksulu:2021crt, Ruiz:2021gsv, Miceli:2022efx, Hayashi:2024jwt}.

At the same time, estimating the frequency of multimessenger detections is of high interest, since these events offer significant scientific potential.
Among other results, GW170817 has proven how joint analysis of multimessenger signals can deliver improved constraints on the \ac{EOS} and Hubble constant compared to inference with one \ac{GW} signal alone~\citep[e.g.][]{LIGOScientific:2017adf, Radice:2017lry, Hotokezaka:2018dfi, Dietrich:2020efo,  Nicholl:2021rcr, Raaijmakers:2021uju, Wang:2020vgr, Pang:2022rzc}.
It is expected that because of the increasing detection frequency and overall better sensitivity, next-generation \ac{GW} interferometers can provide even better constraints on the \ac{BNS} population~\citep{Singh:2021zah}, the \ac{EOS}~\citep{Breschi:2021xrx, Gupta:2022qgg, Walker:2024loo, Finstad:2022oni,Iacovelli:2023nbv}, and the Hubble constant~\citep{Zhang:2020axa, Mitra:2020vzq, Chatterjee:2021xrm, Califano:2022syd, Han:2025fii}.
Reconstructing these universal physical parameters from an ensemble of observed events is not trivial though, as it requires careful statistical analysis to hierarchically infer them from the measured data.
Since this method involves several computational challenges, it is not just the detection rates that are uncertain, but also the projections about statistical accuracies with which these general parameters can be constrained.

Recently, several studies have assessed the prospects and challenges of multimessenger detections of \acp{BNS} in the next-generation era.
For instance, \citet{Loffredo:2024gmx} investigate the number of optical counterparts that could be identified with the Rubin telescope after \ac{GW} detection from \ac{CE} or \ac{ET}, assuming different \acp{EOS}, mass distributions, and merger rates.
\citet{Colombo:2025sdm} study how many neutron star-black hole and \ac{BNS} systems produce sufficiently bright radio, optical, X-ray, or gamma-ray counterparts for detection.
\citet{Bisero:2025tkw} focus on synergies between wide-field spectroscopic facilities and \ac{GW} observations of \acp{BNS}.
A projection of joint \ac{GW} and high-energy \ac{EM} detections is given in \citet{Ronchini:2022gwk}, analysing also the prospect of mutual localization.
\citet{Kaur:2024yag} assess the efficiency of prompt GRB detections and broadband afterglow searches in the upcoming O5 observing run of the LIGO-Virgo-KAGRA Scientific collaboration and for a next-generation observing period, while \citet{Dobie:2021qya} and \citet{Merfeld:2025els} focus on radio follow-up of \ac{BNS} mergers for next-generation \ac{GW} astronomy.
\citet{Steinle:2025xae} model how stellar population synthesis affects the set of observable multimessenger \ac{BNS} events.

Some works focus on the recovery of hierarchical physical parameters from these multimessenger events.
\citet{Han:2025fii} use \ac{FIM} posteriors from their projected multimessenger population to analyse to what extent cosmological parameters can be inferred.
\citet{Khadkikar:2025ith} assess how well \ac{NS} radii can be constrained after a hundred \ac{BNS} detections with next-generation detectors.
Several works are also concerned with joint Bayesian inference of the \ac{BNS} mass distribution and \ac{EOS}~\citep{Wysocki:2020myz, Golomb:2021tll, Golomb:2024lds, Biswas:2024hja}, but focus exclusively on the \ac{GW} signal.
\citet{Ghosh:2024cwc} perform joint hierarchical inference of the \ac{EOS}, mass distribution, and cosmology from \ac{BNS} signals in O5 without an \ac{EM} counterpart.

In this work, we provide a comprehensive overview of how many and which types of \ac{BNS} multimessenger signals can be detected with next-generation \ac{GW} detectors.
Subsequently, we also show how well hierarchical parameters such as the \ac{EOS}, mass distribution, or $H_0$ can be inferred from them.
To this end, we create different mock catalogues of the merging \ac{BNS} population for a one-year observation run.
Based on various assumptions, we forward-model the \ac{EM} observables, i.e. generate predictions for the prompt \ac{GRB}, \ac{KN}, \ac{GRB} afterglow, and \ac{KN} afterglow for each \ac{BNS}, and construct a mock observation algorithm to mimic how different \ac{EM} telescopes would perform targeted follow-up observations.
We compare the expected rates for two different \ac{BNS} mass distributions, assuming different scenarios for the \ac{ET} and \ac{CE} detectors.
Thereafter, we use the resulting joint \ac{GW} and \ac{KN} `detections' in an injection-recovery campaign to project the possible constraints on the \ac{EOS}, mass distribution, and cosmological parameters using full Bayesian hierarchical inference.

The article is structured as follows:
Section~\ref{sec:events} describes how we generate a mock population of \ac{BNS} multimessenger events.
In Sec.~\ref{sec:observations}, we implement a detection algorithm to assess how many joint \ac{GW} and \ac{EM} detections can be expected from our catalogues with different detector and telescope configurations.
Section~\ref{sec:hierarchical_inference} then focuses on reconstructing general physical parameters from multimessenger events using hierarchical inference.
We provide a summary and discussion of our results in Sec.~\ref{sec:conclusion}.
Throughout the paper, we write $m_1, m_2$ to refer to the \textit{source frame} component masses of a \ac{NS} binary; \textit{detector frame} masses carry a superscript $m_1^{\text{det}}, m_2^{\text{det}}$.

\section{Multimessenger event catalogues}
\label{sec:events}
We create two different catalogues for the \ac{BNS} merger population. 
The catalogues are identical except for the population models used to draw the \acp{NS} masses and spins.
We use fitting and ad hoc prescriptions to deduce the \ac{KN}, \ac{GRB}, and afterglow emission.
The following subsections provide the details on how the \ac{BNS} catalogues and \ac{EM} observables were generated.

\subsection{BNS catalogues}
\label{subsec:bns_catalogues}
\begin{figure}
    \centering
    \includegraphics[width=\linewidth]{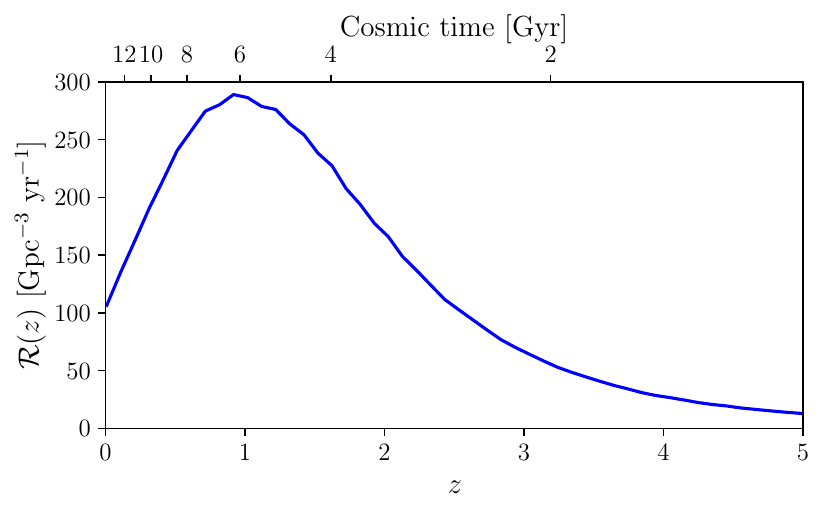}
    \caption[BNS merger rate density for mock catalogues]{\ac{BNS} merger rate density $\mathcal{R}(z)$ used to generated our \acp{BNS}.
    Its evolution as a function of redshift follows from Eqs.~\eqref{eq:merger_rate_density} and \eqref{eq:delay-time_distribution} with $\mathcal{R}_{\rm n} = 50$~Gpc$^{-3}$~yr$^{-1}$ and $\tau=3$~Gyr.
    The local merger rate density is $\mathcal{R}(0) = 107$~Gpc$^{-3}$~yr$^{-1}$.}
    \label{fig:merger_rate}
\end{figure}
\begin{figure}
    \centering
    \includegraphics[width=\linewidth]{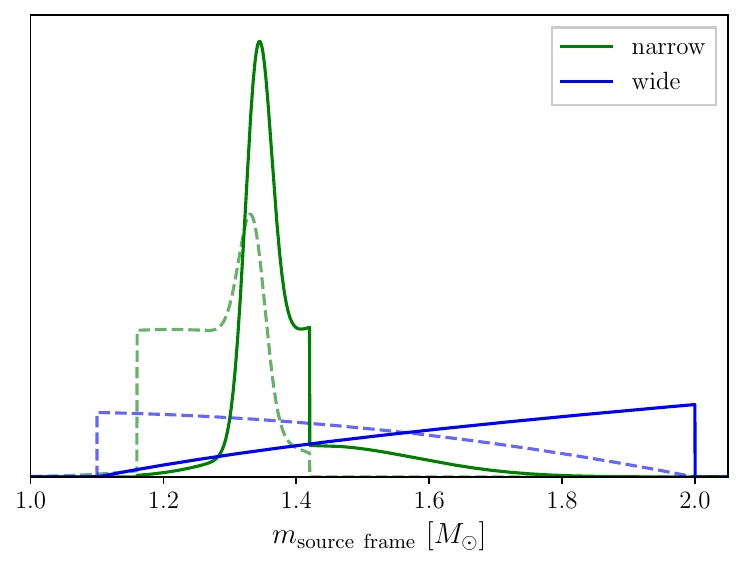}
    \caption[BNS mass distributions for mock catalogues]{Narrow (green) and wide (blue) mass distributions used to generate our \acp{BNS}. 
    The solid lines show the marginalized mass distribution of the heavier \ac{NS} and the dashed lines of the lighter one.}
    \label{fig:mass_distribution}
\end{figure}
We model the \ac{BNS} merger rate density as
\begin{align}
    \dot{n}(z, m_1, m_2, \chi_1, \chi_2) = R_{\text{obs}}(z) p(m_1, m_2, \chi_1, \chi_2)\ .
\end{align}
Here, the last factor represents the intrinsic mass and spin distribution, assumed to be independent of redshift, and $R_{\text{obs}}(z)$ is the merger rate in observer frame, i.e., the number of \ac{BNS} mergers at redshift $z$ per observer frame year.
This quantity is related to the comoving source-frame merger rate density $\mathcal{R}(z)$, i.e., the number of \ac{BNS} mergers at redshift $z$ per comoving volume per source frame year, via 
\begin{align}
    \label{eq:BNS_observation_rate}
    R_{\text{obs}}(z) &= \frac{1}{1+z} \frac{d V_c}{dz} \mathcal{R}(z)\ .
\end{align}
It is commonly assumed that the source frame merger rate density is a convolution of the cosmic star formation rate $\psi(z)$~\citep{Madau:2014bja} with a delay-time distribution $\mathcal{D}(z\mid z')$ that quantifies how many stars formed at redshift $z'$ will result in merging \acp{BNS} at redshift $z$:
\begin{align}
        \mathcal{R}(z) &= \mathcal{R}_{\rm n} \times \int_z^{\infty} \psi(z') \mathcal{D}(z \mid z')\ dz'\ .
        \label{eq:merger_rate_density}
\end{align}
Several forms for the delay-time distribution have been suggested in the literature~\citep{Wanderman:2014eza, Safarzadeh:2019znp}.
We adopt a simple exponential expression~\citep{Vitale:2018yhm, Yi:2021wqf, Du:2025wto}
\begin{align}
    \mathcal{D}(z\mid z') = \frac{1}{\tau} \exp\left(-\frac{t(z) - t(z')}{\tau} \right) \frac{dt}{dz}\ ,
    \label{eq:delay-time_distribution}
\end{align}
where $t(z)$ is the cosmic time as a function of redshift, and $\tau$ and $\mathcal{R}_{\rm n}$ are hyperparameters that we set to $\tau = 3$~Gyr, and $\mathcal{R}_{\rm n} = 50$~Gpc$^{-3}$~yr$^{-1}$~\citep{Du:2025wto}.
We further assume a \ac{LCDM} cosmology consistent with the Planck 2018 results, i.e., $H_0 = 67.66$ km s$^{-1}$ Mpc$^{-1}$ and $\Omega_m =0.31$~\citep{Planck:2018vyg}.
The resulting $\mathcal{R}(z)$ is shown in Fig.~\ref{fig:merger_rate} and corresponds to a local merger rate of $\mathcal{R}(0) = 107$~Gpc$^{-3}$~yr$^{-1}$, in agreement with the current estimate 5.1--154.7~Gpc$^{-3}$~yr$^{-1}$ from \ac{GW} observations~\citep{LIGOScientific:2026ctl}, as well as constraints from short GRB observations~\citep{Salafia:2022xjd, Du:2025wto, DeSantis:2026wfw}, and the Milky Way \ac{NS} population~\citep{Pol:2020tfz, Grunthal:2021kqg}.
All results and absolute event numbers quoted in subsequent sections are thus relative to our specific choice for the merger rate density.
In particular, when we draw the population from the merger rate, we receive 90\,510~\ac{BNS} signals from the entire observable universe that reach Earth in one year.

For the \ac{BNS} mergers drawn this way, we use two different mass and spin distributions.
In one instance, we use a mass-population model inferred from the Galactic double pulsars~\citep{Farrow:2019xnc}.
We refer to this mass model as `narrow' mass distribution.
This model assumes that one \ac{NS} of the \ac{BNS} is a recycled component, spun up by accretion from the progenitor star of the other \ac{NS}.
The other \ac{NS} is called the slow component.
The masses are denoted by $m_r$ and $m_s$ respectively and follow different distributions
\begin{align}
    p(m_r) &= w \frac{\exp\left(-\frac{(m_r-\mu_1)^2}{2\sigma_1^2}\right)}{\sqrt{2\pi} \sigma_1}+ (1-w) \frac{\exp\left(-\frac{(m_r-\mu_2)^2}{2\sigma_2^2}\right)}{\sqrt{2\pi} \sigma_2}\ , \label{eq:narrow_mass_model_r}\\
    p(m_s) &= \frac{1}{m^u  - m^l} \quad \text{if } m^l \leq m_s \leq m^u\ ,
\label{eq:narrow_mass_model_s}
\end{align}
where the hyperparameters $\mu_1 = 1.34~\msun$, $\mu_2 = 1.47~\msun$, $w=0.68$, $\sigma_1= 0.02~\msun$, $\sigma_2=0.15~\msun$, $m^l = 1.16~\msun$, $m^u=1.42~\msun$ are the median values from a Bayesian fit to the masses of 18 galactic \acp{BNS}~\citep{Farrow:2019xnc}.
It has been shown that the spin of the slow component at merger is negligible~\citep{Zhu:2017znf, Zhu:2020zij}, thus we set $\chi_s = 0$.
On the other hand, since the spin of the recycled \ac{NS} arises mainly from accretion of the companion star in the orbital plane, it is aligned to the orbital axis.
If the ensuing supernova of the companion star is not too asymmetric and does not produce large kicks, the spins of the \acp{NS} will stay aligned with the orbit~\citep{Podsiadlowski:2003py, Miller:2014aaa}.
The spin inclinations from the known Galactic \acp{BNS} support that this is the case~\citep{Stairs:2004ye, Ferdman:2013xia}.
Following \citet{Zhu:2020zij}, we draw the aligned spin component from a gamma distribution with a shape parameter of 2 and a scale parameter of $0.012$:
\begin{align}
   \chi_r \sim \Gamma(\alpha=2, \lambda=0.012)\ .
\end{align}
Although the recycled pulsar is often the heavier one, the distributions in Eqs.~\eqref{eq:narrow_mass_model_r} and~\eqref{eq:narrow_mass_model_s} do not enforce this.
To adhere to common \ac{GW} nomenclature, we label the larger component mass $m_1$, and the smaller one $m_2$
\begin{align}
    m_1,\ m_2 &= \max(\{m_r, m_s\}),\ \min(\{m_r, m_s\})\ ,
\end{align}
and relabel the spins accordingly.

The other mass model is taken from \citet{Landry:2021hvl}, which is based on the two confirmed \acp{BNS} candidates from \acp{GW} observations, and takes the form
\begin{align}
    p(m_1, m_2) \propto (m_2/m_1)^\alpha \quad \text{if} \quad m_{\text{min}} \leq m_2 \leq m_1 \leq m_{\text{max}}\ ,
\label{eq:wide_mass_model}
\end{align}
with hyperparameters $m_{\text{min}} = 1.1~\msun$, $m_{\text{max}}=2.0~\msun$, and the power-law index $\alpha=2$.
This population model is referred to as `wide' mass distribution.
Additionally, we employ a wider spin distribution for the component's spins, while at the same time sticking to the assumption of orbital spin alignment for simplicity and reduction of computational cost.
Specifically, we draw $\chi_1$ and $\chi_2$ from a normal distribution
\begin{align}
    \chi_j \sim \mathcal{N}(\mu=0, \sigma=0.05)\ .
\end{align}
In this sense, the narrow mass model represents a scenario in which the merging \ac{BNS} population follows trends from the galactic \ac{BNS} population and isolated binary evolution, whereas the wide mass model represents a more agnostic view on the origin and properties of \acp{BNS}.
The resulting marginalized mass distributions for our narrow and wide catalogue are shown in Fig.~\ref{fig:mass_distribution}.

To obtain values for the \ac{NS} radii and tidal deformabilities, we use a Quantum-Monte-Carlo relativistic mean field model~\ac{EOS}, namely QMC-RMF3 from \citet{Alford:2022bpp, Alford:2023rgp} and \citet{Chatterjee:2025_eos}.
The \ac{TOV} mass is $M_{\text{TOV}} =2.14$~$\msun$ and its radius and tidal deformability for a canonical 1.4~$\msun$ \ac{NS} are $R_{1.4}=12.16$~km and $\Lambda_{1.4}=383$.
This \ac{EOS} was recently used in \ac{NR} simulations~\citep{Neuweiler:2025klw} and is compatible with common nuclear and astrophysical constraints.
Other \ac{GW} parameters such as inclinations and sky positions are drawn isotropically, coalescence phases $\phi_c$ uniformly in $(0, 2\pi)$, and polarization angles $\psi$ uniformly in $(0, \pi)$.
The merger times are spread uniformly throughout the year 2050.
We emphasize that our two catalogues differ only in the values drawn from the \ac{NS} mass and spin distribution, all remaining parameters, e.g. EOS, distances, inclinations, etc., are kept identical.

\subsection{Kilonova light curves}
\label{subsec:kilonova}
\begin{figure}
    \centering
    \includegraphics[width=\linewidth]{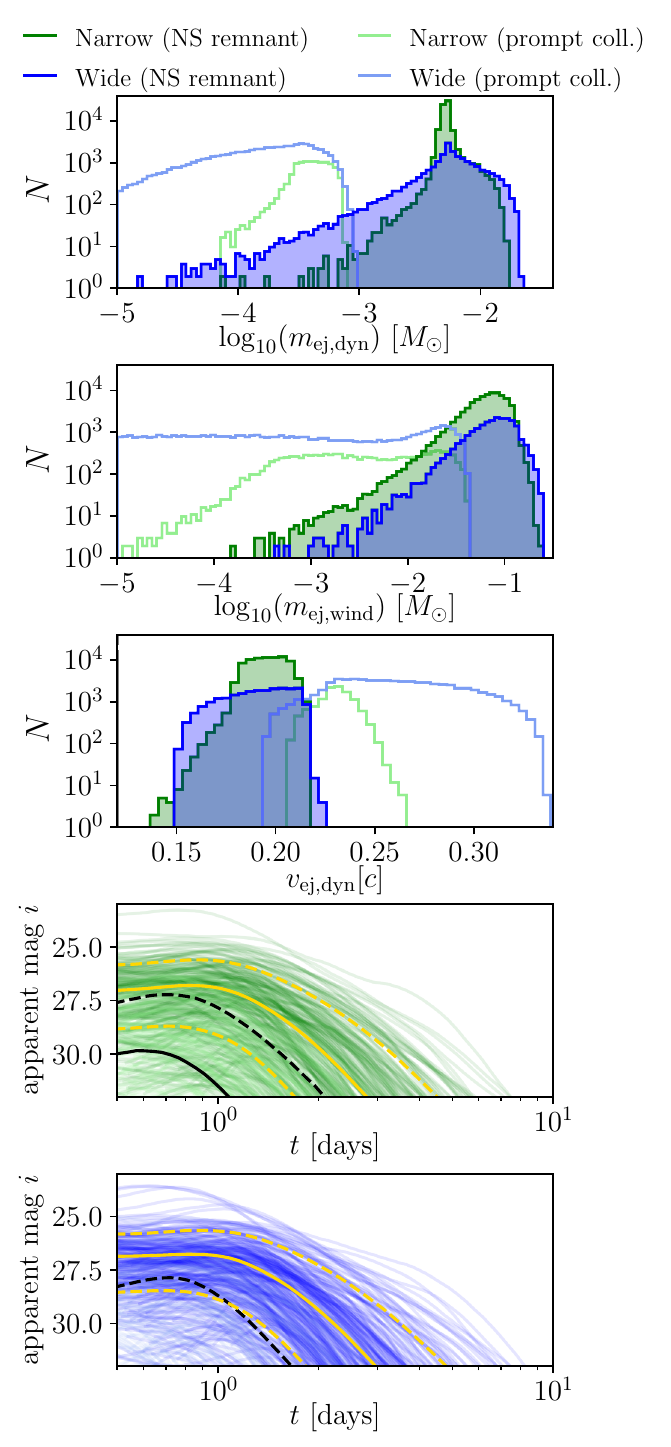}
    \caption[Ejecta properties of BNS mock catalogues]{Ejecta properties for the \ac{BNS} catalogues with the narrow (green) and wide (blue) mass distributions. 
    The histograms with no filling and lighter colours represent \acp{BNS} that undergo prompt collapse.
    The distributions of 300 apparent $i$ band light curves for the narrow (penultimate panel) and wide (last panel) distributions are also shown.
    There, black (orange) solid lines are the median magnitude for \acp{KN} within $z\leq1.5$ and with (without) prompt collapse, and the dashed lines are the corresponding upper and lower 68\% magnitude quantiles.}
    \label{fig:kn_population}
\end{figure}
To calculate the resulting \ac{KN} light curves from the \ac{BNS} mergers in our catalogues, we use a machine-learning surrogate trained on flux densities from the 3D-Monte-Carlo radiation transport code \textsc{possis}~\citep{Bulla:2019muo, Bulla:2022mwo}.
Specifically, we extend the parameter range of the existing surrogate from \citet{Koehn:2025zzb} to accommodate a wider range for the dynamical and wind ejecta masses and higher dynamical velocities.
In Appendix~\ref{app:possis_geometry}, we explain how we adapted the idealized geometry implemented in \textsc{possis}~\citep{Kawaguchi:2019nju, Bulla:2022mwo} to cover this extended parameter space.
The inputs for our \ac{KN} model are the masses, mass-averaged velocities, and electron fractions of the dynamical and wind ejecta components, as well as the inclination of the binary.
To determine these input parameters for each of our \acp{BNS}, we use two different \ac{NR}-informed fitting prescriptions.
Which set of fitting formulas is used depends on whether the binary is expected to undergo prompt collapse.
Prompt collapse is defined here as collapse to a black hole within 2~ms after merger~\citep{Kolsch:2021lub, Puecher:2024dhl}, but see~\citet{Ecker:2024kzs} for an alternative criterion.
\ac{NR} simulations have shown that if the total mass of a \ac{BNS} exceeds a certain \ac{EOS}-dependent threshold, the remnant system will collapse promptly~\citep[e.g.][]{Bauswein:2013jpa, Kolsch:2021lub, Cokluk:2023xio, Kashyap:2021wzs, Schianchi:2024vvi, Ecker:2024kzs, Dhani:2025yxf} and produce significantly less ejecta than if a (meta-)stable remnant \ac{NS} was formed~\citep{Sarin:2020gxb}.
However, the condition for prompt collapse also depends on other parameters such as mass ratio $q$ and spin.
Therefore, we use the machine-learning classifier from~\citet{Puecher:2024dhl}, trained on a set of \ac{NR} simulations, to assess whether a binary promptly collapses.
If the classifier predicts prompt collapse, we use Eqs.~(9)--(11) from \citet{Loffredo:2024gmx}
\begin{subequations}
\label{eq:pc_fittings}
\begin{align}
    m_{\text{ej, dyn}} &= a \tilde{\Lambda} (q^{-1}- b)e^{\frac{c}{q}} \ ,\\
    v_{\text{ej, dyn}} &= a' \frac{m_1}{m_2} \left(1 + b' \frac{G m_1}{c^2 R_1} \right) + (1 \leftrightarrow 2) + c' \ ,\\
    \log_{10}(m_{\text{disc}}) &= \min(-1, a'' + b''q +c''\tilde{\Lambda}q^2) \ ,
\end{align}
\end{subequations}
with numerical coefficients
\footnote{In the published version of \citet{Loffredo:2024gmx}, there was a typo for $b''$ in their Eq.~(10). We confirmed $b''=-13.4$ through author correspondence.}
$a=1.25\times10^{-4},\ b=0.982,\ c=-2.44,\ a'=-0.395,\ b'=-1.627,\ c'=0.798,\ a''=7.70,\ b'' =-13.4,\ c''=8.16\times10^{-3}$.
Otherwise, we assume that the \ac{BNS} at least briefly produces some (meta-)stable \ac{NS} remnant and we use the following relations from \citet{Radice:2018pdn} and \citet{Kruger:2020gig} to determine the dynamical ejecta properties
\begin{subequations}
\label{eq:nopc_dyn}
\begin{align}
    \frac{m_{\text{ej, dyn}}}{10^{-3}~\msun}&= m_1 \left( a \frac{c^2 R_1}{G m_1} +b \frac{G m_1}{c^2 R_1} + c \left(\frac{m_2}{m_1}\right)^n  \right) + (1\leftrightarrow2) \\
    v_{\text{ej, dyn}} &= a' \frac{m_1}{m_2} \left( 1 + b' \frac{G m_1}{c^2 R_1} \right) + (1\leftrightarrow2) + c'\ ,
\end{align}
\end{subequations}
where $a=-9.33,\ b=-337.56,\ c=114.17,\ n=1.55,\ a'=-0.287,\ b'=-3.0,\ c'=0.49$.
The disc mass fit is taken from \citet{Dietrich:2020efo} and \citet{Pang:2022rzc}
\begin{subequations}
\label{eq:nopc_disc}
\begin{align}
    &\log_{10}(m_{\text{disc}}) = \text{max}\left(-3, a \left[1+ b \tanh\left(\frac{c-(m_1+m_2)/M_{\text{coll}}}{d}\right)\right]\ \right),\\
    &M_{\text{coll}} = \left(-3.606 \frac{G M_{\text{TOV}}}{c^2 R_{1.6}} +2.38 \right) \mtov \ ,\\
    &a = a_0 + \Delta a\xi,\ b=b_0 + \Delta b\xi,\ \xi =\frac{1}{2} \tanh\left(\beta (q-q_{\text{trans}})\right)\ ,
\end{align}
\end{subequations}
where $\mtov=2.14~\msun$ is the TOV mass of our \ac{EOS}, $R_{1.6}=12.08$~km is the corresponding radius of a 1.6~$\msun$ \ac{NS}, and $a_0=-1.725,\ \Delta a=-2.337,\ b_0=-0.564,\ \Delta b=-0.437,\ c=0.958,\ d=0.057,\ \beta=5.879,\ q_{\text{trans}}=0.886$.
Wind ejecta arise when parts of the disc are blown away by neutrino interaction, thermal advection, as well as magnetic, hydrodynamic, and viscous effects~\citep{Dessart:2008zd, Fernandez:2015use, Perego:2014fma, Siegel:2017nub, Nedora:2019jhl, Sarin:2020gxb, Ciolfi:2020wfx}. 
However, \ac{NR} simulations struggle to quantify how much of the disc will be ejected, mainly because of the difficult treatment of the relevant physical effects and the long time-scales ($\gtrsim$100~ms) on which these processes can take place.
Therefore, we employ an ad hoc prescription 
\begin{align}
    m_{\text{ej, wind}} = \zeta m_{\text{disc}}
    \label{eq:disc_to_wind_mass}
\end{align}
where $\zeta$ is drawn from a truncated normal distribution $\mathcal{TN}$.
If the \ac{BNS} collapsed promptly, we use~\citep{Siegel:2017nub, Fernandez:2018kax, Fahlman:2022jkh},
\begin{align}
    \zeta\sim\mathcal{TN}(\mu=0.2, \sigma=0.1, \zeta_{\min}=0., \zeta_{\max}=0.4)\ ,
\end{align} 
whereas if there is a \ac{NS} remnant, we assume that the higher neutrino flux enhances matter ejection from the disc through various mechanisms~\citep{Nedora:2019jhl, Nedora:2020qtd, Bernuzzi:2024mfx, Neuweiler:2025klw}, setting 
\begin{align}
    \zeta\sim\mathcal{TN}(\mu=0.5, \sigma=0.2, \zeta_{\min}=0., \zeta_{\max}=0.8)\ .
\end{align}
For the dynamical (wind) electron fraction, we simply draw uniform values between 0.15 and 0.35 (0.2 and 0.4). 
The wind ejecta velocities are drawn uniformly between 0.05 and 0.15~$c$.

In Fig.~\ref{fig:kn_population}, we show the resulting distribution of ejecta parameters and the associated KN light curves as determined from the \textsc{possis} surrogate.
Not surprisingly, \acp{BNS} undergoing prompt collapse produce significantly fainter \acp{KN}.
For the narrow mass distribution, about 12\% of \acp{BNS} undergo prompt collapse, while for the wide mass distribution it is 75\%. 
For both mass distributions, the subset of \acp{BNS} without prompt collapse is confined to the mass range $\lesssim 1.5~\msun$ and as a consequence the resulting \ac{KN} light curves from this subset have similar brightness.
This is indicated by the median magnitudes in the bottom panels of Fig.~\ref{fig:kn_population}.
However, the wide mass model overall produces much fewer bright \acp{KN}.
Out of the total of 90,510 \acp{BNS}, only $\approx4300$ light curves are brighter than 26~mag in $i$ band, compared to $\approx12,000$ light curves from the narrow mass model.

While Eqs.~(\ref{eq:pc_fittings})--(\ref{eq:nopc_disc}) allow us to predict the \ac{KN} light curves associated to our \ac{BNS} population, we point out that there are significant differences between different fitting prescriptions available in the literature~\citep{Kruger:2020gig, Raaijmakers:2021slr, Henkel:2022naw}.
Moreover, ejecta masses are sensitively dependent on the initial spins of the \acp{NS}~\citep{Dietrich:2016lyp, Most:2019pac, Rosswog:2023rqa, Neuweiler:2025klw}, which is not accounted for by our fitting procedure. 
Further, the ad hoc prescription for the ratio between wind and disc mass in Eq.~\eqref{eq:disc_to_wind_mass} is especially problematic.
The real wind ejecta distribution of merging \acp{BNS} might deviate substantially from the one shown in Fig.~\ref{fig:kn_population}, if our values for $\zeta$ are misestimated.
Because most of the bright detectable KN light curves in our catalogues arise from large wind ejecta (the dynamical ejecta play a subordinate role in most cases), this would affect our projected numbers for the \ac{KN} detections in later sections significantly.
In Appendix~\ref{app:reduced_wind_ejecta}, we therefore quantify how the \ac{KN} count is affected when the amount of wind ejecta is halved for each \ac{BNS}.

\subsection{GRB emission and GRB afterglow}
\label{subsec:grb}
\begin{figure}
    \centering
    \includegraphics[width=1\linewidth]{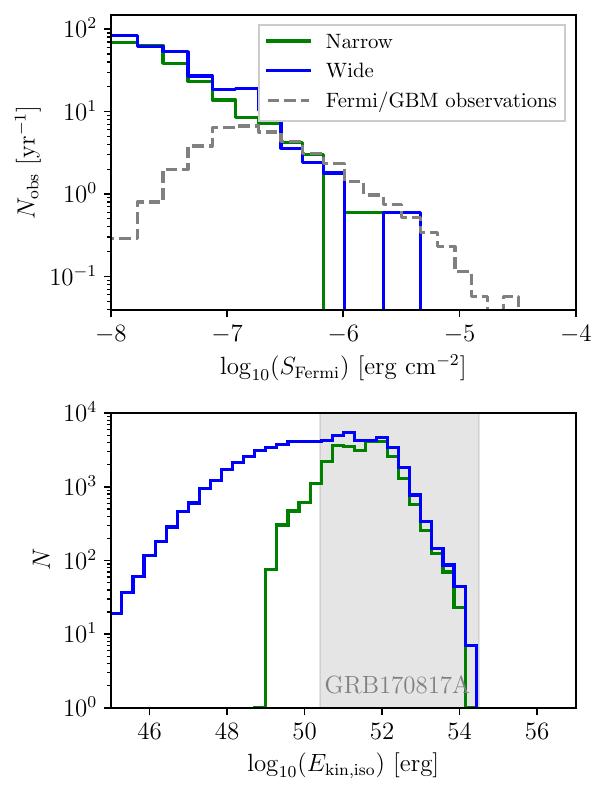}
    \caption[GRB properties of the BNS mock catalogues]{Top: Distribution of the observable bolometric fluence between 0.1 and 1~MeV for the narrow (green) and wide (blue) \ac{BNS} mass distribution.
    For comparison, we show the fluence distribution of the short \acp{GRB} observed by \textit{Fermi}/GBM as grey dashed line. 
    The rates from our catalogues are scaled to a duty cycle factor of 0.6.
    Bottom: Isotropic kinetic energy-equivalent. The grey band indicates the 95\% credibility interval from GRB170817A~\citep{Koehn:2025zzb}.}
    \label{fig:grb_population}
\end{figure}

We assume that a \ac{BNS} produces a short \ac{GRB} when a black hole central engine is able to launch a jet.
Hence, two conditions have to be fulfilled by a particular \ac{BNS} in our catalogues to produce a \ac{GRB}.
First, we require that the remnant eventually collapses to a \ac{BH}, so we check if the \ac{BNS} either promptly collapses or has a total baryonic mass larger than the baryonic Kepler mass limit $M_{\text{Kep, bar}} = 2.918~\msun$ of our QMC-RMF3 \ac{EOS}.
This criterion serves to exclude cases, where the remnant never collapses to a \ac{BH}~\citep{Murguia-Berthier:2016fys, Metzger:2018szx, Beniamini:2020adb, vanPutten:2022xyx, Kalinani:2025itu}, although it has been suggested that magnetar remnants can also power long \acp{GRB}~\citep{DallOsso:2023gdk, Gottlieb:2024mwu}.
Secondly, the jet energy must exceed the threshold required for a breakout.
We use Eq.~(20) from \citet{Duffell:2018iig} with an initial opening $\theta_{\text{init}}=0.4$~rad,
\begin{align}
    E_{\text{jet}} > 0.05\, \theta_{\text{init}}^2 \times \frac{1}{2}m_{\text{ej, dyn}} v_{\text{ej, dyn}}^2  \ .
\end{align}
To estimate the jet energy $E_{\text{jet}}$, we assume a certain fraction of the disc mass is converted through the Blandford-Znajek process~\citep{Blandford:1977ds, Hayashi:2024jwt, Kalinani:2025itu}, though see for instance \citet{Salafia:2020jro} invoking a discussion about neutrino-annihilation-powered jets.
Thus, we can write the jet energy as
\begin{align}
    E_{\text{jet}} = \frac{1}{2} \eta_{\text{BZ}}(\chi_\text{BH}) \times (m_{\text{disc}} - m_{\text{ej, wind}})c^2\ ,
\end{align}
where the factor of $1/2$ accounts for the counter jet and we used the jet efficiency~\citep{Tchekhovskoy:2009ba, Colombo:2022zzp, Colombo:2025sdm}
\begin{subequations}
\begin{align}
    \eta_{\text{BZ}}(\chi_\text{BH}) &= \eta_0\ \Omega_H^2 \left(1 + 1.38\Omega_H^2 -9.2\Omega_H^4\right) \label{eq:blandford_znajek_a} \\
    \Omega_H &= \frac{\chi_{\text{BH}}}{2\left(1+\sqrt{1-\chi_{\text{BH}}^2}\right)}\ .
\end{align}
\label{eq:blandford_znajek}
\end{subequations}
The efficiency depends on the black hole spin $\chi_{\text{BH}}$~\citep{Coughlin:2018fis}
\begin{align}
    \chi_{\text{BH}} = \tanh\left(a \frac{m_1^2m_2^2}{(m_1+m_2)^4} (m_1 + m_2 + b \tilde{\Lambda}) +c\right)\ ,
\end{align}
with\footnote{In the published version of \citet{Coughlin:2018fis}, there was a typo for $a$ and $b$ in their Eq.~(4).
We confirmed $a=0.537/0.25^2=8.592$ and $b=-0.185/400=-4.625\times10^{-4}$ through author correspondence.}
$a=8.592$, $b=-4.625\times10^{-4}$, $c=-0.514$.
Moreover, we have tuned the fudge factor in Eq.~\eqref{eq:blandford_znajek_a} to $\eta_0=0.016$, so that the resulting fluence distribution matches the fluences of the short \acp{GRB} observed with the Gamma-ray Burst Monitor (\textit{Fermi}/GBM), following ~\citet{Colombo:2022zzp} and \citet{Loffredo:2024gmx}.
More details are provided below.

The two conditions to produce a GRB are met by 30\% (74\%) for \acp{BNS} from the narrow (wide) mass distribution.
If the jet is able to break out of the ejecta, we place some energy $E_\gamma$ into the prompt emission, whereas the remaining energy goes into the kinetic energy $E_{\text{kin}}$ of the jet
\begin{align}
\label{eq:grb_energy_budget}
    E_{\text{jet}} = \eta_\gamma E_\gamma + (1-\eta_\gamma) E_{\text{kin}} \ .
\end{align}
Based on observations from long \acp{GRB}, it has been suggested that the gamma-ray efficiency $\eta_\gamma$ clusters around 0.15~\citep{Beniamini:2015eaa, Beniamini:2016hzc}. 
However, the efficiency of short \acp{GRB} might be smaller than for long \acp{GRB}~\citep{Aksulu:2021crt, Dreas:2026jte}.
For this reason, we randomly draw $\eta_\gamma$ uniformly from 0.01 to 0.15.
We further assume that after the breakout, the jet has a Gaussian structure
\begin{align}
    \frac{dE}{d\Omega} = 
        \begin{cases}
        \epsilon_c \exp\left(-\frac{\theta^2}{2\theta_c^2}\right)\qquad &\text{if $\theta\leq\theta_{\rm w}$} \\
        0 \qquad \qquad \qquad \qquad &\text{otherwise}
        \end{cases}\ ,
\label{eq:jet_energy_structure}
\end{align}
\begin{align}
    \Gamma(\theta) = 
        \begin{cases}
        1 + (\Gamma_c-1) \exp\left(-\frac{\theta^2}{2\theta_c^2}\right)\qquad \quad &\text{if $\theta\leq\theta_{\rm w}$} \\
        1 \qquad \qquad \qquad \qquad \qquad  &\text{otherwise}
        \end{cases}\ .
\end{align}
In doing so, we adopt a Gaussian structure for both the jet energy and bulk Lorentz factor $\Gamma$. 
The central Lorentz factor is always fixed at $\Gamma_c= 500$.
We note further that in our setup, $E_{\text{kin}}$ and $E_{\gamma}$ are assumed to both follow the geometry from Eq.~\eqref{eq:jet_energy_structure}.
The normalization $\epsilon_c$ in Eq.~\eqref{eq:jet_energy_structure} is then chosen such that the integrated energy is equal to the value determined from Eq.~\eqref{eq:grb_energy_budget}.
The jet core angles $\theta_c$ are drawn from the range reported in \citet{Escorial:2022nvp}, and we set the wing angle to $\theta_w = 2\theta_c$.
We further assume that the jet inclination aligns with the binary's angular momentum, although there are scenarios in which there can be a misalignment~\citep{Muller:2024wzl}.

To assess the observed energy of the prompt emission, we closely follow the method of \citet{Colombo:2022zzp, Colombo:2025sdm}.
In particular, we assume that the shape of the comoving spectral emissivity $J(\nu')$ is angle-independent and determine the isotropic prompt energy measured in the observer frequency band $(\nu_{\min}, \nu_{\max})$ as~\citep{Salafia:2015vla, Salafia:2019off}
\begin{subequations}
\begin{align}
    &E_{\gamma, \text{iso}} = \int_{\nu_{\min}}^{\nu_{\max}} d\nu \int d\Omega\ J\left(\nu'(\nu)\right)\frac{\delta(\theta, \varphi, \iota)^2}{\Gamma(\theta)}\frac{dE_\gamma}{d\Omega}\ , \\
    &J(\nu') = 0.368\ \nu'^{0.24} e^{-1.24\ \nu'/ \nu_p}\ , \\
    &\nu'(\nu) = \frac{(1+z)}{\delta(\theta, \varphi, \iota)} \nu\ , \\
    &\delta(\theta, \varphi, \iota) = \frac{1}{\Gamma(\theta)-  \sqrt{\Gamma(\theta)^{2}-1}\ \cos(\alpha(\theta, \varphi,\iota))} \ , \\
    &\cos(\alpha(\theta, \varphi,\iota)) = \cos(\varphi) \sin(\theta) \sin(\iota) + \cos(\theta) \cos(\iota)\ ,
\end{align}
\end{subequations}
where $J(\nu')$ is normalized, $\nu_p=3$~keV, and $\delta(\theta, \varphi, \iota)$ is the special-relativistic Doppler factor that depends on the angle between the line of sight and the fluid element at $(\varphi, \theta)$.
At a redshift $z$ and luminosity distance $d_L$, the bolometric fluence is then determined as
\begin{align}
S_{\nu_{\min} -\nu_{\max}} = \frac{1+z}{4\pi d_L^2} E_{\gamma, \text{iso}}\ .
\end{align}

To determine the GRB afterglow emission, we use a machine-learning surrogate based on \textsc{pyblastafterglow}~\citep{Nedora:2024vrv, Koehn:2025zzb}.
In \textsc{pyblastafterglow}, the emission is modelled as synchrotron radiation from accelerated electrons in the forward shock when the jet hits the ambient, constant-density interstellar medium.
The interstellar medium densities are drawn from an ad hoc uniform distribution~\citep{Fong:2015oha, Escorial:2022nvp}
\begin{align}
\label{eq:interstellar_density_distribution}
    \log_{10}(n_{\text{ism}}/\text{cm}^{-3}) \sim \mathcal{U}(-5, 0)\ .
\end{align}
The kinetic energy entering the afterglow calculation comes from Eq.~\eqref{eq:grb_energy_budget}, but because of restrictions in the training data set for the machine-learning surrogate, we restrict the central isotropic kinetic energy equivalent $E_{\text{kin,iso}} = 4\pi \frac{dE_{\text{kin}}}{d\Omega}_{|\theta=0}$ to be at least $10^{47}$~erg.
The microphysical parameters are kept fixed for each afterglow computation, i.e., we set the electron power law index to $p=2.15$, electron energy fraction to $\varepsilon_e =0.1$, and magnetic electron fraction to $\varepsilon_B = 10^{-3}$, in agreement with GRB170817A~\citep{Ghirlanda:2018uyx, Pang:2022rzc, Koehn:2025zzb}, although smaller values of $\varepsilon_B\sim10^{-4}$ have been suggested \citep{sadeh_synchrotron_2024,sadeh_nonthermal_2025}.

In Fig.~\ref{fig:grb_population}, we show the distribution of the fluence $S_{\text{Fermi}}$ that would be observed by \textit{Fermi}/GBM instrument and the distribution of $E_{\text{kin,iso}}$. 
To ensure consistency between our ad hoc prescriptions for the gamma-ray energy $E_\gamma$ and the observed short \ac{GRB} population, we tuned $\eta_0$ in such a way that our results roughly agree with the fluence distribution of short \textit{Fermi}/GBM \acp{GRB}~\citep{FERMI-GBM-catalog, vonKienlin:2020xvz}, the latter is shown as grey dashed line in the top panel of Fig.~\ref{fig:grb_population}.
The \textit{Fermi}/GBM instrument has a wide \ac{FOV} that is essentially only impaired by the occultation from Earth, therefore we assume a duty cycle of 0.6~\citep{Burns:2015fol} and rescale our fluence distributions accordingly.
Similarly, our kinetic energies do not exceed the credible interval inferred from GRB170817A~\citep{Koehn:2025zzb}, although we also note that compared to most short \acp{GRB} its kinetic energy is rather high~\citep{Fong:2015oha, Lamb:2020ccz, Escorial:2022nvp}.

Despite these modelling attempts, we emphasize that GRB launch mechanisms and the prompt emission origin are largely uncertain. 
In our setup, this is most tangible through the fudge factor $\eta_0$ for the Blandford-Znajek efficiency in Eq.~\eqref{eq:blandford_znajek} and the GRB prompt efficiency $\eta_\gamma$ in Eq.~\eqref{eq:grb_energy_budget}.
Since details about the prompt emission are currently unknown, tuning these parameters at least enforces some consistency with the observed GRB population. 
Nevertheless, the geometry of the emitting region and the comoving spectral shape could be very different from the kinetic energy profile~\citep{Rudolph:2023auv} and microphysical parameters for the afterglow shock are not necessarily confined to the canonical values from GRB170817A~\citep{Aksulu:2021crt}.
Additionally, the ad hoc distribution for the circumstellar densities in Eq.~\eqref{eq:interstellar_density_distribution} does impact the peak flux and peak time of the \ac{GRB} afterglows, thus having a direct impact on our projected number of detectable \ac{GRB} afterglows.
A more accurate picture of the \ac{BNS}-associated \ac{GRB} population might only be obtained after more observations have taken place.

\subsection{KN afterglow light curves}
The non-thermal radio afterglow produced by the mildly relativistic \ac{KN} ejecta \citep{Nakar:2011cw} is computed using the model from \citet{sadeh_non-thermal_2023}. 
The model input is the velocity distribution of the ejecta described in detail in Appendix~\ref{app:possis_geometry} together with the \ac{KN} parameters determined in Sec.~\ref{subsec:kilonova}.
For each angular bin, we first compute the cumulative mass profile
\begin{equation}
M(>\gamma\beta,\theta)\ ,
\end{equation}
obtained by integrating the mass histogram over all velocity bins exceeding a given value of $\gamma\beta$. 
To account for the anisotropic geometry of the ejecta, each angular bin is converted to an isotropic-equivalent mass profile by scaling with the solid angle of the bin,
\begin{equation}
M_{\rm iso}(>\gamma\beta,\theta)=
\frac{4\pi}{\Delta\Omega(\theta)}\,M(>\gamma\beta,\theta)\ .
\end{equation}
We then construct an effective spherical ejecta profile by averaging these isotropic-equivalent distributions over polar angle using weights proportional to the physical mass in each angular bin. 
This procedure yields a single cumulative mass distribution $M_{\rm iso,avg}(>\gamma\beta)$ that characterizes the ejecta relevant for the afterglow emission \citep[for non-spherical corrections, see][]{sadeh_late-time_2024}.
The resulting distribution is fitted with a continuous broken power-law profile of the form
\begin{equation}
M(>\gamma\beta)=
\begin{cases}
M_0\left(\frac{\gamma\beta}{\gamma_0\beta_0}\right)^{-s_{\rm ft}}, & \gamma\beta>\gamma_0\beta_0, \\
M_0\left(\frac{\gamma\beta}{\gamma_0\beta_0}\right)^{-s_{\rm KN}}, & \gamma\beta\leq\gamma_0\beta_0 ,
\end{cases}
\end{equation}
where $M_0$ is the cumulative ejecta mass at the break velocity $\gamma_0\beta_0$, $s_{\rm ft}$ describes the high-velocity fast-tail component, and $s_{\rm KN}$ characterizes the slower KN ejecta. 
The fit is performed in logarithmic space over the range $0.1<\gamma\beta\leq0.7$, which captures the mildly relativistic portion of the ejecta that dominates the afterglow emission.

The fitted parameters $(M_0,\gamma_0\beta_0,s_{\rm ft},s_{\rm KN})$ are then used as input to the synchrotron afterglow model from \citep{sadeh_non-thermal_2023}, which describes the interaction of the ejecta with the surrounding interstellar medium. 
The model predicts the synchrotron flux density from the ejecta-driven forward shock. 
The analytic approximations are valid for mildly relativistic ejecta with $\gamma\beta\lesssim 1$, consistent with the velocity range in our \acp{BNS}.

For each simulated event, we compute the afterglow light curve assuming fiducial microphysical parameters $\varepsilon_e=0.1$, $\varepsilon_B=10^{-3}$, and electron index $p=2.2$.
We point out that the microphysical parameters for the KN afterglow do not have to coincide with their analogues from the GRB afterglow. 
Hence, the electron index $p$ here was simply set to a canonical value of 2.2~\citep{Kathirgamaraju:2019xwu} and differs from $p=2.15$ in Sec.~\ref{subsec:grb} that was based on GRB170817A.
On the other hand, the draws for the interstellar medium densities are taken over from the previous computations of the GRB afterglows to maintain consistency.
The synchrotron flux is evaluated in the spectral regime $\nu_a < \nu_m < \nu < \nu_c$, corresponding to the slow-cooling segment of the synchrotron spectrum, as expected for KN afterglow \citep{sadeh_non-thermal_2024}. 
In this regime, the flux density scales as $F_\nu \propto \nu^{-(p-1)/2}$, which is the spectral dependence adopted in the analytic model. 
It is evaluated over a logarithmically spaced time grid spanning $10$ to $10^{4}$ days after merger. 
This way we create the KN afterglow light curves for the synthetic BNS populations.

\section{Multimessenger detections}
\label{sec:observations}
\subsection{GW detections}
We consider four scenarios for the next-generation \ac{GW} detector landscape.
In the first one, there is one triangle-shaped \ac{ET} detector~\citep{Punturo:2010zz, ET:2019dnz, Branchesi:2023mws, ET:2025xjr} with 10~km arm lengths located in the Meuse-Rhine Euregion.
In the second scenario, there are two L-shaped \ac{ET} detectors with 15~km-long arms each, one located in Sardinia, Italy, and the other in Lusatia, Germany. 
The corresponding \acp{PSD} are taken from \citet{ET_asd_files}.
In the third and fourth scenarios, the very same \ac{ET} configurations are joined by \ac{CE}~\citep{Reitze:2019iox, Evans:2021gyd} with 40~km arm-length located in Hanford, USA, with the \ac{PSD} taken from \citet{CE_asd_files}.
We label these configurations as \ett, ETL, \ett+CE, and ETL+CE, respectively.
A GW signal from a binary is considered as detected if its optimal network \ac{SNR} $\rho_\text{opt}$ exceeds $12$.
For the injection of the signal in the detector we use the IMRPhenomXAS\_NRTidalv3 waveform model~\citep{Pratten:2020fqn, Abac:2023ujg}.
We estimate the 90\% credibility sky localization area $\Delta\Omega^{\text{GW}}$ through the \ac{FIM} formalism as implemented in \textsc{gwfish}~\citep{Dupletsa:2022scg}.
The likelihood is evaluated from a minimum frequency of 5~Hz and takes Earth's rotation into account.

\subsection{EM counterparts}
\label{subsec:EM counterparts detection}

\begin{figure*}
    \centering
    \includegraphics[width=\linewidth]{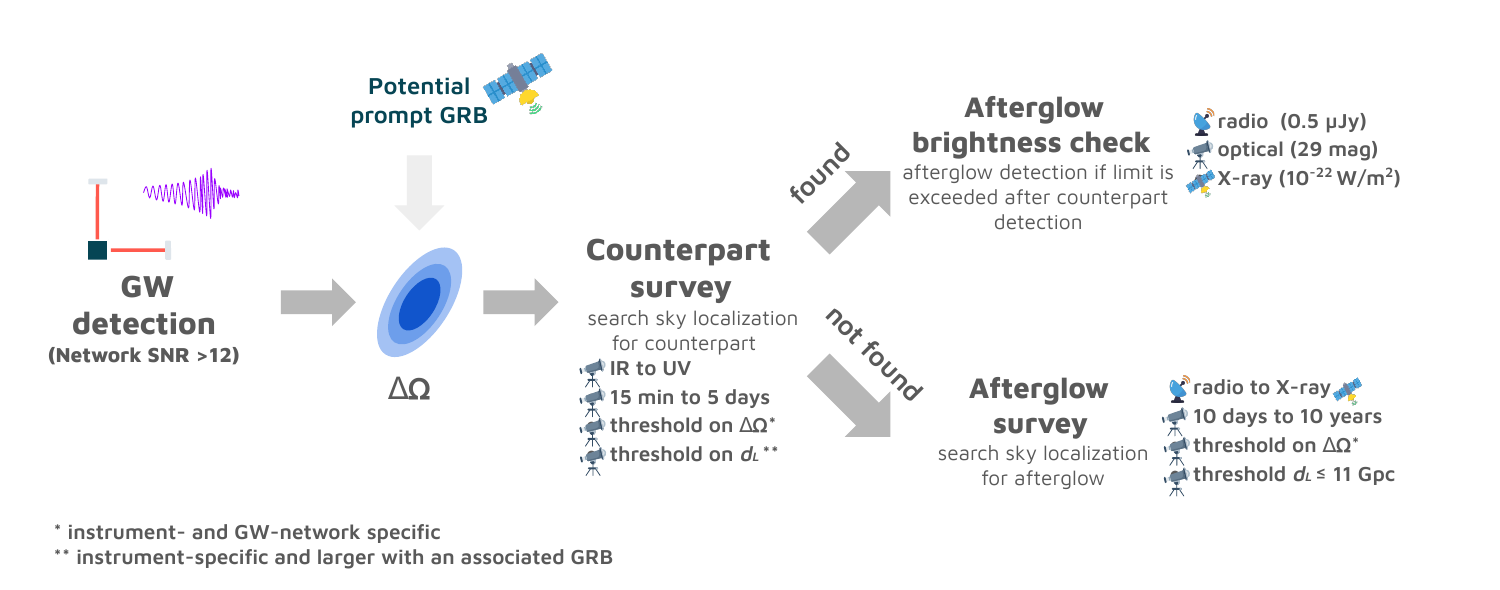}
    \caption[Sketch of mock detection algorithm for multimessenger events]{
    Schematic overview of the mock detection algorithm that we use to determine for which \ac{BNS} \ac{GW} signals an \ac{EM} counterpart will be detected.
    Starting from the GW detection, we check whether a KN or UVOIR afterglow can be found in the first days, and, depending on the result, search for the remaining afterglow.}
    \label{fig:detection_algorithm}
\end{figure*}
\begin{table}
\centering
\caption{Instruments considered for \ac{GRB} detection together with a bolometric fluence threshold $S_{\text{thr}}$ to claim detection.}
\label{tab:grb_instruments}
\begin{tblr}{
  colspec = {Q[c,2.2cm] Q[c, 1.6cm] Q[c, 0.9cm] Q[c, 1.4cm] Q[c,0.8cm]}
}
\toprule
\toprule
Instrument & Energy [MeV] & Detection prob. & $S_{\text{thr}}$\newline[erg~cm$^{-2}$] & $\Delta\Omega^{\text{GRB}}$ [deg$^2$] \\
\midrule
\textit{Fermi}/GBM & 0.008--40 & 0.6 &$2 \times 10^{-7}$ & 100 \\ 
\textit{Swift}/BAT & 0.015--0.15 & 0.1 & $2 \times 10^{-8}$ & 0.1\\ 
\textit{GECAM} & 0.006--5 & 0.8 & $7 \times 10^{-8}$ & 10\\ 
\textit{SVOM}/ECLAIRs & 0.004-0.15 & 0.14 & $7.2\times10^{-8}$ & 0.1\\ 
\bottomrule
\end{tblr}
\end{table}

\begin{table}
\centering
\caption{Observatories considered for the detection of UVOIR counterparts in targeted follow-up campaigns.}
\label{tab:kn_instruments}
\begin{tblr}{
  colspec = {Q[c,2cm] Q[c, 1.3cm] Q[c, 1cm] Q[c, 1.3cm] Q[c,1cm]}
}
\toprule
\toprule
Observatory & Filter & Exposure time~[s] & FOV [deg$^2$] & Limiting mag \\
\midrule
Rubin & $i$ & 600 & 9.6 & 25.6 \\ 
Rubin & $g$ & 600 & 9.6 & 26.5 \\
Pan-STARRS & $i$ & 400 & 7 & 24 \\
Pan-STARRS & $g$ & 400 & 7 & 24 \\ 
ZTF & $i$ & 300 & 47 & 22 \\
ZTF & $g$ & 300 & 47 & 22 \\ 
\textit{Roman} & F158 & 60 & 0.28 & 24.9 \\ 
\textit{Roman} & F213 & 60 & 0.28 & 23.7 \\
\textit{ULTRASAT} & near-UV & 900 & 204 & 22.5 \\
\bottomrule
\end{tblr}
\end{table}

\begin{table}
\centering
\caption{Radio and X-ray instruments considered for long term wide-field surveys to find afterglows when no short-term UVOIR counterpart was found. 
In this case, Rubin is also assumed to join this effort in the UVOIR range, with the same specifications as in Table~\ref{tab:kn_instruments} (see text for details).}
\label{tab:afterglow_instruments}
\begin{tblr}{
    colspec = {Q[c,2cm] Q[c, 1.3cm] Q[c, 0.9cm] Q[c, 1.cm] Q[c,1.4cm]}
}
\toprule
\toprule
Instrument & Frequency & Exposure time~[s] & FOV [deg$^2$] & Limiting flux \\
\midrule
SKA2/MID & 1.4 GHz & 300 & 5 & 5~$\mu$Jy \\
DSA & 1.4 GHz & 300 & 10.6 & 3.5~$\mu$Jy \\
\textit{Einstein Probe}/WXT & 0.5--4~keV & 1000 & 3600 &  2.6$\times10^{-11}$ erg~s$^{-1}$~cm$^{-2}$\\
\bottomrule
\end{tblr}
\end{table}
The search for an EM counterpart to a \ac{BNS} \ac{GW} signal with next-generation detectors can be divided into three stages.

\begin{enumerate}
    \item The inspiral of the \acp{BNS} might stay in the frequency band of the next-generation \ac{GW} detectors for many hours, and thus in certain cases there is the opportunity to provide a localization volume before the merger takes place. This is relevant for identifying precursors and very early parts of the \ac{KN}~\citep{Metzger:2014yda}, very high energy emission ($\gtrsim10$~GeV), and GRB prompt emission~\citep{Banerjee:2022gkv}.
    Even if the pre-merger sky localization is poor, \ac{GRB} prompt emission can be detected with wide-field gamma-ray observatories.
    
    \item After the merger has taken place, dedicated wide-field searches in the UV, optical, and infrared (UVOIR) can scan the \ac{GW} sky localization volume to find a transient within the first days. 
    The aim in this phase is to detect the \ac{KN} and, if the inclination is small, possibly the \ac{GRB} afterglow. 
    The specific instruments and observation strategies used in this phase can have a significant impact on the number of identified counter parts~\citep[e.g.][]{Camisasca:2023dqx, Barna:2024bcd, DES:2023hft, Loffredo:2024gmx}.
    
    \item On longer time scales (weeks to years), targeted observations, especially in the radio or X-ray, might reveal afterglow features from the GRB or \ac{KN}~\citep{Nakar:2011cw}.
    The observation strategy in this phase depends on whether previously an \ac{EM} counterpart was identified.
    If so, deep follow-up observations by the most powerful instruments will likely find the afterglow, provided it is luminous enough.
    Otherwise, wide-field searches with less sensitive instruments but large \ac{FOV} are needed to find the remaining afterglow~\citep{Morsony:2023afu, Kaur:2024yag}.
\end{enumerate}

To assess for which \acp{BNS} in our catalogues a counterpart would be detected, we adopt several simplifying assumptions, owing to the unknown availability of telescopes in the next-generation era and the complexity of simulating different observing strategies that cannot be reasonably covered in this article.
The basic outline of our mock detection algorithm is visualized in Fig.~\ref{fig:detection_algorithm} and reflects the three stages listed above.

In the first stage, the goal is to detect the prompt GRB.
With our setup, we ignore pre-merger alerts and assume that the GRB prompt emission is detected with a certain detection probability (the relative field-of-view times the duty cycle) if the fluence exceeds an instrument-specific threshold.
We consider four wide-field \ac{GRB} instruments, namely \textit{Fermi}/GBM~\citep{Meegan:2009qu}, the \textit{Swift} Burst Alert Telescope \citep[\textit{Swift}/BAT;][]{Barthelmy:2005hs}, the \textit{GECAM} satellite~\citep{Li:2021pns}, and the ECLAIRs telescope aboard \textit{SVOM}~\citep[\textit{SVOM}/ECLAIRs;][]{Wei:2016eox}, each with their own energy range, detection probability, detection threshold, and angular localization $\Delta\Omega^{\text{GRB}}$ listed in Table~\ref{tab:grb_instruments}.
If the GRB is detected and $\Delta\Omega^{\text{GRB}}$ is smaller than $\Delta\Omega^{\text{GW}}$, we use its value as input for the next observation stage, i.e.,
\begin{align}
    \Delta\Omega = \min(\Delta\Omega^{\text{GW}}, \Delta\Omega^{\text{GRB}})\ .
    \label{eq:sky_loc}
\end{align}

The second stage represents the hunt for the short-term UVOIR counterpart.
For the search in the hours and days after the merger, the most important parameters are the sky localization $\Delta\Omega$ and the UVOIR light-curve brightness of the counterpart.
Our light curves are generated with the \textsc{possis} machine-learning surrogate presented in Appendix~\ref{app:possis_geometry} and also contain the optical contribution from the \ac{GRB} afterglow using \textsc{pyblastafterglow}, although the latter is only relevant at small inclinations.
The \ac{KN} afterglow is not included at this stage, as it will take months to years to become visible.
Further, we account for Milky Way extinction~\citep{Pei:1992ub} but neglect any extinction by the host galaxy.
Although it is unclear which telescopes will accompany the next-generation \ac{GW} searches, we consider specific observatories that are either expected to remain in operation until then or for which comparable successors could be available.
In the optical and infrared, the target-of-opportunity programme of the Vera Rubin telescope~\citep[Rubin;][]{Andreoni:2024pkp} is expected to be one of the main drivers of counterpart detection from \ac{BNS} mergers due to its wide \ac{FOV} combined with its large aperture.
We additionally consider the Pan-STARRS telescope~\citep{Hodapp:2004}, the Zwicky Transient Facility~\citep[ZTF;][]{Bellm:2019}, and the Nancy Grace Roman Space Telescope~\citep[\textit{Roman};][]{Spergel:2015sza, Andreoni:2023xlv}, and \textit{ULTRASAT}~\citep{Shvartzvald:2023ofi} as participants in the detection campaigns.
The limiting magnitudes, exposure times, and \ac{FOV} assumed for these instruments are listed in Table~\ref{tab:kn_instruments}.
The values there are taken from the instrument-specific references and additionally inspired by the recent follow-up campaign of AT2025ulz~\citep{Kasliwal:2025keb, Gillanders:2025fwf, Hall:2025qsm}.
For all our instruments, we adopt an observing strategy similar to \citet{Loffredo:2024gmx}.
We simply assume that observers decide for or against a targeted follow-up campaign by enforcing instrument-specific thresholds on $\Delta\Omega$ and the luminosity distance.
We consider a counterpart detected if it has been seen at least twice, either across two different epochs, filters, or instruments, ignoring any issues related to classification and impostors~\citep{Barna:2025xxn}.
Observation epochs are instrument-specific but take place between 15 minutes and 5 days after the merger.
A more detailed outline of our algorithm is given in Appendix~\ref{app:observation_strategy_UVOIR}.
We point out that this algorithm only checks whether an UVOIR counterpart is found. 
It does not classify the transient type, i.e., whether a \ac{KN} or a GRB afterglow is seen.
The latter would require additional observations that are not accounted for by our algorithm.

The third stage of our mock observation algorithm concerns the long-term observations to identify potential afterglows.
We always consider the joint light curve of the \ac{GRB} and \ac{KN} afterglow.
Here, the strategy depends on whether a UVOIR counterpart was found in the second stage.
If so, the sky position is known up to arc-second level and the search for remaining afterglow signatures would be driven by deep long-exposure observations.
Hence, in this case, we assume radio and X-ray afterglows are detected if their fluxes exceed thresholds representative of the most sensitive instruments expected to operate in the next-generation era.
Specifically, these thresholds are set to $0.5$~$\mu$Jy at 1.4~GHz (achievable e.g. with the radio telescopes SKA2~\citep{Braun:2019gdo, Bonaldi:2020ukl, SKA2} or ngVLA~\citep{ngVLA_limits, Corsi:2019xxn}) or $10^{-19}$~erg~cm$^{-2}$~s$^{-1}$ in the 0.5--2~keV range (achievable e.g. with the X-ray satellites \textit{Athena}~\citep{Piro:2021oaa}, \textit{AXIS}~\citep{Marchesi:2020smf, Reynolds:2023vvf}\footnote{The \textit{AXIS} proposal was recently cancelled~\citep{AXIS:2026}.}
, or \textit{Lynx}~\citep{Vikhlinin:2022sde}).
Likewise, we consider the possibility of a late-time optical afterglow detection, setting a threshold of 29~mag in the $g$ band (achievable e.g. with the ELT~\citep{ELT_MICADO} or \textit{JWST}~\citep{Gardner:2023}).
These thresholds must be exceeded after the identification of the UVOIR counterpart from the previous stage and before ten years have passed.

On the other hand, if no UVOIR counterpart was found in the first place, we assume that another wide-field follow-up survey in the radio, UVOIR, and X-ray takes place.
In our conception of this scenario, the instruments carrying out such a survey are SKA2, DSA~\citep{Hallinan:2019qyo}, Rubin, and the WXT instrument aboard \textit{Einstein Probe}~\citep[\textit{Einstein Probe}/WXT;][]{Yuan:2025cbh}.
The presumptive telescope properties are listed in Table~\ref{tab:afterglow_instruments}.
The detection algorithm is similar to the one for the immediate counterpart search, though the observation epochs are spread out from 10 days up to 10 years.
More details on it are provided in Appendix~\ref{app:observation_strategy_afterglow}.
As before, our algorithm does not determine which type of afterglow is found.
To classify whether an afterglow arises from a GRB or \ac{KN}, additional observations would be needed.

\subsection{Detection prospects}
\label{subsec:detection_prospects}
\begin{figure*}
    \centering
    \includegraphics[width=1\linewidth]{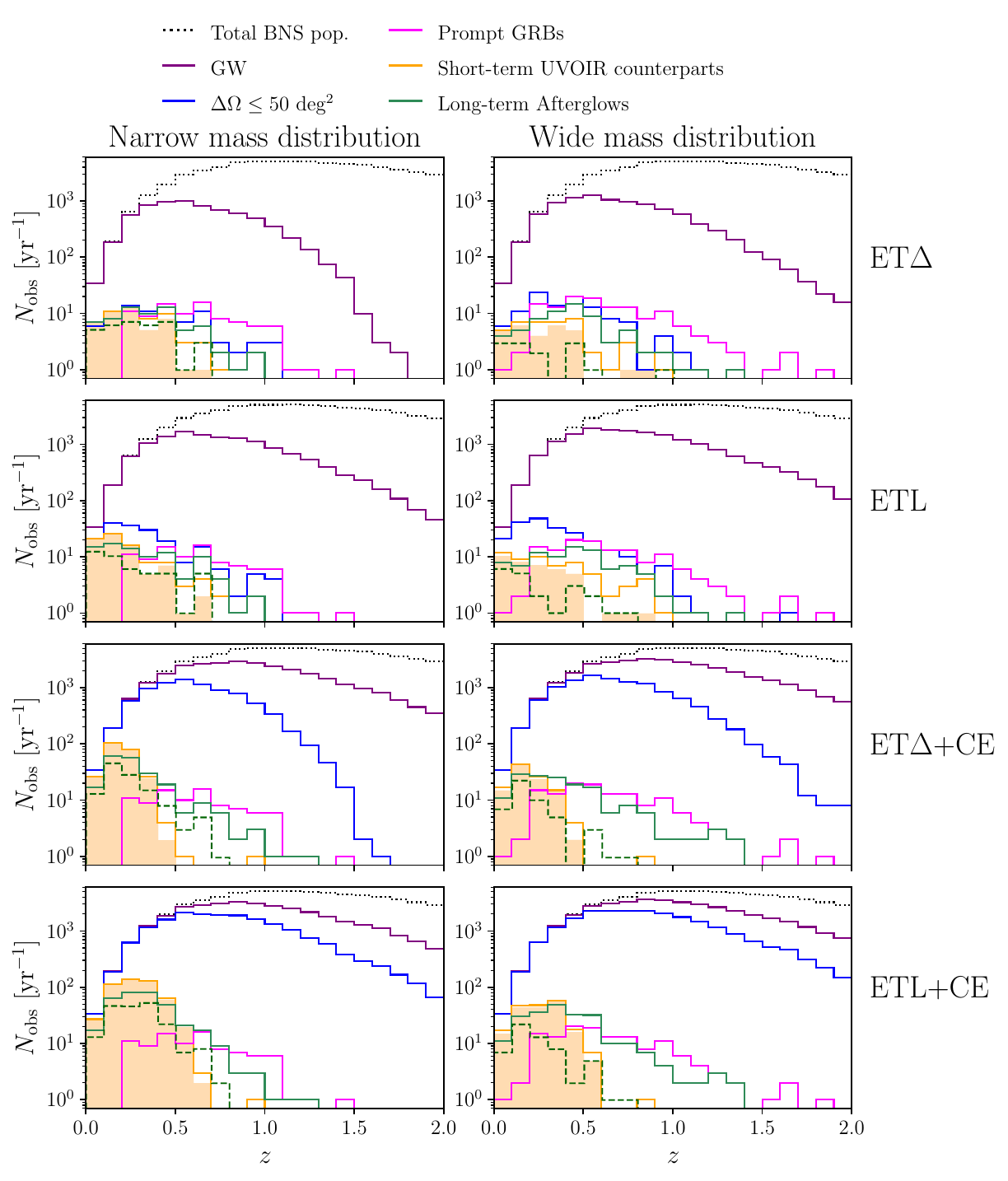}
    \caption[Multimessenger detections from mock catalogues across redshift]{Redshift distribution for our projection of  \ac{BNS} detections in one year.
    The rows correspond to a specific next-generation \ac{GW} detector network.
    The left and right columns stand for the narrow or wide mass distribution, respectively.
    Different types of detections (\ac{GW} signal, \ac{GW} signal with short-term UVOIR counterpart, etc.) are colour-coded according to the legend.
    The dotted line corresponds to the entire cosmic \ac{BNS} population.
    Orange patches mark UVOIR counterparts with a noticeable \ac{KN}, while green dashed line marks afterglows at least partially visible as \ac{KN} afterglow (see text for details).}
    \label{fig:detections}
\end{figure*}

\begin{table*}
\centering
\caption{Number of multimessenger detections in one year of next-generation \ac{GW} observations assuming the merger rate from Fig.~\ref{fig:merger_rate}. 
The columns represent the different \ac{BNS} mass distributions and \ac{GW} detector configurations. 
GRB detections are first reported independently of any \ac{GW} observation.
The other sections display the short-term UVOIR counterparts (within the first 5 days) and long-term afterglows (up to 10 years) found by the follow-up detection algorithm presented in Fig.~\ref{fig:detection_algorithm}.
The bracketed values in the afterglow section are those afterglows found through follow-up surveys without an earlier UVOIR counterpart.
We also list how many \ac{GW} events will have joint detections, e.g., how many will have both a short-term UVOIR and an afterglow (afg.) counterpart.
The last rows reveal which physical types of transients are, in principle, observable from the \acp{BNS} with at least one UVOIR or afterglow counterpart.
This classification is not determined by the detection algorithm itself, but rather in a post-processing step; see the text for a detailed discussion.
}
\begin{tblr}{
    colspec = {Q[c, 1.6cm] Q[c, 2cm] Q[c, 1.2cm]  Q[c, 1.2cm] Q[c, 1.2cm] Q[c, 1.2cm] Q[c, 1.2cm] Q[c, 1.2cm] Q[c, 1.2cm] Q[c, 1.2cm]},
    vline{7} = {3-Z}{solid},
}
\toprule 
\toprule 
 & & \SetCell[c=4]{c} Narrow mass distribution& & & & \SetCell[c=4]{c} Wide mass distribution & & &\\ 
 & & ET$\Delta$ & ETL & ET$\Delta$+CE & ETL+CE & ET$\Delta$ & ETL & ET$\Delta$+CE & ETL+CE \\ 
 \midrule 
\SetCell[r=5]{c} Prompt\linebreak GRB & \textit{Fermi}/GBM & \SetCell[c=4]{c} 21 & & & & \SetCell[c=4]{c} 36 & & & \\ 
 & \textit{Swift}/BAT & \SetCell[c=4]{c} 7 & & & & \SetCell[c=4]{c} 12 & & & \\ 
 & \textit{GECAM} & \SetCell[c=4]{c} 82 & & & & \SetCell[c=4]{c} 122 & & & \\ 
 & \textit{SVOM}/ECLAIRs & \SetCell[c=4]{c} 1 & & & & \SetCell[c=4]{c} 3 & & & \\ 
 & \textbf{Total} & \SetCell[c=4]{c} 91 & & & & \SetCell[c=4]{c} 134 & & & \\ 
\midrule 
 GW & - & 6984 & 13622 & 30456 & 35868 & 9556 & 17683 & 36636 & 42482 \\ 
 GW+GRB & -  & 67& 78& 91& 91& 92& 115& 134& 134 \\ 
\midrule 
\SetCell[r=10]{c} Short-term UVOIR counterparts & Rubin $g$ & 54 & 75 & 212 & 500 & 43 & 58 & 98 & 212 \\ 
 & Rubin $i$ & 53 & 75 & 218 & 486 & 37 & 52 & 96 & 190 \\ 
 & PSTARRS~$g$ & 10 & 28 & 80 & 82 & 9 & 14 & 35 & 35 \\ 
 & PSTARRS~$i$ & 12 & 35 & 100 & 102 & 10 & 19 & 47 & 47 \\ 
 & ZTF $g$ & 5 & 12 & 12 & 12 & 2 & 4 & 4 & 4 \\ 
 & ZTF $i$ & 6 & 13 & 13 & 13 & 2 & 4 & 5 & 4 \\ 
 & \textit{Roman} F158 & 13 & 17 & 14 & 40 & 11 & 13 & 13 & 23 \\ 
 & \textit{Roman} F213 & 13 & 17 & 14 & 40 & 11 & 13 & 13 & 23 \\ 
 & \textit{ULTRASAT} & 4 & 12 & 12 & 12 & 2 & 2 & 3 & 4 \\ 
 & \textbf{Total}  & 59 & 91 & 242 & 497 & 43 & 61 & 107 & 195  \\ 
\midrule 
\SetCell[r=4]{c} Long-term afterglows & Radio  & 52~(21) & 65~(23) & 116~(70) & 168~(96) & 64~(34) & 80~(43) & 123~(96) & 172~(126)  \\ 
 & Optical  & 3~(1) & 7~(1) & 8~(1) & 13~(1) & 1~(0) & 1~(0) & 3~(0) & 3~(0)  \\ 
 & X-ray  & 46~(0) & 66~(0) & 142~(0) & 253~(0) & 33~(0) & 45~(0) & 63~(0) & 105~(0)  \\ 
 & \textbf{Total}  & 67 & 89 & 212 & 349 & 67 & 88 & 159 & 231  \\ 
\midrule 
\SetCell[r=4]{c} Joint detections & UVOIR+afg.  & 48& 68& 142& 254& 36& 48& 64& 106 \\ 
 & GRB+UVOIR  & 27& 26& 10& 14& 29& 35& 11& 16 \\ 
 & GRB+afg.  & 42& 43& 51& 53& 55& 67& 70& 74 \\ 
 & GRB+UVOIR+afg.  & 27& 26& 10& 14& 29& 35& 11& 16 \\ 
\midrule 
\SetCell[r=3]{c} \textit{Actual transient type} & \textit{KN} & \textit{46} & \textit{78} & \textit{236} & \textit{489} & \textit{28} & \textit{39} & \textit{98} & \textit{181} \\ 
 & \textit{GRB afg.} & \textit{53} & \textit{66} & \textit{131} & \textit{199} & \textit{64} & \textit{83} & \textit{136} & \textit{200} \\ 
 & \textit{KN afg.} & \textit{35} & \textit{44} & \textit{118} & \textit{198} & \textit{13} & \textit{21} & \textit{49} & \textit{59} \\ 
\bottomrule

\end{tblr}
\label{tab:detections}
\end{table*}

In Fig.~\ref{fig:detections}, we show the distribution of \ac{BNS} multimessenger detections across redshift for the different mass models and for the different next-generation detector networks.
Further, in Table~\ref{tab:detections} we list the total number of detections per instrument.
For brevity, we will refer to the event catalogues by their mass distribution and detector network, e.g., `narrow+\ett+CE' for the detection catalogue with \acp{NS} drawn from the narrow mass distribution and \ett+CE as \ac{GW} network.
Each catalogue thereby refers to one year of observations with the corresponding detectors.

As expected from previous studies~\citep{Evans:2021gyd, Branchesi:2023mws, Regimbau:2012ir, Singh:2021zah, Walker:2024loo}, next-generation \ac{GW} detectors will detect almost every merging \ac{BNS} within $z\lesssim0.4$ and occasionally catch signals out to $z\approx 2$--$3$.
Apart from the sensitivity of the detectors, the redshift cutoff additionally depends on the mass distribution.
For instance, the wide distribution contains heavy \acp{BNS} with total source frame mass $m_1 + m_2 \gtrsim 3$~\msol that can still be detected at $z\gtrsim2$.

The most important difference between the detector configurations with respect to prospective multimessenger detections is the number of events with tight sky localization.
For ET alone, the number of \ac{BNS} signals with $\Delta\Omega_{\text{GW}}\leq100$~deg$^2$ is $\sim70$ and $\sim350$ respectively for \ett and ETL.
If CE is included, the increased base line between the detectors causes \textsc{gwfish} to estimate almost every sky localization to be within 100~deg$^2$, even for \acp{BNS} at high redshifts.
In fact, for \ett+CE 10\% of all detected \ac{GW} events have $\Delta\Omega_{\text{GW}} \leq 22$~deg$^2$, 50\% of events have $\Delta\Omega_{\text{GW}} \leq 80$~deg$^2$. 
These limits are reached by only 0.1\% or 0.6\% of events for a single \ett detector.
Correspondingly, for ETL+CE, the 10\% quantile in the $\Delta\Omega_{\text{GW}}$ distribution is at 13~deg$^2$, the median at 48~deg$^2$.
In an ETL only network, again just 0.1\% and 0.7\% of events stay below these limits.
The choice of mass distribution only has a minor effect on the number of well-localized \ac{BNS} signals, although the wide mass distribution tends to create slightly more such events, because heavier \acp{NS} tend to produce signals with higher \ac{SNR}.

While less relevant for good sky localization, the mass distribution has a large impact on the detectability of the \ac{EM} counterpart because it determines how many \acp{BNS} undergo prompt collapse.
We emphasize again that we count an early UVOIR counterpart as detected if there are at least two different detection flags, either in different epochs, instruments, or bands and that the light curve could be from a \ac{KN} or near-optical GRB afterglow, or both.
Overall, the narrow mass distribution leads to more immediate counterparts, as we project 59 detections for \ett, 91 for ETL, 242 for \ett+CE, and even 497 for ETL+CE.
In case of the wide mass distribution, the corresponding numbers drop to 43, 61, 107, and 195.
The difference is mainly driven by the different number of bright \acp{KN}.
Even though some \acp{BNS} from the wide mass model can occasionally produce very bright \acp{KN} when the component masses are asymmetric and not too large, the narrow mass model has a lower prompt collapse rate and thus produces more visible \acp{KN}.
As a consequence of the better sky localization in the CE networks, the narrow+ETL+CE catalogue therefore provides the most successful follow-ups, whereas wide+\ett delivers the fewest events.

In any case, the Vera~Rubin telescope dominates the counterpart search and provides the largest share of detections.
However, there remain certain detections that are not observed by Vera~Rubin at all, either because of sky accessibility or because of the more stringent threshold on $\Delta\Omega$ we have set for this telescope to start a survey (see Appendix~\ref{app:observation_strategy_UVOIR}).
Especially for the narrow mass model and with CE, there are about $\sim30$ observed UVOIR counterparts not identified through Rubin, though for the other catalogues, this number ranges between $5$ and $20$.
Conversely, our observation strategy causes Vera~Rubin to sometimes flag transients just once and so these are not included as a full detection, as described in Sec.~\ref{subsec:EM counterparts detection}.
Depending on the mass distribution and \ac{GW} network, this happens $8$--$70$ times and is roughly equally likely in the $g$ or $i$ band.

As mentioned, not all UVOIR counterparts correspond to \acp{KN}, as some just arise from near-optical \ac{GRB} afterglows.
This is visualized in Fig.~\ref{fig:detections}, where the orange line indicates all detected UVOIR counterparts and the orange patches represent those events with a noticeable \ac{KN} contribution.
The latter is a post-processing step, that is performed after running our mock detection algorithm.
To determine which events display a noticeable \ac{KN}, we apply a simplistic criterion that checks whether at least once after detection, the total flux arises at least to 30\% from \ac{KN} emission, i.e., whether there is a point in time after detection where the \ac{KN} is not outshined by more than 1~mag.
In reality, the contribution of the different components would only become distinguishable through model comparison and a larger photometric (or spectroscopic) data set, including for instance radio or X-ray observations.
Generally, most UVOIR counterparts found within $d_L\lesssim3500$~Mpc have a noticeable contribution from a \ac{KN}.
For farther events, the UVOIR counterpart is likely from a GRB afterglow.
In our catalogues, the latter case occurs more often when CE is not part of the \ac{GW} network, because otherwise the stricter limits on $\Delta\Omega$ cause the searches to focus on closer events.
We also note that \ac{KN} detections are primarily correlated by large winds with $\log_{10}(m_{\text{ej, wind}})\gtrsim-2$, meaning apart from sufficiently massive discs, the auxiliary $\zeta$ has to be $\gtrsim0.1$, while the dynamical ejecta have a lesser influence on detectability.
Naturally, prompt collapse \acp{BNS} are under-represented in the final \ac{KN} sample:
For the wide population about 20\% to 25\% of detected \acp{KN} still come from prompt collapse, for the narrow mass distribution it is about 1\% to 6\%.

For the afterglows, our results show that without a previously identified UVOIR counterpart, a dedicated follow-up afterglow search will be most successful in the radio, given our choice of survey instruments from Table~\ref{tab:afterglow_instruments}.
The number of afterglows found this way are listed as bracketed values in Table~\ref{tab:detections}.
While Rubin finds one optical afterglow later than 10 days and SKA2/MID and DSA still discovers tens of late-time radio afterglows, all the late-time X-ray afterglows are too faint for the wide-field \textit{Einstein Probe}/WXT.

In the same way that not all UVOIR counterparts are \acp{KN}, not all identified afterglows are pure GRB afterglows, since at very late times they can also carry a noticeable contribution from the \ac{KN} afterglow.
The green line in Fig.~\ref{fig:detections} shows all afterglows, while the green dashed line represents those afterglows with a noticeable \ac{KN} afterglow.
Analogously to the UVOIR counterpart, we classify the afterglows in post-processing. 
An afterglow is listed as \ac{KN} afterglow at the bottom of Table~\ref{tab:detections} if the contribution of the \ac{KN} component to the total observable afterglow flux is more than 30\% at any point between the first detection and ten years after merger.
This is checked for the radio, optical, and X-ray bands each.

Naturally, the narrow mass distribution results in more detectable \ac{KN} afterglows, since the ejecta masses are higher.
Thus, we project $35$ visible \ac{KN} afterglow contributions for events observed with \ett, increasing to 198 in case of ETL+CE.
But even for the wide mass model, 13 to 59 \ac{KN} afterglows could be observed, depending on the \ac{GW} detector network.
In reality, confident classification of these afterglows would require additional late time observations that are not included in our detection algorithm.
Therefore, the \ac{KN} afterglow numbers quoted here only include those that are noticeable within ten years after the merger.
Since the peak times of the observable \ac{KN} afterglows in our sample range from $\sim$200 to 11000 days after merger, with the median peak time being around $\sim2000$~days,
there is also a chance for discovering more \ac{KN} afterglows when considering observation epochs past ten years.
Interestingly, not all \ac{KN} afterglows have an identified \ac{KN} counterpart, since about 15\% to 35\% are observed in combination with just a GRB afterglow alone.
However, no \ac{KN} afterglows are observed without prior detection of either a proper \ac{KN} or a \ac{GRB} afterglow.

As mentioned, the wider mass distribution leads to slightly more prompt GRBs and GRB afterglows, because more \acp{BNS} collapse to \acp{BH} and produce successful jets, even though disc masses are often smaller compared to the narrow mass distribution.
However, given the ad hoc nature of our \ac{GRB} launch model, these different \ac{GRB} detection rates should not be interpreted as robust predictions.
Despite the larger number of GRB afterglows, the narrow mass distribution leads to an equal or even slightly higher number of observable afterglows overall, because of a larger number of \ac{KN} afterglows.
We also mention that only about 40\% to 50\% of observed \acp{GRB} have a detected afterglow.
Conversely, $\sim$80\% of GRB afterglows have a detected prompt emission, when considering just \ett or ETL.
If CE is in the network, more \ac{GRB} afterglows are detected overall and this share drops to $\sim35$\%.

Another observation is that \textit{GECAM} enhances the short GRB detection rate by a factor of a few compared to rates from the \textit{Fermi}/GBM telescope.
\textit{Fermi}/GBM currently detects about 40 short GRBs per year, whereas in our projection the increased sensitivity from \textit{GECAM} leads to $\sim100$ additional short GRBs per year.
With CE, virtually every GRB has an accompanying \ac{GW} signal, while in an ET-only network this share drops to $\sim70$\%.
In our algorithm, \ac{GRB} detection is independent of the \ac{GW} network, although alerts from the early inspiral in the detector could lead to additional \ac{GRB} detections from instruments with smaller \ac{FOV} but higher sensitivity~\citep{Banerjee:2022gkv, Sachdev:2020lfd}.

In addition to the previously discussed modelling systematics for the \ac{EM} counterparts, our projection for the number of multimessenger detections presented here suffers from further short-comings.
One issue is that for the large search areas in the follow-up, other transients will pollute the survey~\citep{DES:2017dgt, Barna:2025xxn, Stevenson:2025fqt, Fulton:2025hsx}. 
Additional pointings would be needed to confidently identify the one transient corresponding to the \ac{BNS} merger.
In our approach, we neglect this circumstance and simply claim detection once the transient has been in one of the tilings, thus assuming more confident detections than might be possible in a real follow-up scenario.
This effect naturally applies more severely to GW detector configurations with worse sky localizations.
On the other hand, there are some assumptions in our detection algorithm that might counteract an overestimation of possible counterparts.
For one, we always search the tiles in random order, while other algorithms suggest focusing on the tiles with the highest posterior likelihood first, especially if there are multiple alerts or telescopes~\citep{Coughlin:2019qkn, Dobie:2019eeg, Gupte:2020mfp}.
Moreover, a biased search towards the known galaxies within the search volume might enhance the detection probability as well, provided complete galaxy catalogues are available in direction and up to the distance of the event~\citep{Gehrels:2015uga, Coughlin:2018lta, Dalya:2018cnd, Coughlin:2019qkn, Dobie:2021qya, Dalya:2021ewn, DES:2025key}.

Beyond the systematics associated with the identification of \ac{EM} counterparts, uncertainties also arise from the treatment of the \ac{GW} detection.
Specifically, by assuming a duty cycle of $100\%$ for each detector, we overestimate the number of detections with tight sky localization. 
While the ETL configuration usually provides tighter sky localizations due to its larger baseline than \ett, a single L-shaped ET detector is considered almost blind to sky location~\citep{Branchesi:2023mws}, except in cases of long high-\ac{SNR} signals when earth rotation and finite arm lengths are taken into account~\citep{Baral:2023xst}.
Therefore, in case only one L-shaped interferometer operates, e.g., through asynchronous duty cycles etc., the efficacy of \ac{EM} follow-ups could be severely impaired, such that, over a full observation cycle, the \ett design might deliver better relative detection prospects than indicated by our work here~\citep{Negri:2026clm}.
On the other hand, if contemporary \ac{GW} detectors continue to operate in the next-generation era, they could assist sky localization efforts for close events and thus improve the overall multimessenger detection prospects~\citep{Finstad:2022oni, Pandey:2024mlo, Baral:2025geo}.
We leave a more detailed investigation of syntheses between current and next-generation detectors to future work.
Another problem is that the sky localization might be overestimated due to known shortcomings and limitations in the \ac{FIM} formalism~\citep{Vallisneri:2007ev,Rodriguez:2013mla,Dhani:2024jja}.
Together with further complications like \ac{PSD} estimation~\citep{Goncharov:2022dgl}, glitch mitigation~\citep{Narola:2024qdh}, detector calibration~\citep{Wong:2025iaf}, correlated noise~\citep{Janssens:2022xmo}, overlapping \ac{GW} signals~\citep{Baka:2025yqx}, or the breakdown of the long wavelength approximation~\citep{Virtuoso:2024kyp, Baker:2025taj}, this should be addressed in the future.

\section{Hierarchical recovery of physical parameters}
\label{sec:hierarchical_inference}

Having constructed catalogues of multimessenger \ac{BNS} mock detections in Sec.~\ref{sec:observations}, the aim in this section is to project how well these events could be used to constrain the \ac{EOS}, mass distribution, and cosmology.
Since these parameters apply to each \ac{BNS} simultaneously, they are hyperparameters that need to be inferred in a hierarchical setup.
We introduce the method of hierarchical sampling in the first subsection, and in the second subsection, we apply it to jointly constrain the \ac{EOS} and \ac{BNS} mass distribution.
In the third subsection, we additionally incorporate cosmological parameters in our hierarchical inference setup.
To limit the scope, we restrict ourselves to ET-only catalogues and ignore multimessenger events without a noticeable \ac{KN}.
That means we conduct four analyses: one for narrow+\ett, combining 46 \ac{KN} events; one for narrow+ETL, combining 78 events; one for wide+\ett, combining 28 events; and one for wide+ETL, combining 39 events, as indicated in the third-to-last row of Table~\ref{tab:detections}.
Including CE events and \ac{EM}-bright events without a \ac{KN} may be investigated in future work.

\subsection{Hierarchical inference}
Quantities such as $H_0$, the \ac{EOS}, or population parameters are hyperparameters, because they equally apply to all \acp{BNS}. 
These hyperparameters $\lambda$ prescribe a conditional probability $p(\theta\mid\lambda)$ for the binary specific parameters $\theta$, such as mass, tidal deformability, or distance.
For instance, parameters for the mass distribution $\lambda_{\text{pop}}$ naturally define a conditional probability for the source-frame component masses $p(m_1, m_2 \mid \lambda_{\text{pop}})$, whereas the \ac{EOS} fixes the tidal deformabilities
\begin{align}
\begin{split}
    &p(\Lambda_1, \Lambda_2 \mid m_1, m_2, \text{EOS}) = \\ 
    &\qquad\delta(\Lambda_1 - \Lambda^{\text{EOS}}(m_1))\ \delta(\Lambda_2 - \Lambda^{\text{EOS}}(m_2))\ .
\end{split}
\end{align}
Then, the final posterior on $\lambda$ from a series of independent events $\{d_1, d_2,\dots, d_n\}$ can be written as
\begin{align}
    p(\lambda \mid \{d_1,\dots, d_n\}) &= \frac{\mathcal{L}(\lambda \mid \{d_1,\dots, d_n\}) \pi(\lambda)}{\int d\lambda\ \mathcal{L}(\lambda \mid \{d_1,\dots, d_n\}) \pi(\lambda)}\ ,
    \label{eq:hierarchical_posterior}
\end{align}
with a hyperprior $\pi(\lambda)$ and the joint hyperlikelihood~\citep{Mandel:2018mve, Golomb:2021tll, Golomb:2024lds}
\begin{align}
\begin{split}
        &\mathcal{L}(\lambda \mid \{d_1,\dots, d_n\}) = \\
        &\qquad p_{\text{det}}(\lambda)^{-n} \times  \prod_{j=1}^n \int \mathcal{L}(\theta^{(j)}\mid d_j)\ p(\theta^{(j)}\mid\lambda) \ d\theta^{(j)}\ .
    \label{eq:hierarchical_likelihood}
\end{split}
\end{align}
Here, the $p_{\text{det}}(\lambda)$ counteracts the selection bias in the observed population so that the inferred $\lambda$ corresponds to the astrophysical population as opposed to the observed one.
In order to sample the hierarchical posterior (also called hyperposterior) in Eq.~\eqref{eq:hierarchical_posterior} through standard sampling techniques like \ac{MCMC}~\citep{MCMC_handbook} or nested sampling~\citep{Skilling:2004pqw,Skilling:2006gxv}, one must be able to efficiently evaluate Eq.~\eqref{eq:hierarchical_likelihood}. 
However, evaluating the $n$ multi-dimensional integrals directly at every sampling step in $\lambda$ is computationally intractable.
While recent methods are opening up the possibility of end-to-end hierarchical analysis through simulation-based inference~\citep{Leyde:2026hvm}, at this stage, these techniques are limited to black hole mergers.
For this reason, we first sample a parameter posterior $p(\theta^{(j)}\mid d_j)$ for each event separately using a wide agnostic prior $\pi(\theta)$ on the individual \ac{BNS} parameters. 
We can then write 
\begin{align}
\begin{split}
        &\mathcal{L}(\lambda \mid \{d_1,\dots, d_n\})\\
        &\qquad = p_{\text{det}}(\lambda)^{-n} \times \prod_{j=1}^n \int p(d_j) \frac{p(\theta^{(j)}\mid d_j)}{\pi(\theta^{(j)})} p(\theta^{(j)}\mid \lambda)\ d\theta^{(j)} \\
        &\qquad\propto p_{\text{det}}(\lambda)^{-n} \times \prod_{j=1}^n \int \frac{p(\theta^{(j)} \mid d_j)}{\pi(\theta^{(j)})} p(\theta^{(j)} \mid \lambda)\ d\theta^{(j)} \, .
\end{split}
\label{eq:hierarchical_likelihood2}
\end{align}
The difficulty in the implementation then reduces to finding an efficient way of evaluating the posterior density from the numerical samples for $p(\theta^{(j)} \mid d_j)$.
This problem can be tackled in multiple ways~\citep{Golomb:2021tll}.
In this work, we resort to training a \acl{NF}~\citep[\acs{NF};][]{Kobyzev:2019ydm,Papamakarios:2019fms} on the numerical posterior samples, as first proposed in~\cite{Ruhe:2022ddi}.
This makes the hyperlikelihood in Eq.~\eqref{eq:hierarchical_likelihood2} tractable and enables us to sample the hyperposterior in Eq.~\eqref{eq:hierarchical_posterior} using standard sampling algorithms.

\subsection{Hierarchical inference of the EOS and mass distribution}
\label{subsec:eos_inference}
\begin{table}
\centering
\caption{
    Hyperpriors for the \ac{EOS}, mass distribution parameters, and cosmological parameters. 
    Uniform priors are denoted by $\mathcal{U}$. 
    The random base points for the speed-of-sound extension scheme are indexed as $n_j$ and $c^2_{s,j}$, and in total, six of these base points were used.
    The cosmological hyperparameters are added to the inference in Sec.~\ref{subsec:cosmo_inference}.
}
\label{tab:hyperpriors}
\begin{tblr}{
    colspec = {Q[c, 1.7cm] Q[c, 2.2cm]  Q[c, 2cm]},
    vline{7} = {3-Z}{solid},
}
\toprule 
\toprule
 & Symbol & Prior \\
\midrule
\SetCell[r=12]{c} EOS & $E_{\sat}$ [MeV] & fixed to $-16$ \\
 & $K_{\sat}$ [MeV] & $\mathcal{U}(150,300)$\\
 & $Q_{\sat}$ [MeV] & $\mathcal{U}(-500,1100)$\\
 & $Z_{\sat}$ [MeV] & $\mathcal{U}(-2500,1500)$\\ 
 & $E_{\sym}$ [MeV] & $\mathcal{U}(28,45)$ \\
 & $L_{\sym}$ [MeV] & $\mathcal{U}(10,200)$ \\
 & $K_{\sym}$ [MeV] & $\mathcal{U}(-400,100)$ \\
 & $Q_{\sym}$ [MeV] & $\mathcal{U}(-1000,1500)$ \\
 & $Z_{\sym}$ [MeV] & $\mathcal{U}(-2000, 1500)$\\
 & $n_{\text{break}}$ [$\nsat$] & $\mathcal{U}(1,2)$ \\
 & $n_j$ [$\nsat$] & $\mathcal{U}(n_{\text{break}}, 12)$ \\
 & $c_{s,j}^2$ & $\mathcal{U}(0,1)$ \\
\midrule
\SetCell[r=7]{c} Narrow~mass distribution & $\mu_1$ [$\msun$] & $\mathcal{U}(1, 2)$\\
& $\sigma_1$ [$\msun$] & $\mathcal{U}(0.01, 0.1)$ \\
& $\mu_2$ [$\msun$] & $\mathcal{U}(1, 2)$ \\
& $\sigma_2$ [$\msun$] & $\mathcal{U}(0.01, 0.2)$ \\
& $w$ & $\mathcal{U}(0.3, 1)$ \\
& $m^l$ [$\msun$] & $\mathcal{U}(1, 1.25)$ \\
& $m^u$ [$\msun$] & $\mathcal{U}(1.25, 1.5)$ \\
\midrule
\SetCell[r=3]{c} Wide~mass distribution & $m_{\text{min}}$ [$\msun$] & $\mathcal{U}(1, 2)$ \\
& $m_{\text{max}}$ [$\msun$] & $\mathcal{U}(1.5, 2.5)$ \\
& $\alpha$ & $\mathcal{U}(0, 4)$ \\
\midrule
\SetCell[r=2]{c} Cosmology \linebreak (Sec.~\ref{subsec:cosmo_inference}) & $H_0$~[km~s$^{-1}$~Mpc$^{-1}$] & $\mathcal{U}(50, 100)$ \\
& $\Omega_0$ & $\mathcal{U}(0,1)$\\
\bottomrule
\end{tblr}
\end{table}
Since \ac{EOS} and mass distribution potentially impact each other during inference for $\gtrsim10$ events~\citep{Wysocki:2020myz, Golomb:2021tll}, we infer them simultaneously using the multimessenger events from our narrow+\ett, narrow+ETL, wide+\ett, and wide+ETL catalogues.
As stated above, we restrict ourselves to events with a noticeable \ac{KN}, so the third-to-last row of Table~\ref{tab:detections} lists how many events are analysed per catalogue.
Note that the catalogues always contain one year of \ac{GW} observations.
For each catalogue, we perform the hierarchical analysis twice.
The first time we only use \ac{GW} signals, the second time we also include information from the host galaxy redshift and transient light curve.

\subsubsection{Method and setup}
To obtain the individual \ac{GW} posteriors $p(\theta^{(j)} \mid \text{GW}_j)$, we first perform standard injections and recoveries of the \ac{GW} events using IMRPhenomXAS\_NRTidalv3.
A detailed description of our injection and recovery setup is given in Appendix~\ref{app:gw_injections}.
Using \textsc{flowjax}~\citep{ward2023flowjax}, we train \acp{NF} on the source frame component masses and tidal deformability samples of each \ac{GW} posterior.
When considering only the \ac{GW}-data for hierarchical inference, Eq.~\eqref{eq:hierarchical_likelihood2} takes the form
\begin{align}
\begin{split}
& \mathcal{L}(\mathrm{EOS}, \lambda_{\mathrm{pop}} \mid \mathrm{GW}_j)
=  \\
&\int \biggl[
\frac{p(m_1, m_2, \Lambda_1^{\mathrm{EOS}}(m_1), \Lambda_2^{\mathrm{EOS}}(m_2)
\mid  \mathrm{GW}_j)}{\pi(m_1, m_2)}\ \times \\
& \qquad\ p(m_1, m_2 \mid \lambda_{\mathrm{pop}}) \biggr]
\ dm_1\,dm_2\ .
\label{eq:gw_eos_likelihood}
\end{split}
\end{align}
Here, we divide by the implicit parameter estimation prior on the source frame masses $\pi(m_1, m_2)$ set by the uniform detector frame mass prior and redshift prior.
This is not necessary for $\Lambda_1$ and $\Lambda_2$, because the corresponding priors are uniform.
There are several options to evaluate the integral in Eq.~\eqref{eq:gw_eos_likelihood} numerically.
One option is to sample $m_1, m_2\sim p(m_1, m_2\mid \lambda_{\text{pop}})$ and evaluate the marginalized \ac{GW} posterior density for these samples through a \ac{NF}.
However, since often the \ac{GW} posterior mass range is narrower than the support of the mass distribution, we found it is more efficient and stable to instead draw
\begin{align}
        m_1,\ m_2 \sim p(m_1, m_2 \mid \text{GW}_j)\ ,
\end{align}
evaluate the conditional density $p(\Lambda_1, \Lambda_2~\mid~m_1, m_2, \text{GW}_j)$ through a conditional normalizing flow, and then average over all drawn $(m_1, m_2)$ values
\begin{align}
\begin{split}
    & \mathcal{L}(\text{EOS}, \lambda_{\text{pop}}\mid \text{GW}_j) = \\
    & \sum_{(m_1, m_2)\sim p(\theta\mid \text{GW}_j)}  \biggl[
    \frac{p(\Lambda_1^{\text{EOS}}(m_1), \Lambda_2^{\text{EOS}}(m_2) \mid m_1, m_2, \text{GW}_j)}{\pi(m_1, m_2)} \times \\
    & \qquad \qquad \qquad \qquad  p(m_1, m_2 \mid \lambda_{\text{pop}}) \biggr]\ .
\end{split}
\end{align}
Typically, we take 1000~mass pairs for a balance between numerical stability and performance.

For an \ac{EM}-bright \ac{BNS} merger, we assume that the host galaxy redshift is available.
This has a direct impact on the extraction of the source frame masses.
For a \ac{GW}-only inference, the source frame masses are just determined from the luminosity distance posterior samples and the injected Planck 2018 cosmology.
When incorporating the host galaxy redshift, we use the same \ac{GW} posterior and the same likelihood from Eq.~\eqref{eq:gw_eos_likelihood} as before, the only difference is that the conversion to source-frame masses is performed using the host galaxy redshift rather than the luminosity distance samples from the \ac{GW} posterior.
We assume that the host galaxy redshift $z$ can be measured with a Gaussian uncertainty of 1\% around the injected value.
This measurement error of 1\% represents the uncertainty in cosmological redshift from peculiar velocities~\citep{Hjorth:2017yza, Blake:2025etn}, though for very close events it can also be higher~\citep{Cozzumbo:2024vxw}.
If the merger takes place in a fainter galaxy at larger redshifts, finding the host galaxy might not be guaranteed.
However, as better galaxy catalogues will become available in the future~\citep{Dalya:2018cnd, Dalya:2021ewn, DES:2025key}, the risk of missing the complementary redshift is not too large and we neglect this issue for now.

To additionally incorporate the information from the transient light curve, we first obtain a separate light curve posterior on the ejecta parameters. 
This is different from e.g.~\citet{Dietrich:2020efo} and \\citet{Pang:2022rzc}, where \ac{GW} and \ac{EM} source parameters are sampled from a joint posterior.
Since we transform our completed posteriors back to marginalized likelihoods and combine \ac{GW} signal and light curve as independent measurements, these two approaches are essentially equivalent, although in practice numerical differences could arise.
For each event, we inject a mock light curve $\text{EM}_j$ and recover the source parameters to get a posterior on $\log_{10}(m_{\text{ej, dyn}})$, $\log_{10}(m_{\text{ej, wind}})$, and the other \ac{EM} model parameters.
A detailed description of our injection and recovery method for the \ac{EM} counterpart is given in Appendix~\ref{app:em_injections}.
Since the priors on the dynamical ejecta mass are uniform, we can drop them and write the hierarchical multimessenger likelihood in Eq.~\eqref{eq:hierarchical_likelihood2} as
\begin{align}
\begin{split}
&\mathcal{L}(\text{EOS}, \lambda_{\text{pop}} \mid \text{GW}_j, \text{EM}_j)
= \\
&\int  \biggl[ \frac{p(m_1, m_2, \Lambda_1^{\text{EOS}}(m_1), \Lambda_2^{\text{EOS}}(m_2), d_L, \iota \mid \text{GW}_j)}{\pi(m_1, m_2) \pi(d_L) \pi(\iota)}\times \\
& \qquad \ \frac{p(\log_{10}m_{\text{ej, dyn}}^{\text{fit}} , d_L, \iota \mid \text{EM}_j)}{\pi(d_L) \pi(\iota)}\ \times\\
& \qquad\ \ p(m_1, m_2 \mid \lambda_{\text{pop}})  \biggr]\ dm_1\,dm_2\,d[d_L]\,d\iota\ .
\label{eq:mm_eos_likelihood}
\end{split}
\end{align}
The `fit' superscript means that the dynamical ejecta mass is evaluated by using the fitting relations in Eqs.~(\ref{eq:pc_fittings}) and (\ref{eq:nopc_dyn}) when inserting $m_1$, $m_2$, and the sampled \ac{EOS}. 
In order to decide whether to use the prompt collapse or non-prompt collapse formulae for a particular $(m_1$, $m_2)$ draw, we use the criterion
\begin{align}
    m_1 + m_2 \leq k_{\text{coll}} \times \mtov\ ,
    \label{eq:pc_simple_criterion}
\end{align}
where $k_{\text{coll}}$ is drawn uniformly between 1.2 and 1.6.
This differs from criterion used to generate the \ac{KN} population, which instead relied on the classifier from \citet{Puecher:2024dhl}, as described in Sec.~\ref{subsec:kilonova}.
However, the available version of the classifier is incompatible with our \textsc{jax}-based hierarchical inference framework, preventing direct use of the classifier during sampling.
Nevertheless, Eq.~\eqref{eq:pc_simple_criterion} serves as an adequate approximation to the classifier for $k_{\text{coll}}=1.3$ and $\mtov =2.14~\msun$~\citep{Puecher:2024dhl}.
The inclination and luminosity distance parameters are shared by the \ac{GW} and \ac{EM} inference, thus we integrate over these parameters, so the measurements mutually inform each other.
In practice, this is done by sampling $m_1$, $m_2$, $d_L$, and $\iota$ from the \ac{GW} posterior, converting them to $\log_{10}(m_{\text{ej, dyn}})$ via the sampled \ac{EOS} and the fitting relations and evaluating $p(\log_{10}m_{\text{ej, dyn}} , d_L, \iota \mid \text{EM}_j)$ through a \ac{NF}.
The \ac{GW} likelihood is then evaluated through a conditional normalizing flow for $p(\Lambda_1, \Lambda_2~\mid~m_1, m_2, d_L, \iota)$.
We also point out that we always include the host galaxy redshifts when using the light curve in our hierarchical inferences.
Equation~\eqref{eq:mm_eos_likelihood} carries information from the dynamical ejecta but ignores any information that could be obtained from the wind ejecta mass.
In Appendix~\ref{app:wind_ejecta_likelihood}, we conduct an additional analysis incorporating this information and find that the the general outcome is unchanged.

Equations~\eqref{eq:gw_eos_likelihood} and \eqref{eq:mm_eos_likelihood} are implemented in \textsc{jester}~\citep{Wouters:2025zju}, which enables us to sample directly on the microphysical representation of the \text{EOS}.
Our parametrization uses a fixed crust~\citep{Douchin:2001sv}, and parametrizes the low-density region above the crust-core transition through the meta-model up to fourth order~\citep{Margueron:2017eqc, Margueron:2017lup}, where the parameters are the Taylor expansion coefficients, i.e., the saturation and symmetry energy parameters $E_{\text{sat}}, K_{\text{sat}}, \dots, E_{\text{sym}}, L_{\text{sym}}, \dots, etc.$
Specifically, we use the meta-model implementation as described by~\citet{Somasundaram:2020chb} and \citet{Grams:2021lzx}.
Above a density $n_{\text{break}}$, the meta-model is replaced by \ac{CSE}, where the squared speed-of-sound is linearly extrapolated at six random base points $(n_j, c^2_{s,j})$~\citep{Tews:2018iwm, Somasundaram:2021clp}.
The mass distributions for the recovery are the same as the ones in Sec.~\ref{subsec:bns_catalogues} used to generate the population, i.e., the narrow and wide mass distribution.
The priors for the \ac{EOS} and mass distribution parameters are listed in Table~\ref{tab:hyperpriors}.
In addition to the likelihoods from either Eqs.~\eqref{eq:gw_eos_likelihood} or \eqref{eq:mm_eos_likelihood}, we additionally factor in that the \ac{EOS} should reach the mass of PSR~J1614-2230~\citep{NANOGrav:2023hde} by adding the cumulative normal probability
\begin{align}
\begin{split}
    &\mathcal{L}(\text{EOS} \mid \text{PSR J1614}) = \\&\quad \quad {\rm CDF}_{\mathcal{N}}(M_{\text{TOV}}(\text{EOS}) \mid \mu_{\rm PSR}=1.937~\msun, \sigma_{\rm PSR}=0.037~\msun)
\end{split}
\end{align}
to the overall hierarchical likelihood.
This stabilizes the \ac{EOS} inference and improves convergence.
To then sample Eq.~\eqref{eq:hierarchical_posterior}, we use the sequential \ac{MCMC} algorithm~\citep{del2006sequential, chopin2020introduction} as implemented in \textsc{blackjax}~\citep{Cabezas:2024blackjax}.
Because of GPU hardware acceleration, a hierarchical inference run can be completed within hours on an NVIDIA H100, despite our parameter spaces having up to 30 dimensions.
The exact runtime depends on specific sampler settings, the number of events, and the number of layers for the \acp{NF}.
Our longest run was for the hierarchical multimessenger posterior of our largest catalogue narrow+ETL and took 6.5~h.

After sampling, we correct for selection bias by reweighting our hierarchical posterior samples with $p_{\text{det}}(\lambda)^{-n}$.
Estimating the detection probabilities for the hyperparameter samples requires fast predictions for the \ac{GW} and \ac{KN} detectability.
Especially the last part, while necessary for our set of \ac{KN} events, remains challenging to quantify and therefore we here only coarsely approximate $p_{\text{det}}(\lambda)$ and relegate a more sophisticated treatment to future work.
A detailed description of how we determine the detection probability is given in Appendix~\ref{app:detection_probability}.

\subsubsection{Results}
\begin{figure*}
    \centering
    \includegraphics[width=1\linewidth]{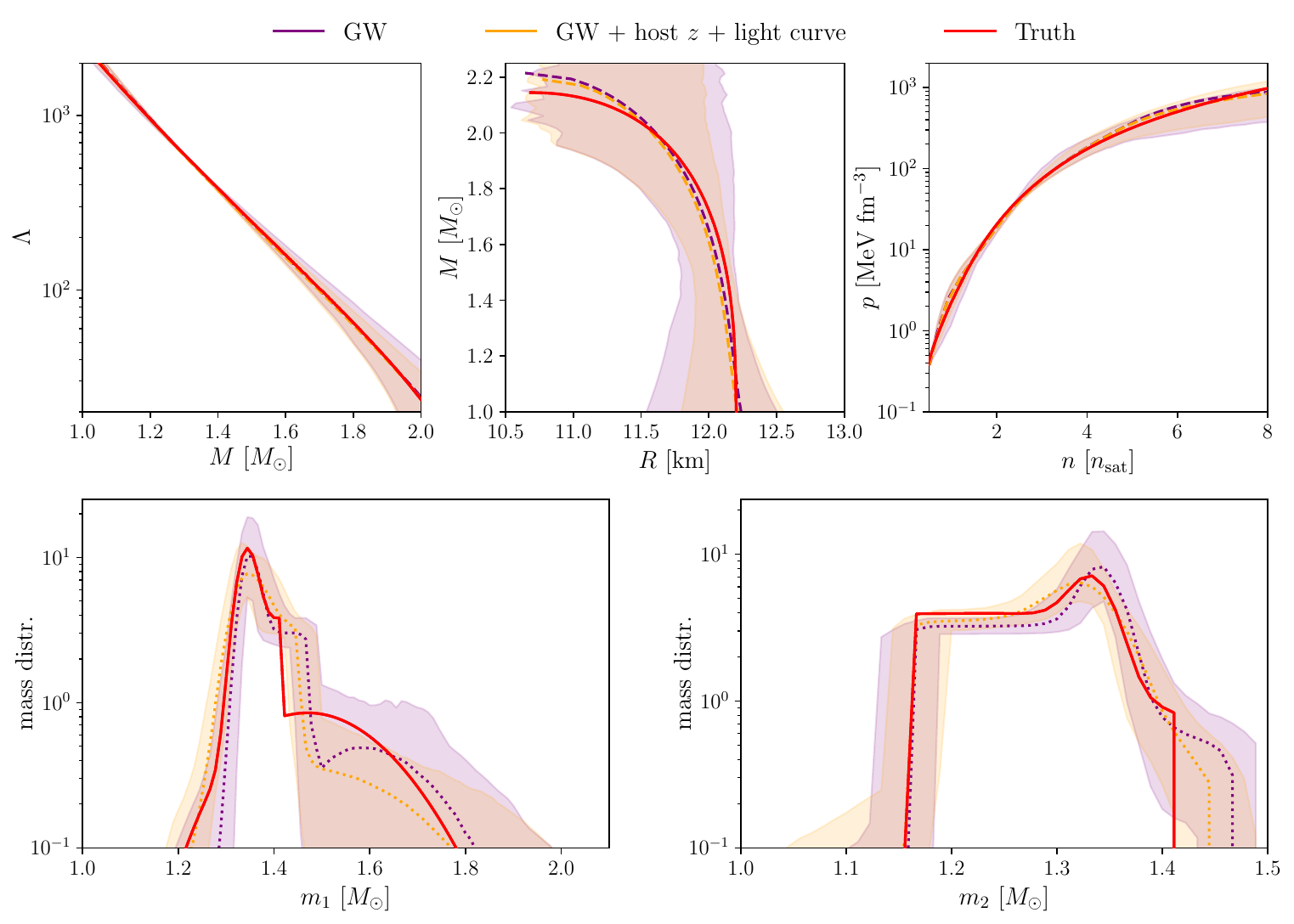}
    \caption[Hierarchical posteriors for EOS and mass distribution from narrow+ETL]{Joint hierarchical inference of the \ac{EOS} and mass distribution from the \ac{KN} events in the narrow+ETL catalogue.
    The posteriors are obtained using either only the \ac{GW} data (purple) or the full multimessenger data (orange).
    The top panels show the \ac{EOS} posterior equivalently in terms of the $M$-$\Lambda$, $M$-$R$, and $n$-$p$ relationship with the 95\% credibility limits at each mass or density (shaded areas) and the highest-likelihood \ac{EOS} (dashed line).
    The bottom panels show the marginalized mass distribution density for primary and secondary mass, 
    with the 95\% credibility limit at each mass (shaded areas) and the median (dotted line). 
    The injected \ac{EOS} and mass distribution are shown as red lines.}
    \label{fig:eos_inference_narrow_ETL}
\end{figure*}

\begin{figure}
    \centering
    \includegraphics[width=1\linewidth]{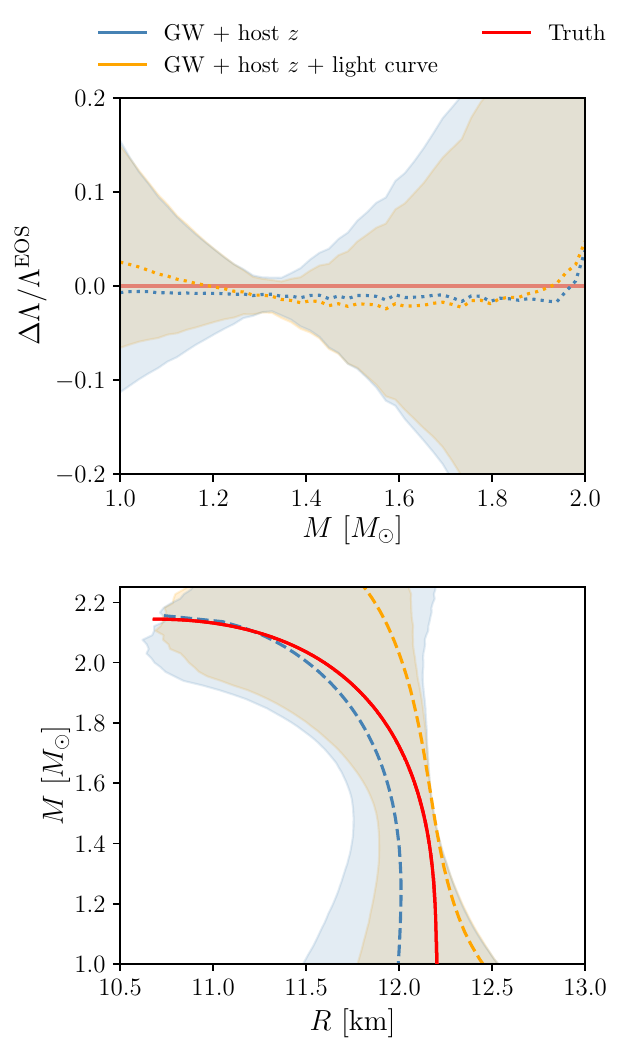}
    \caption[Comparison of EOS posteriors from narrow+ETL]{\ac{EOS} posteriors from the \ac{KN} events in narrow+ETL when the full multimessenger information is used (orange) and when only based on the \ac{GW} signals and host galaxy redshifts (blue).
    The top panel shows the 95\% credibility intervals on the relative error for the $M$-$\Lambda$ relationship (shaded areas) and the median of the relative error (dotted line).
    The bottom panel displays the $M$-$R$ relationships, with the same layout as the corresponding panel in Fig.~\ref{fig:eos_inference_narrow_ETL}.
    For a better comparison, the posteriors in this figure are not reweighted with the detection probabilities.
    }
    \label{fig:eos_inference_narrow_ETL_mm_vs_gw}
\end{figure}

\begin{figure}
    \centering
    \includegraphics[width=1\linewidth]{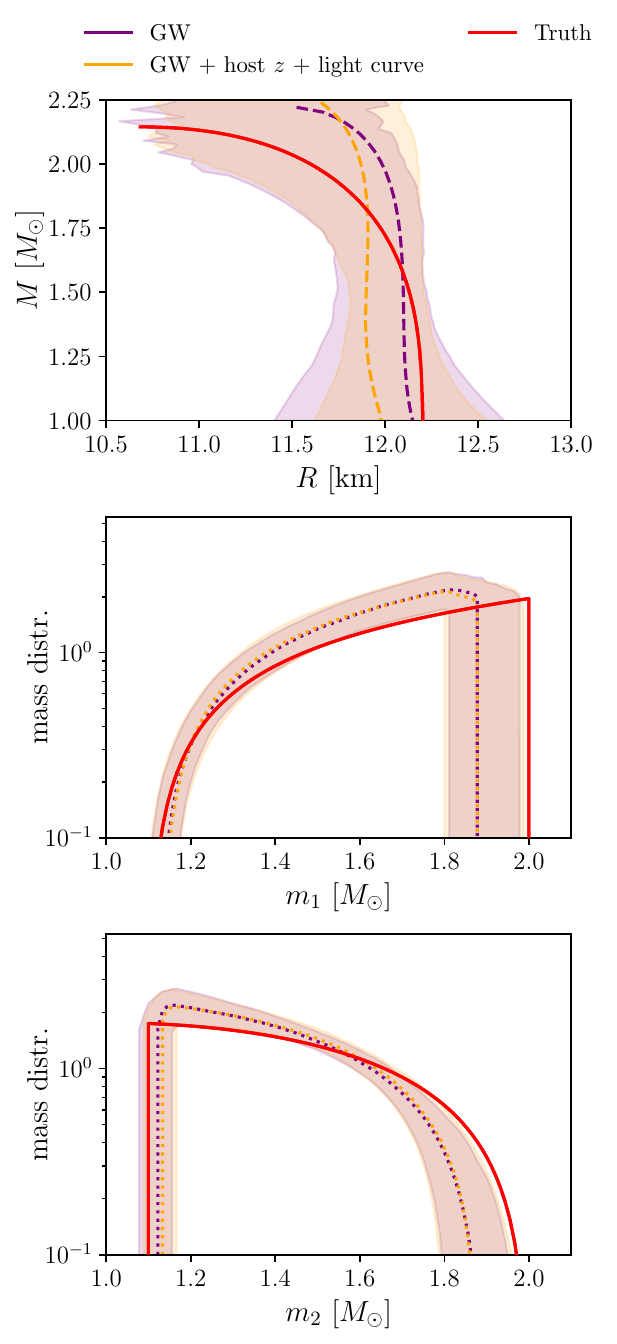}
    \caption[Hierarchical posteriors for EOS and mass distribution from wide+ETL]{\ac{EOS} and mass distribution inference from the \ac{KN} events in the wide+ETL catalogue.
    The layout is equivalent to Fig.~\ref{fig:eos_inference_narrow_ETL}, except we only show the $M$-$R$ representation of the \ac{EOS} posterior in the top panel.
    }
    \label{fig:eos_inference_wide_ETL}
\end{figure}

\begin{figure}
    \centering
    \includegraphics[width=1\linewidth]{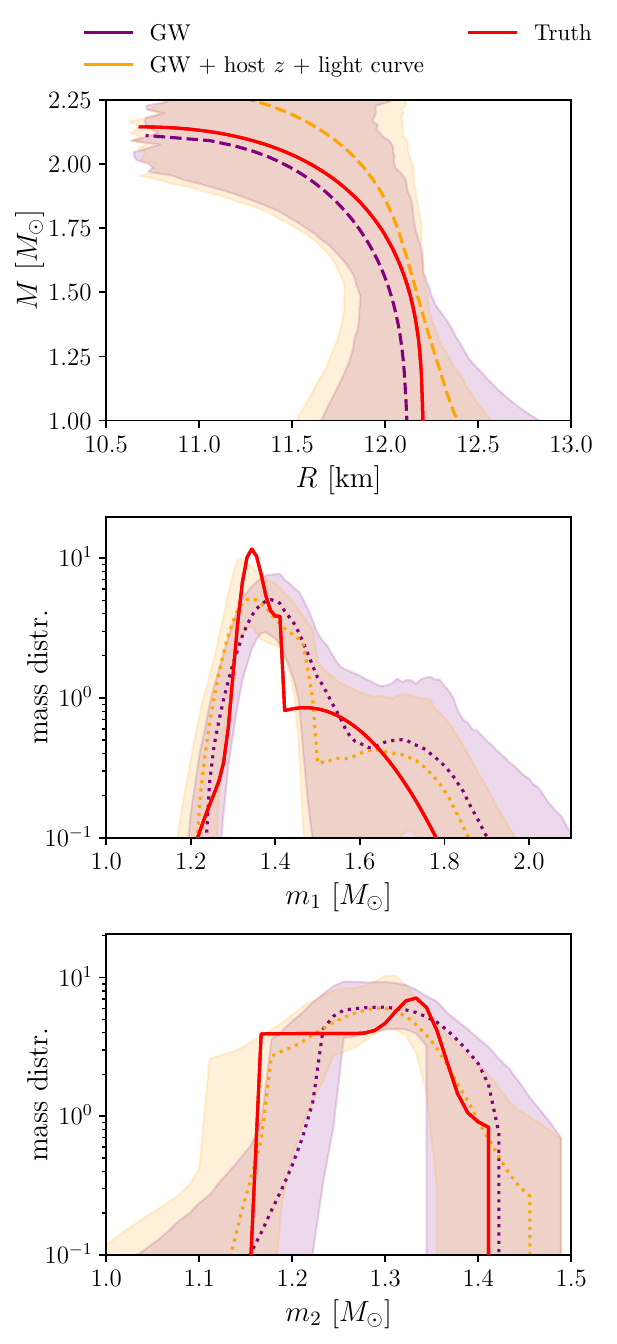}
    \caption[Hierarchical posteriors for EOS and mass distribution from narrow+\ett]{\ac{EOS} and mass distribution inference from the \ac{KN} events in the narrow+\ett catalogue.
    The layout is equivalent to Fig.~\ref{fig:eos_inference_wide_ETL}.
    }
    \label{fig:eos_inference_narrow_ETT}
\end{figure}

\begin{figure}
    \centering
    \includegraphics[width=1\linewidth]{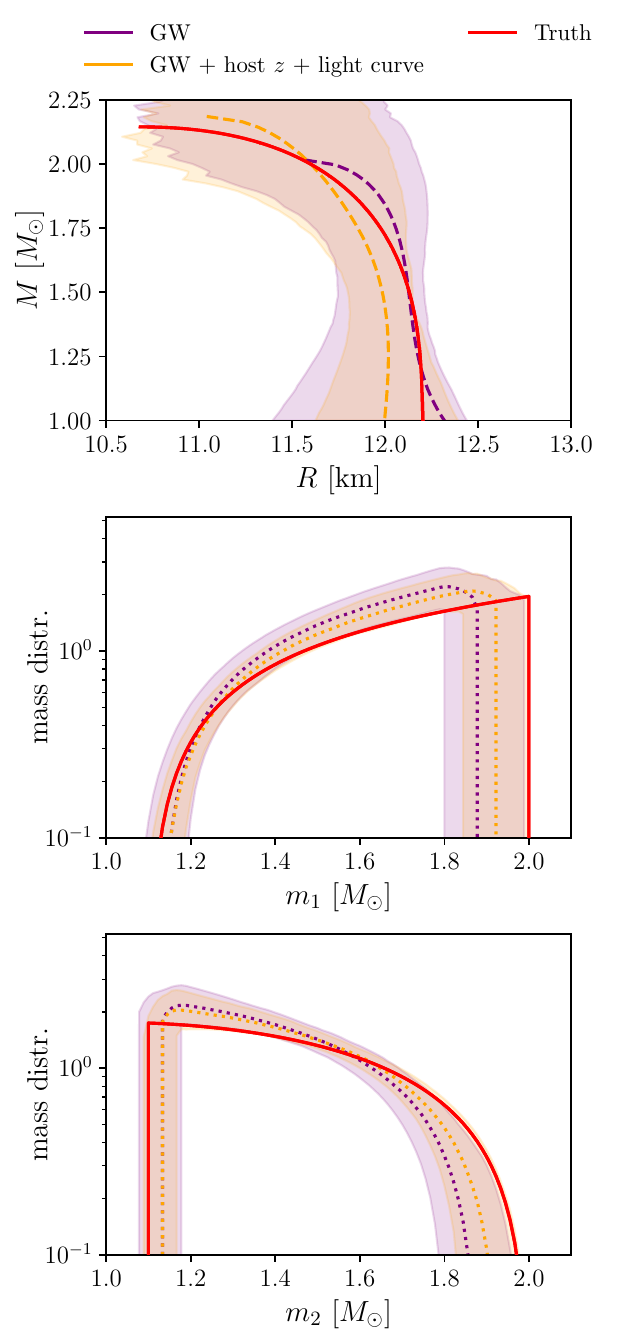}
    \caption[Hierarchical posteriors for EOS and mass distribution from wide+\ett]{\ac{EOS} and mass distribution inference from the \ac{KN} events in the wide+\ett catalogue.
    The layout is equivalent to Fig.~\ref{fig:eos_inference_wide_ETL}.
    }
    \label{fig:eos_inference_wide_ETT}
\end{figure}

\begin{figure}
    \centering
    \includegraphics[width=1\linewidth]{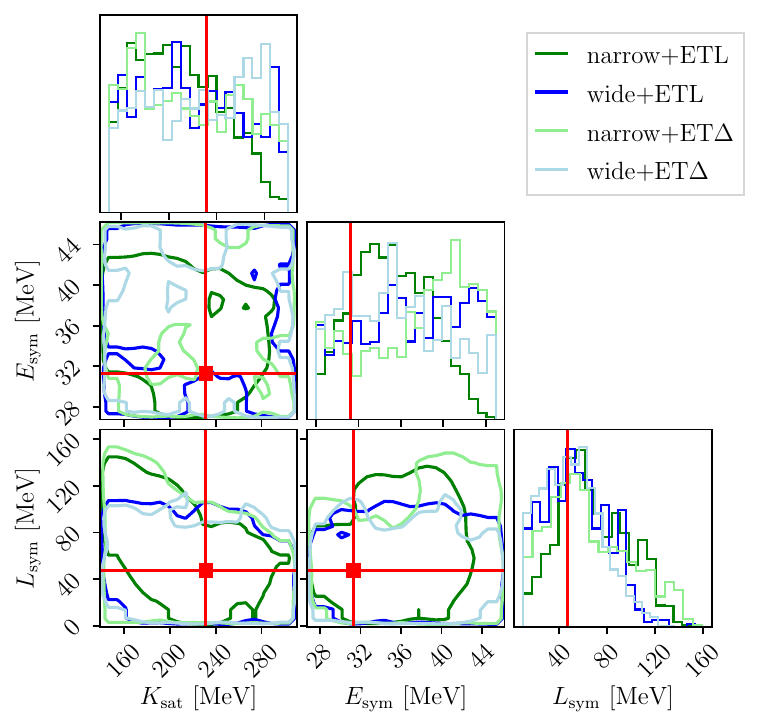}
    \caption[Hierarchical posteriors for nuclear empirical parameters]{Posterior on the nuclear empirical parameters from the multimessenger inferences. 
    We show the 95\% credibility contours in different colours according to the legend.
    The values from our injected QMC-RMF3 \ac{EOS} are marked in red.
    }
    \label{fig:NEP_posterior}
\end{figure}

\begin{table}
\centering
\caption[Credibility intervals for hierarchical posteriors on EOS and mass distribution]{
    Posterior credibility intervals for the inference results from Sec.~\ref{subsec:eos_inference}.
    We show the 95\% credibility limits on \ac{EOS} and mass distribution parameters.
    The first value corresponds to the result from the analysis with \ac{GW} data alone, the bracketed value to the one with multimessenger information.
}
\label{tab:eos_inference_results}
\begin{tblr}{
    colspec = {Q[c, 0.8cm] Q[c, 1.2cm]  Q[c, 1.2cm] Q[c, 1.2cm]  Q[c, 1.2cm] Q[c, 0.6cm]},
}
\toprule 
\toprule
 & Narrow+ETL & Wide+ETL & Narrow+\ett & Wide+\ett  & Truth\\
\midrule 
\SetCell[r=2]{c} $\Lambda_{1.4}$ & ${382}^{+13}_{-13}$ & ${399}^{+19}_{-26}$ & ${381}^{+18}_{-15}$ & ${391}^{+28}_{-30}$ &\SetCell[r=2]{c} 383 \\ 
 & $({376}^{+12}_{-11})$ & $({400}^{+19}_{-22})$ & $({380}^{+16}_{-17})$ & $({380}^{+24}_{-22})$ &\\[0.5em] 
\SetCell[r=2]{c} $R_{1.4}$ \newline [km] & ${12.08}^{+0.14}_{-0.30}$ & ${12.00}^{+0.26}_{-0.31}$ & ${12.06}^{+0.26}_{-0.20}$ & ${11.99}^{+0.24}_{-0.26}$ &\SetCell[r=2]{c} 12.16 \\ 
 & $({12.04}^{+0.18}_{-0.14})$ & $({12.02}^{+0.18}_{-0.19})$ & $({12.03}^{+0.21}_{-0.26})$ & $({11.99}^{+0.18}_{-0.18})$ &\\[0.5em] 
\SetCell[r=2]{c} $M_{\rm{TOV}}$ \newline $[\msun]$ & ${2.11}^{+0.16}_{-0.16}$ & ${2.15}^{+0.09}_{-0.18}$ & ${2.13}^{+0.14}_{-0.18}$ & ${2.14}^{+0.12}_{-0.19}$ &\SetCell[r=2]{c} 2.14 \\ 
 & $({2.09}^{+0.24}_{-0.14})$ & $({2.23}^{+0.14}_{-0.25})$ & $({2.16}^{+0.21}_{-0.21})$ & $({2.12}^{+0.20}_{-0.17})$ &\\[0.5em] 
\SetCell[r=2]{c} $p(3\nsat)$ [MeV~fm$^{{-3}}$] & ${76}^{+25}_{-11}$ & ${85}^{+26}_{-14}$ & ${73}^{+21}_{-10}$ & ${78}^{+26}_{-12}$ &\SetCell[r=2]{c} 75 \\ 
 & $({72}^{+16}_{-7})$ & $({89}^{+28}_{-16})$ & $({75}^{+28}_{-13})$ & $({74}^{+18}_{-11})$ &\\[0.5em] 
\midrule 
\SetCell[r=2]{c} $\mu_1~[\msun]$ & ${1.35}_{-0.02}^{+0.02}$ & - & ${1.37}_{-0.07}^{+0.06}$ & - & \SetCell[r=2]{c} 1.34 \\ 
 & $({1.34}_{-0.02}^{+0.03})$ & - & $({1.32}_{-0.03}^{+0.06})$ & - &  \\[0.5em] 
\SetCell[r=2]{c} $\sigma_1~[\msun]$ & ${0.02}_{-0.01}^{+0.02}$ & - & ${0.07}_{-0.05}^{+0.03}$ & - & \SetCell[r=2]{c} 0.02 \\ 
 & $({0.03}_{-0.02}^{+0.02})$ & - & $({0.05}_{-0.03}^{+0.05})$ & - &  \\[0.5em] 
\SetCell[r=2]{c} $\mu_2~[\msun]$ & ${1.60}_{-0.19}^{+0.18}$ & - & ${1.70}_{-0.28}^{+0.19}$ & - & \SetCell[r=2]{c} 1.47 \\ 
 & $({1.50}_{-0.14}^{+0.25})$ & - & $({1.66}_{-0.27}^{+0.13})$ & - &  \\[0.5em] 
\SetCell[r=2]{c} $\sigma_2~[\msun]$ & ${0.14}_{-0.08}^{+0.06}$ & - & ${0.13}_{-0.10}^{+0.07}$ & - & \SetCell[r=2]{c} 0.15 \\ 
 & $({0.17}_{-0.07}^{+0.03})$ & - & $({0.12}_{-0.08}^{+0.07})$ & - &  \\[0.5em] 
\SetCell[r=2]{c} $w$ & ${0.81}_{-0.22}^{+0.14}$ & - & ${0.81}_{-0.49}^{+0.13}$ & - & \SetCell[r=2]{c} 0.68 \\ 
 & $({0.84}_{-0.18}^{+0.12})$ & - & $({0.84}_{-0.45}^{+0.11})$ & - &  \\[0.5em] 
\SetCell[r=2]{c} $m^l~[\msun]$ & ${1.16}_{-0.03}^{+0.02}$ & - & ${1.23}_{-0.06}^{+0.02}$ & - & \SetCell[r=2]{c} 1.16 \\ 
 & $({1.16}_{-0.02}^{+0.03})$ & - & $({1.17}_{-0.07}^{+0.07})$ & - &  \\[0.5em] 
\SetCell[r=2]{c} $m^u~[\msun]$ & ${1.47}_{-0.03}^{+0.03}$ & - & ${1.42}_{-0.07}^{+0.07}$ & - & \SetCell[r=2]{c} 1.42 \\ 
 & $({1.45}_{-0.04}^{+0.04})$ & - & $({1.47}_{-0.11}^{+0.03})$ & - &  \\[0.5em] 
\midrule 
\SetCell[r=2]{c} $m_{\rm min}$ $[\msun]$ & - & ${1.14}_{-0.05}^{+0.03}$ & - & ${1.13}_{-0.06}^{+0.05}$ & \SetCell[r=2]{c} 1.1 \\ 
 & - & $({1.14}_{-0.04}^{+0.03})$ & - & $({1.13}_{-0.05}^{+0.04})$ &  \\[0.5em] 
\SetCell[r=2]{c} $m_{\rm max}$ $[\msun]$ & - & ${1.89}_{-0.06}^{+0.12}$ & - & ${1.88}_{-0.07}^{+0.11}$ & \SetCell[r=2]{c} 2.0 \\ 
 & - & $({1.90}_{-0.07}^{+0.09})$ & - & $({1.93}_{-0.08}^{+0.08})$ &  \\[0.5em] 
\SetCell[r=2]{c} $\alpha$ & - & ${1.42}_{-1.25}^{+1.91}$ & - & ${1.46}_{-1.42}^{+2.32}$ & \SetCell[r=2]{c} 2.0 \\ 
 & - & $({1.41}_{-1.27}^{+2.24})$ & - & $({1.73}_{-1.65}^{+1.89})$ &  \\[0.5em] 
\bottomrule 

\end{tblr}
\end{table}

In Table~\ref{tab:eos_inference_results}, we list all posterior credibility intervals of our hierarchical posteriors.
Unless otherwise specified, we quote credibility intervals at the 95\% level.

We first discuss the posteriors from narrow+ETL, the largest catalogue.
In Fig.~\ref{fig:eos_inference_narrow_ETL}, we show the joint inference of the \ac{EOS} and mass distribution from the corresponding \ac{KN} events.
Specifically, we compare the \ac{GW}-only posterior (in purple) that just uses information from the \ac{GW} signals to the full multimessenger posterior (in orange), where additional information from host galaxy redshifts and transient light curves is incorporated.
In both cases, the $M$-$\Lambda$ relationship is well constrained.
For $M\leq 1.7~\msun$, the relative uncertainty on $\Lambda$ stays within 20\% and gets as low as 3\% around 1.4~$\msun$.
Interestingly, the relative uncertainty for the $M$-$R$ relationship is minimized at a slightly larger mass of 1.55~$\msun$. 
We attribute this to common correlations between the tidal deformability around 1.4~$\msun$ and the radii at slightly larger masses~\citep{Yagi:2016bkt, Suleiman:2024ztn}.
Despite the fact that higher-mass radii are slightly better determined, we still quote the canonical radius of a 1.4~$\msun$ \ac{NS} $R_{1.4}$, which here is constrained to ${12.08}^{+0.14}_{-0.30}~({12.04}^{+0.18}_{-0.14})$~km from the posterior using GW-only (multimessenger) information.
The injected value lies at $R_{1.4} = 12.16$ km.
Above 1.7~$\msun$, the uncertainty for the \ac{EOS} increases significantly, since only one event contains a \ac{NS} heavier than this limit.
The only other constraint at high masses is from the requirement that our \acp{EOS} must support the mass of PSR J1614-2230.
As a consequence, any constraint on $\mtov$ remains relatively loose.
Similarly, the lightest \ac{NS} from our events is 1.15~$\msun$ and so the posterior uncertainty for the \ac{EOS} increases below 1.2~$\msun$.
However, the relative uncertainty here is smaller than for high masses.
Further, we note that while the $M$-$\Lambda$ relationship is very well constrained, the \ac{NS} radii tend to be slightly underestimated in both the \ac{GW} and multimessenger posterior, and also in inferences from other catalogues that will be discussed below.
It is known how one $M$-$\Lambda$ curve can be well approximated by slightly different \acp{EOS}~\citep{Raithel:2022aee} and hence there are slightly different $M$-$R$ representations for the same tight $M$-$\Lambda$ constraint.
Moreover, our generic \ac{EOS} extrapolation scheme \ac{CSE} can only approximate our injected QMC-RMF3 \ac{EOS} within a certain accuracy.
Given the tight $M$-$\Lambda$ relationship imposed by the \ac{GW} signals, the sampler apparently is able to fit this using \ac{CSE} \acp{EOS} that have slightly smaller radii than our injected \ac{EOS}.
As a consequence, the radii of our injected \ac{EOS} and of our \ac{EOS} samples from the hyperposterior do not align perfectly at every mass interval and exhibit a modest bias.
This effect is inherent to our method, since we sample directly on the microphysical \ac{EOS} description and our likelihood is dominated by tidal deformability measurements.
The $M$-$\Lambda$ relationship is so well constrained because of the relatively high \ac{GW}-\ac{SNR} of our events.
If the error on $\Lambda$ was larger or the radius itself was measured, e.g. through high-precision X-ray pulse profiles~\citep{STROBE-XScienceWorkingGroup:2019cyd}, our method would likely not experience this modest bias towards smaller radii.
Also, when repeating the exact same analyses but with different initial sampling seeds, we observe that the $M$-$\Lambda$ posterior remains unchanged, but the 95\% credibility limit on the radii, especially at lower masses, can shift around as much as 0.1~km.

With regards to the mass distribution, we find that its general shape is well recovered.
Yet, some hypersamples experience large deviations from the true mass distribution, especially with respect to the effective range of $m_1$ and $m_2$.
According to our hierarchical posteriors, the 2$\sigma$ upper cutoff for $m_1$ could lie between $\sim 1.5~\msun$ and $\gtrsim 1.8~\msun$.
In particular, while the parameters of the dominant peak in the double Gaussian distribution are relatively well recovered, see $\mu_1$ and $\sigma_1$ in Table~\ref{tab:eos_inference_results}, the parameters of the second Gaussian peak remain relatively broad.
This is because there are more \acp{NS} from the first peak, especially after selecting only those events with a \ac{KN}.
Hence, the inference also slightly overestimates the dominance of the first peak by finding $w={0.81}_{-0.22}^{+0.14}~({0.84}_{-0.18}^{+0.12})$, the true value being 0.68.
As a consequence, $m^u$, which is correlated with $w$, is slightly overestimated.
Usually, the reweighting the posterior with the detection probabilities should overcome the bias of preferring the first Gaussian peak, but given the imperfect evaluation of the detection probability, a modest bias remains.
This shows how degeneracies in the mass model can make point estimates for specific model parameters unreliable, despite the fact that the general shape of the distribution is well-captured.
Unlike the \ac{EOS}, the mass distribution is rather sensitive to reweighting the samples with the detection probability.
Therefore, when analysing real data, modelling choices in $p_{\text{det}}(\lambda)$ could easily bias the inferred mass distribution.

When comparing the posteriors using \ac{GW} data alone and the ones containing the full multimessenger information, Fig.~\ref{fig:eos_inference_narrow_ETL} illustrates how both deliver very similar \ac{EOS} constraints, though small differences remain.
To determine whether these differences arise from the measured host redshifts and consequently tighter source-frame masses or the transient light-curve data, we run a third inference that uses the host galaxy redshifts and the \ac{GW} signal, but not the light curves.
The resulting \ac{EOS} posteriors are compared in Fig.~\ref{fig:eos_inference_narrow_ETL_mm_vs_gw}.
The top panel shows the relative error on the tidal deformability, the bottom panel the inferred $M$-$R$ relationship.
Both results are very similar, although the 95\% limits for both $\Lambda$ and radii are slightly narrower for the full multimessenger posterior.
However, for masses above 1.2~$\msun$, the difference in the radii is at most 0.15~km and therefore comparable to the aforementioned sampling uncertainties.
Additionally, the 95\% limits move more than the median or other quantiles, hence the radius posterior distributions in this mass regime are more similar than the representation in Fig.~\ref{fig:eos_inference_narrow_ETL_mm_vs_gw} might suggest.
Therefore, these differences are likely artefactual.
Below 1.2~$\msun$, the shift in radii and tidal deformability becomes more pronounced, but it does not exceed 0.3~km or a few percent in tidal deformability.
Out of the 78 events in total, only 11 have a component mass of less than 1.2~$\msun$, so it seems possible that light-curve data could be more impactful to the inference in this mass regime where the tidal deformability constraints from the \acp{GW} are not so stringent as for higher masses.
However, this conclusion is not robust from the results presented here.
Overall, Figs.~\ref{fig:eos_inference_narrow_ETL} and \ref{fig:eos_inference_narrow_ETL_mm_vs_gw} indicate how the information from the transient light curves, when added to a series of relatively high-\ac{SNR} events, has only a negligible impact on the \ac{EOS}, if any.
Similarly, we find that the recovery of the mass distribution is enhanced by the host galaxy redshifts, but not by adding the light-curve data.

In Fig.~\ref{fig:eos_inference_wide_ETL}, we show the corresponding \ac{EOS} and mass distribution posteriors from the wide+ETL catalogue.
The overall constraint on the \ac{EOS} is slightly worse compared to the narrow+ETL inference, although the difference is marginal.
The GW-only and multimessenger posteriors are again very similar.
Small differences in radii and tidal deformabilities likely arise from the previously mentioned effects and the transient light-curve data continues to have neglible impact.
Despite the fact that the wide mass distribution contains heavier \acp{BNS} than the narrower one, the \ac{EOS} posterior is still best constrained around 1.4~$\msun$. 
This is because by selecting only events with a \ac{KN}, \acp{BNS} with moderate masses around 1.4~$\msun$ or smaller mass ratios are preferred.
The heaviest \ac{NS} in wide+ETL reaches $1.97~\msun$, but heavier \acp{BNS} have less clear tidal signatures in their signals, which then automatically translates into higher relative uncertainties on $\Lambda$ at higher masses.
With regards to the mass distribution, the limits on $m_{\text{min}}$ and $m_{\text{max}}$ are relatively well constrained for both the \ac{GW} and multimessenger posterior.
However, unlike for the narrow mass distribution, the host galaxy redshifts do not yield improved constraints on the mass distribution parameters.
Because the parametric form of the wide mass model is rather simple, 
the parameters $m_{\text{max}}$ and $m_{\text{min}}$ do not suffer from any degeneracies and are thus relatively easy to constrain.
On the other hand, the mass ratio power $\alpha$ is slightly harder to infer and its posterior support remains broad, suggesting more than 40 events are needed to confidently constrain this parameter.
We also briefly note that the upper limit $m_{\text{max}}$ is slightly underestimated because of the selection effect and imperfect reweighting.

In Fig.~\ref{fig:eos_inference_narrow_ETT} and Fig.~\ref{fig:eos_inference_wide_ETT}, we show the results from the inferences with the narrow+\ett and wide+\ett catalogues.
In general, they show similar behaviour to the corresponding inferences with ETL.
Despite the smaller number of events in the \ett catalogues, the \ac{EOS} inferences provide similar accuracy compared to the ETL inferences.
This is because the \ac{EOS} inference is dominated by a few events with relatively high \ac{SNR} and clear tidal signatures.
Many low-\ac{SNR} events essentially carry no tidal information and thus the constraint on the \ac{EOS} does not scale directly with the number of events.
Although the \ett detector usually observes lower \acp{SNR} than ETL, narrow+\ett, and wide+\ett still contain five and three events with an \ac{SNR} above 80, the highest \acp{SNR} are 326 and 235, respectively.
The best tidal deformabilities are constrained within 38\% and 45\% (half of the 95\% credibility interval).
These same events are also in the corresponding ETL catalogues with higher \ac{SNR}, but the posterior uncertainty on $\Lambda_1$ and $\Lambda_2$ is comparable.
Regarding the mass distribution, we again find that the inference of the wide mass distribution does not benefit much from the host galaxy redshifts, while some parameters of the narrow mass model improve in the multimessenger posterior.
This is the same mechanism as for the ETL inferences, where the tighter source frame masses help to alleviate some of the narrow mass model's degeneracies.
The wide mass distribution's parametrization is simpler and thus the result remains relatively unaffected by additional data.
In fact, even with multimessenger information the posteriors from wide+\ett in Fig.~\ref{fig:eos_inference_wide_ETT} miss the true $m_{\text{max}}$ of 2.0~$\msun$.
This is completely analogous to wide+ETL, and the displacement from the true upper mass limit arises again from heavy \ac{BNS} undergoing prompt collapse and being underrepresented.

The results presented here are based on injections that rely exclusively on the QMC-RMF3 \ac{EOS} introduced in Sec.~\ref{sec:events}.
Naturally, the choice for the injected \ac{EOS} itself impacts the uncertainties with which it can be recovered, because stiffer \acp{EOS} will lead to clearer tidal signals.
To assess the robustness of our method, we repeat our hierarchical inference for the wide+\ett catalogue in Appendix~\ref{app:wide_ETT_stiffer} with a stiffer \ac{EOS}.

As a final remark, we comment on the possibility to constrain the microphysical description of the \ac{EOS}.
At densities below 2$\nsat$ and above 5$\nsat$, the relative 95\% credibility limits on the pressure usually cover a factor of 2.
The pressure is most constrained at densities between 2$\nsat$ and 3$\nsat$, at 3$\nsat$, we find that the relative error on $p(3\nsat)$ spans roughly 20\%, see Table~\ref{tab:eos_inference_results}.
While this would certainly represent a large improvement over current estimates for the pressure-density relationship in \ac{NS} matter, it also shows how reconstructing the microscopic description of (supra-)nuclear matter from \ac{BNS} observables is not straightforward and, for low and high densities, will require additional input from nuclear theory and experiment.
This is confirmed by Fig.~\ref{fig:NEP_posterior}, where we show the posterior on the meta-model parameters around saturation density from our multimessenger inferences.
The only parameter that experiences some robust notable constraint is $L_{\rm sym}$, but even for narrow+ETL, the uncertainty remains broad with $L_{\rm sym} = {64}_{-45}^{+60}~$MeV. 
The value from our injected QMC-RMF3 \ac{EOS} is 47~MeV.
The inability to constrain individual meta-model parameters, even with many precise multimessenger observations, mainly originates from degeneracies in its parametrization~\citep{Mondal:2021vzt,Iacovelli:2023nbv,Wouters:2025zju}.

\subsection{Hierarchical inference including cosmology}
\label{subsec:cosmo_inference}
\begin{figure}
    \centering
    \includegraphics[width=\linewidth]{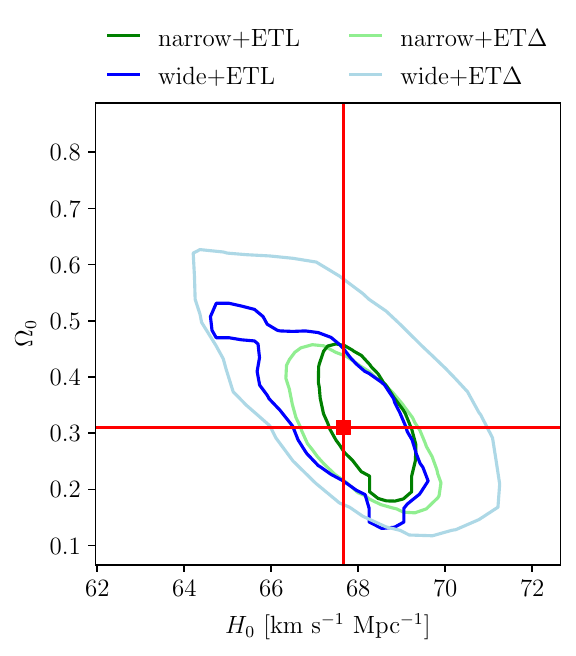}
    \caption[Hierarchical posteriors for the cosmology from different mock catalogues]]{Cosmological constraints from the \ac{KN} events in all our catalogues, compared in different colours according to the legend.
    We show the 95\% credibility region for $H_0$ and $\Omega_0$, with the injected Planck18 parameters shown marked by red lines.}
    \label{fig:cosmo_inference_all}
\end{figure}

\begin{figure*}
    \centering
    \includegraphics[width=1\linewidth]{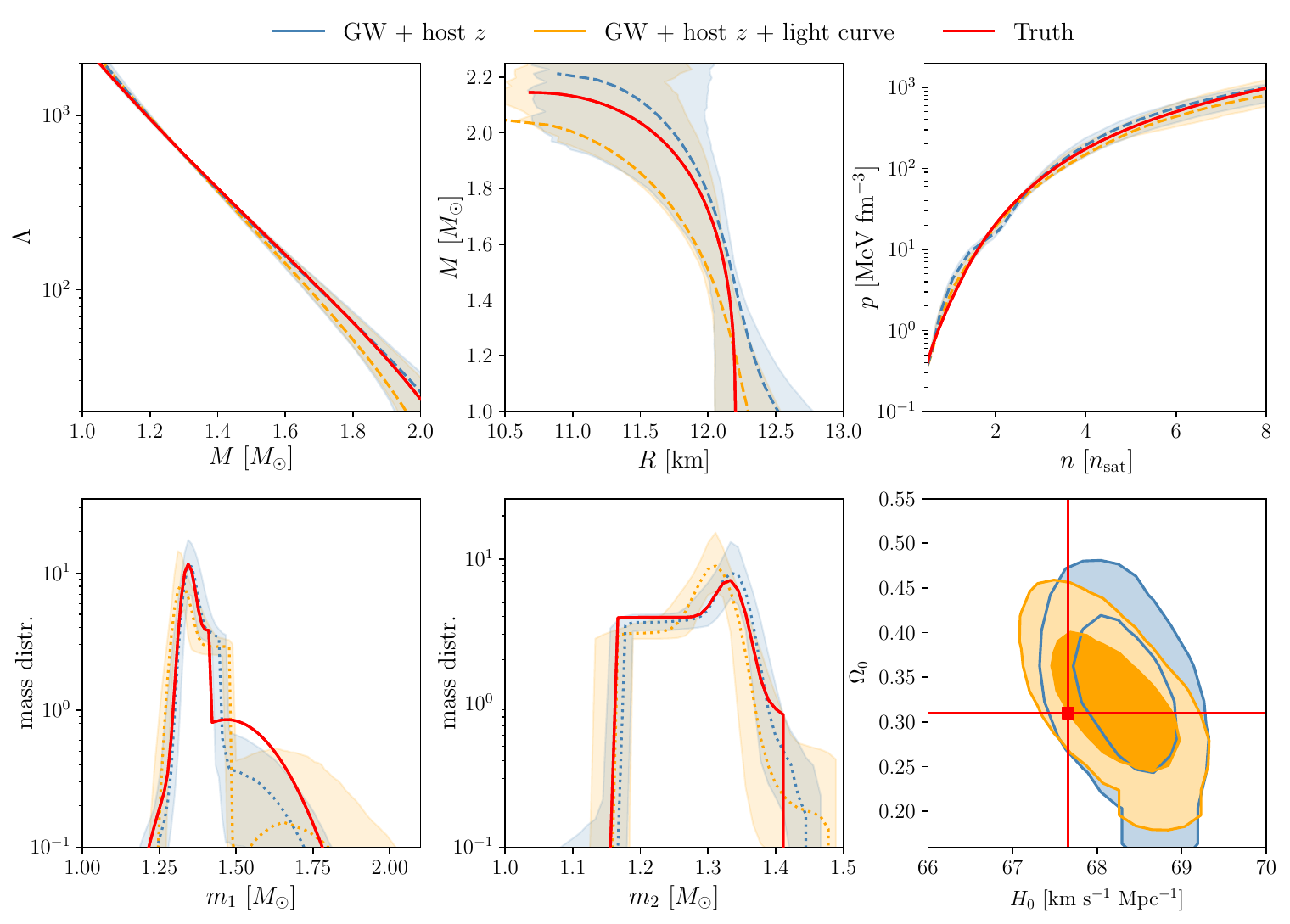}
    \caption[Hierarchical posteriors for EOS, mass distribution, and cosmology from narrow+ETL]{Joint hierarchical inference of the \ac{EOS}, mass distribution, $H_0$, and $\Omega_0$ from the \ac{KN} events in the narrow+ETL catalogue.
    We show both the multimessenger posterior (orange) and the posterior using the \ac{GW} data and host galaxy redshift, but no light curve posteriors (blue).
    The layout is equivalent to Fig.~\ref{fig:eos_inference_narrow_ETL}, 
    except that in the bottom right panel, we show the 68\% and 95\% credibility regions for the cosmological parameters.
    }
    \label{fig:cosmo_inference_narrow_ETL}
\end{figure*}

\begin{table}
\centering
\caption{
    Posterior credibility intervals for the inference results from Sec.~\ref{subsec:cosmo_inference}.
    We show the 95\% credibility limits on \ac{EOS} and cosmological parameters from the multimessenger posteriors.
    For brevity, we only list the mass distribution parameters in Appendix~\ref{app:detection_probability}.
}
\label{tab:cosmo_inference_results}
\begin{tblr}{
    colspec = {Q[c, 0.8cm] Q[c, 1.2cm]  Q[c, 1.2cm] Q[c, 1.2cm]  Q[c, 1.2cm] Q[c, 0.6cm]},
}
\toprule 
\toprule
 & Narrow+ETL & Wide+ETL & Narrow+\ett & Wide+\ett  & Truth\\
\midrule 
$\Lambda_{1.4}$ & ${371}^{+12}_{-11}$ & ${395}^{+18}_{-22}$ & ${380}^{+16}_{-14}$ & ${379}^{+20}_{-23}$ &383 \\[0.5em] 
$R_{1.4}$ \newline [km] & ${12.12}^{+0.08}_{-0.11}$ & ${12.05}^{+0.14}_{-0.17}$ & ${11.95}^{+0.25}_{-0.20}$ & ${11.99}^{+0.15}_{-0.21}$ &12.16 \\[0.5em] 
$M_{\rm{TOV}}$ \newline $[\msun]$ & ${2.11}^{+0.15}_{-0.13}$ & ${2.21}^{+0.14}_{-0.24}$ & ${2.08}^{+0.25}_{-0.13}$ & ${2.13}^{+0.21}_{-0.18}$ &2.14 \\[0.5em] 
$p(3\nsat)$ [MeV~fm$^{{-3}}$] & ${70}^{+14}_{-6}$ & ${86}^{+24}_{-15}$ & ${73}^{+27}_{-9}$ & ${74}^{+18}_{-11}$ &75 \\[0.5em] 
\midrule 
$H_0$ [km~s$^{-1}$ Mpc$^{-1}$] & ${68.2}_{-0.9}^{+0.9}$ & ${67.5}_{-1.7}^{+1.4}$ & ${68.1}_{-1.2}^{+1.4}$ & ${68.1}_{-3.2}^{+2.3}$ & 67.7 \\[0.5em] 
$\Omega_0$ & ${0.32}_{-0.07}^{+0.11}$ & ${0.34}_{-0.13}^{+0.15}$ & ${0.30}_{-0.11}^{+0.11}$ & ${0.34}_{-0.15}^{+0.22}$ & 0.31 \\[0.5em] 
\bottomrule 

\end{tblr}
\end{table}
Aside from information about the \ac{EOS} and mass distribution, \ac{EM}-bright \ac{BNS} mergers provide joint measurements of the luminosity distance and redshift if their host galaxy is identified.
Thus, it is also possible to deduce constraints on the the local cosmology.
In this section, we expand the previous analyses from Sec.~\ref{subsec:eos_inference} to additionally include the Hubble constant $H_0$ and matter density parameter $\Omega_0$ as hyperparameters in our hierarchical inference. 
The conditional distribution linking the relevant source properties, i.e., $d_L$ and $z$, to the cosmological parameters $H_0$ and $\Omega_0$ is simply
\begin{align}
    p(d_L \mid z, H_0, \Omega_0) = \delta(d_L(z) - d_L)\ ,
\end{align}
where $d_L(z)$ is a well-known function for a flat \ac{LCDM} universe
\begin{align}
    d_L(z) = \frac{c (1+z)}{H_0} \int_0^z \frac{1}{\sqrt{\Omega_0 (1+z')^3 + (1-\Omega_0)}}\ dz'.
\end{align}
We can therefore write the full hierarchical likelihood as 
\begin{align}
\begin{split}
& \mathcal{L}(\text{EOS}, \lambda_{\text{pop}}, H_0, \Omega_0 \mid \text{GW}_j, \text{EM}_j) = \\
&\int  \biggl[ \frac{p(m_1^{\text{det}}, m_2^{\text{det}}, \Lambda_1^{\text{EOS}}(m_1), \Lambda_2^{\text{EOS}}(m_2), d_L(z), \iota \mid \text{GW}_j)}{\pi(d_L) \pi(\iota)}\ \times \\
& \qquad \ \frac{p(\log_{10}m_{\text{ej, dyn}}^{\text{fit}}, d_L(z), \iota \mid \text{EM}_j)}{\pi(d_L)\pi(\iota)}\times\\
& \qquad \ p(m_1, m_2 \mid \lambda_{\text{pop}}) \times p(z \mid \text{HG}_j) \biggr]\ dm_1\,dm_2\,dz\,d\iota\ .
\end{split}
\end{align}
Here, $p(z \mid \text{HG}_j)$ stands for the redshift measurement of the host galaxy, which as before is assumed to be Gaussian with an uncertainty of 1\%.
Technically the redshift measurement already entered the light curve inference as a prior and thus the redshift should appear as a parameter in the \ac{EM} posterior, but in practice it remains completely independent of any other inferred \ac{EM} parameters and we can therefore keep its normal distribution a separate factor.
Note, that this time we do not need to divide the \ac{GW} posterior by the prior for $m_1$ and $m_2$, because we are explicitly integrating over $z$ and can therefore evaluate the posterior in detector frame masses, for which the priors on $m_1^{\text{det}}$ and $m_2^{\text{det}}$ were uniform.
Thus, to evaluate the integrals, we simply draw
\begin{align}
\begin{split}
    m_1^{\text{det}},\ m_2^{\text{det}},\ \iota &\sim p(m_1^{\text{det}}, m_2^{\text{det}}, \iota \mid \text{GW}_j)\ , \\
    z &\sim p(z \mid \text{HG}_j)\ ,
\end{split}
\end{align}
and then evaluate the (conditional) posterior densities through the appropriate (conditional) \acp{NF}.

The resulting posterior credibility intervals from all different catalogues are presented in Table~\ref{tab:cosmo_inference_results}, the cosmological credibility contours are visualized in Fig.~\ref{fig:cosmo_inference_all}.
We generally find very good recovery of the injected Planck18 cosmology.
While there is a small degeneracy between $H_0$ and $\Omega_0$, the injected values are all well enclosed in the 95\% credibility regions, as we have sufficiently many close events at low redshifts to break the degeneracy.
The $H_0$-uncertainty of $\pm 1$~km~s$^{-1}$~Mpc$^{-1}$ is comparable to the projections from other studies when considering a similar number of events~\citep[e.g.][]{Chen:2017rfc, Feeney:2018mkj, Califano:2022syd, Han:2025fii}.
Unlike the \ac{EOS}, the recovery of $H_0$ clearly scales with the number of events, because even low \ac{SNR} signals have some constraining power and contribute to the statistical accumulation, whereas for the \ac{EOS}, many low \ac{SNR} events carry essentially no measurable tidal information.

In Fig.~\ref{fig:cosmo_inference_narrow_ETL}, we show the dedicated results for our largest \ac{KN} catalogue, i.e., narrow+ETL.
We also show a hierarchical posterior from narrow+ETL without light-curve data, i.e., just from the \ac{GW}+host galaxy redshift data.
The \ac{EOS} constraints are similar to the multimessenger posterior shown in Fig.~\ref{fig:eos_inference_narrow_ETL}, although they appear slightly narrower than before.
Very small improvements on the \ac{EOS} can also be observed for the inferences from other catalogues.
The reason for this is not entirely clear, as it could either arise from a better exploitation of the correlation between luminosity distance and component masses, given that only luminosity distances from the proposed cosmology are evaluated, or from issues during sampling.
We checked that it is not the result of a stronger gradient in detection probability when including $H_0$ and $\Omega_0$, since even without reweighting the samples, the \ac{EOS} posteriors in Fig.~\ref{fig:cosmo_inference_narrow_ETL} are narrower than in Sec.~\ref{subsec:eos_inference}.
The posteriors with and without light-curve data are similar, although the posterior without light curve information is slightly biased towards higher values for $H_0$.
This is mainly because of two rather close events, where the luminosity distances are slightly biased towards smaller distances.
The corresponding light curve posteriors manage to counteract this offset by providing a stricter lower limit on the luminosity distance, demonstrating how luminosity distance and inclination constraints from light curve posteriors can be relatively impactful in cosmological inference, even with a strong \ac{GW} data set.
We also note that there is a inclination bias in our event set.
The possibility of such a bias was also pointed out by \citet{Chen:2020dyt} and \citet{Chen:2023dgw} and arises when the \ac{EM} emission from \acp{BNS} is anisotropic.
This is the case in our setup, because farther \ac{GW} signals were only followed up if a prompt \ac{GRB} was detected and because the \textsc{possis} \ac{KN} model is intrinsically brighter along the poles due to the higher average electron fraction in that region.
As a result, most of our multimessenger events at $z\gtrsim0.2$ are observed close to or on axis.
In the \ac{GW} posteriors, $\iota$ and $d_L$ are highly degenerate and for many on-axis events there is a significant tail at closer distances, meaning on average, high redshift posteriors underestimate the true luminosity distance.
In the \ac{GW}-only inferences, the sampler will fit these smaller luminosity distances through larger values for $\Omega_0$.
This is especially noticeable before including the detection probability, where we find $\Omega_0={0.42}_{-0.13}^{+0.17}~({0.48}_{-0.18}^{+0.28})$ from the multimessenger (GW+host redshift) posterior, compared to $\Omega_0 = {0.32}_{-0.07}^{+0.11}~({0.33}_{-0.08}^{+0.10})$ after reweighting.
Since our detection probability algorithm is imperfect, the light curve posteriors are useful in providing additional constraints on $\iota$ and $d_L$ to mitigate the inclination bias and thus provide a better cosmological posterior.
Future work could investigate if and how an inclination bias could be addressed in the absence of strong constraints on $\iota$ and $d_L$ from light curves.

\subsection{Discussion}

Naturally, several caveats apply to the hierarchical analyses presented here.
Most importantly, we assumed perfect knowledge about the \ac{GW} waveform and \ac{KN} emission by injecting and recovering the signals with the same models.
Systematic uncertainties and hitherto missing physical features in current waveform models could lead to biased posteriors for tidal deformabilities, especially at next-generation accuracy~\citep{Wade:2014vqa, Samajdar:2018dcx, Samajdar:2019ulq,  Gamba:2020wgg, Kunert:2021hgm, Williams:2026jqv, Gittins:2026ntx}.
Although potentially this could be handled by introducing systematic waveform parameters~\citep{Kumar:2026ckp}, a biased \ac{EOS} recovery remains possible and might lead to tensions with other \ac{NS} observations or nuclear physics constraints.
Even more prone to systematic modelling issues is the determination of the ejecta from \ac{KN} observations.
\citet{King:2025tqo} demonstrate how a mismatch between the modelled and true ejecta morphology can lead to significantly biased dynamical ejecta estimates.
\citet{Brethauer:2024zxg} conclude that uncertainties in the atomic data and thermalisation lead to typical ejecta uncertainties of $\gtrsim 50\%$.
Another layer of systematics is introduced when linking the ejecta to \ac{BNS} component masses via semi-empirical fitting formulae from \ac{NR} simulations.
The difference between different fitting formulae is quantified in \citet{Henkel:2022naw}.
\citet{Henkel:2025ayq} also show how different \ac{KN} models can over- or underestimate \ac{NS} masses, when the ejecta are used to trace back the binary's components.
Although we broadened our \ac{KN} posteriors by sampling over a systematic uncertainty parameter $\sigma_{\text{sys}}$ (see Appendix~\ref{app:em_injections}), a significant bias in the inferred ejecta masses cannot be excluded when analysing real data~\citep{Hussenot-Desenonges:2025gik}.
While introducing more systematic parameters might mitigate biases, it also comes with an increase of the computational sampling cost~\citep{Jhawar:2024ezm}.
At the same time, it is not only the \ac{EOS}, but also the inference of the mass distribution and the cosmology that could be affected by wrong a-priori assumptions or model mismatches.
\citet{Biscoveanu:2021eht} demonstrate how mismodelling the spin distribution can lead to biases in the recovered mass distribution because of the well-known degeneracy between effective spin and mass ratio.
Our results also show that the mass distribution is rather sensitive to the modelling choices used to account for the selection effect. 
Regarding the cosmological parameters, \citet{Canevarolo:2023dkh} investigate systematic biases caused by lensing and matter effects during the propagation of the \ac{GW}.
\citet{Muller:2024wzl} and \citet{Salvarese:2024jpq} study how $H_0$-inference from \ac{EM} bright \acp{BNS} could be biased if the orbital inclination and \ac{GRB} jet are misaligned or the \ac{EM} inference for the viewing angle is biased.
Further, the bright siren approach requires accurate estimation of the peculiar galactic velocities to correctly extract the cosmological redshift~\citep{Mukherjee:2019qmm, Howlett:2019mdh}, although this becomes less important for farther events.
We also note that inclination and distance measurements for $H_0$ and $\Omega_0$ depend on accurate \ac{GW} detector strain calibration~\citep{Capote:2024rmo} and noise characterization~\citep{Mozzon:2021wam, Kumar:2022tto}.

\section{Conclusion}
\label{sec:conclusion}
In this study, we have estimated how many and which types of multimessenger \ac{BNS} events can be expected in the era of next-generation \ac{GW} detectors. 
We also assessed how well these observations constrain important physical parameters like the \ac{EOS}, mass distribution, and Hubble constant.
We considered two different mass distributions and overall four possible \ac{GW} detector networks, finding relatively large differences between these distinct setups.

\subsection*{Detection prospects}
Regarding the projected number of detections, we find that one can expect to observe 40--100 immediate UVOIR counterparts per year by following up \ac{BNS} \ac{GW} signals from \ett or ETL alone, assuming our specific merger rate with a local value of 107~Gpc$^{-3}$~yr$^{-1}$.
This number could increase to 100--500 if \ac{ET} is joined by \ac{CE} in a global \ac{GW} detector network.
Bearing in mind the spread of the local merger rate, this is in broad agreement with many other studies~\citep[e.g.][]{Branchesi:2023mws, Loffredo:2024gmx, Steinle:2025xae, Colombo:2025sdm}.
Further, we show that many multimessenger \acp{BNS} produce an observable afterglow, either from a \ac{GRB} or from the \ac{KN}.
Some afterglows might even be found without an immediate UVOIR counterpart through dedicated late-time afterglow surveys, especially in the radio.
For the setups with ET alone, the total number of afterglows varies between 70 and 90, with \ac{CE} it can go up as high as 160--350.
We remark that these counts include long-term follow-ups that can span up to 10 years.

Apart from the design for the \ac{GW} detector network, the number of successfully identified \ac{EM} counterparts also depends on which telescopes will be used and which detection strategy they adopt.
Out of the observatories considered in this study, Rubin generally identifies the most UVOIR counterparts.
However, the observation strategy for Rubin's target-of-opportunity strategy is still being finalized~\citep{Andreoni:2024pkp} and therefore the detection counts are subject to uncertainty. 
In particular, our observation strategy involved only the $g$ and $i$ filters, whereas \citet{Andreoni:2021epw} argue that observation strategies with five filters and longer exposure times improve the chance of distinguishing \acp{KN} from background transients.
The results from \citet{Stevenson:2025fqt} show that \ac{KN} detection is most likely in the $r$ and $i$ band.
Moreover, additional telescopes could be deployed to follow up on \ac{GW} alerts in the next-generation era, or some of the existing telescopes we listed in Table~\ref{tab:kn_instruments} will stop operating without a comparable successor.
We refer to \citet[][Figure 4.16]{ET:2025xjr} for a comprehensive list on other possible telescopes, which could alter the multimessenger detection numbers notably~\citep{Salafia:2017ebv}.

\subsection*{Multimessenger constraints}
Using the \ac{KN} events detected by our mock observation algorithm, we conducted hierarchical inferences to recover the input hyperparameters of our catalogues, i.e., \ac{EOS}, mass distribution, and cosmology.
We highlight that this was done in a fully Bayesian framework without any \ac{FIM} approximations, as the posteriors of the individual sources were evaluated through \acp{NF}.
Moreover, by using \textsc{jester} we were able to evaluate the hierarchical likelihood function with exact solutions of the \ac{TOV} and Love equation for the $M$-$R$-$\Lambda$ relationships from our \ac{EOS} prior.
Yet, our projected uncertainties represent a best-case scenario in absence of modelling systematics and under the specific assumptions of our merger rate.
For narrow+ETL, we analysed the largest set of 78 events and found that the canonical \ac{NS} radius is constrained to $R_{1.4} = {12.04}^{+0.18}_{-0.14}$~km at 95\% credibility when using the full multimessenger information.
For our smallest set with 28 events, wide+\ett, we obtain $R_{1.4} = {11.99}_{-0.18}^{+0.18}$~km.
Moreover, the degree to which the \ac{EOS} can be constrained also depends on the \ac{SNR} range in the respective event set.
The impact of the light-curve data on the \ac{EOS} appears negligible, though they might provide additional constraints where the $M$-$\Lambda$ relationship inferred from \acp{GW} is not as tight as for the very high \ac{SNR} signals considered here.
However, when the inference is extended to include cosmological parameters, the light-curve posteriors mitigate the inclination bias and can lead to slightly improved posteriors.
With the 78 events observed in narrow+ETL, we find $H_0 = 68.2_{-0.9}^{+0.9}$~km~s$^{-1}$~Mpc$^{-1}$, while for the 28 sirens of wide+\ett one obtains $H_0={68.1}^{+2.3}_{-3.2}$~km~s$^{-1}$~Mpc$^{-1}$.
The recovery of the mass distribution depends on the mass range and parametrization of the model.
We observe tighter constraints for the parameters of the wide mass distribution, mainly because it is inherently easier to fit.
But for both mass models, individual distribution parameters might be inferred with a bias arising from selection events, since our analyses focus only on events with a \ac{KN}.

There remain several challenges to hierarchical \ac{BNS} inference combining ten or more events.
For one, this method requires accurate modelling of the detection probability. 
This is especially relevant when inferring mass distribution and cosmological parameters. 
While \ac{GW} detection probabilities can be handled comparatively well~\citep{Farr:2019rap, Gerosa:2020pgy, Gerosa:2024isl}, addressing multimessenger selection effects remains challenging.
In \citet{Mancarella:2024qle}, it has been suggested to infer the detection probability from the data itself, but for simplicity, we have adopted a coarse approach and post-process our hyperposterior samples by reweighting them with a simplistic model for the detection probability.
This systematically reduces the effective sample size of our hyperposteriors and also proves insufficient to reliably recover the correct mass distribution parameters.
Future studies have to investigate improved methods to assess selection effects and how they might be applied directly during sampling.
Furthermore, when analysing real data, the impact of outliers in the posterior set should be addressed.
Even with improved sampling methods, posteriors will occasionally contain computational sampling errors and biases.
We observed this when sampling the individual \ac{GW} and \ac{EM} parameter posteriors from our events.
We would find that the true mass ratio, tidal deformability, or inclination was sometimes not contained within the posterior.
We alleviated such pathologies by running with a different sampling seed or occasionally cutting some prior to a narrower range (without impairing the posterior).
Obviously, this is not an option for a real data catalogue where the true values are not known.
Future hierarchical inference algorithms could instead include statistical classification to flag certain posteriors as outliers during sampling~\citep[e.g.][]{Martin:2025a, Leeney:2025ajx}.
This seems especially relevant for inferring $H_0$, which is particularly sensitive to flawed inclination and distance posteriors.

\subsection*{Concluding remarks}
Despite systematic issues and the general challenges discussed throughout the present article, multimessenger \ac{BNS} events provide uniquely powerful opportunities to obtain empirical constraints about the \ac{EOS}, \ac{BNS} mass distribution, and Hubble constant.
Bayesian hierarchical inference provides the natural framework to recover the hyperparameters of interest from the observed data by utilizing the parameter posteriors of the individual sources.
The use of \acp{NF} and advanced sampling techniques enables us to efficiently sample the hyperposterior within hours on modern GPUs, further strengthening the case for adopting hardware acceleration in hierarchical sampling for \ac{GW} problems at scale~\citep{Mastrogiovanni:2023zbw, Borghi:2023opd,Talbot:2024yqw,Mould:2025dts,Tagliazucchi:2025ofb,Papadopoulos:2026puy,Tagliazucchi:2026dpr}.
Our results imply that there are going to be multiple occasions in the next-generation era where these methods could be applied to analyse real data from joint \ac{GW} and \ac{EM} observations, although the exact number of multimessenger events will depend on the true mass distribution, \ac{EOS}, \ac{GW} network sensitivity, and most notably merger rate.
Finally, we remark that although measuring the \ac{EOS} and Hubble parameter arguably represent the most prominent science goals with respect to multimessenger observations of \acp{BNS}, other highly relevant applications exist.
For instance, identifying the host galaxies could deliver crucial insights on structure formation and stellar history~\citep{Rose:2021ykg, Stevance:2023byv}, while joint \ac{GRB} and \ac{GW} detections would shine light on the nature of the \ac{GRB} central engine~\citep{Beniamini:2020adb, vanPutten:2022xyx, DallOsso:2023gdk, Gottlieb:2024mwu}.
Similarly, it has been suggested that type Ia supernovae could be calibrated independently of cepheids when they take place in a galaxy before or after \acp{NS} merge there~\citep{Zhao:2017imr, Gupta:2019okl}.
Thus, various avenues exist in which future multimessenger \ac{BNS} mergers have the potential to reshape our current understanding of astrophysics.

\section*{Acknowledgments}
We thank the anonymous referee for their constructive feedback.
We thank A.~Puecher for her assistance with the prompt collapse classifier.
We thank E.~Loffredo for assistance with the prompt collapse disc mass fit.
We thank H.~Cheng for discussing the inclination bias of multimessenger events.
We thank S. Ronchini for many helpful comments.

TD and HK acknowledge funding from the EU Horizon under European Research Council (ERC) Starting Grant, no.\ SMArt-101076369.
Views and opinions expressed are those of the authors only and do not necessarily reflect those of the European Union or the European Research Council. Neither the European Union nor the granting authority can be held responsible for them.

TW and CVDB are supported by the research programme of the Netherlands Organization for Scientific Research (NWO) under grant number OCENW.XL21.XL21.038.
PTHP is supported by the research programme of the Netherlands Organization for Scientific Research (NWO) under grant number VI.Veni.232.021.
The INAF PLEIADI computing resources (http://www.pleiadi.inaf.it) were used.
We thank SURF (www.surf.nl) for the support in using the National Supercomputer Snellius under project number EINF-14622.
Additional computations were performed on the DFG-funded research cluster Jarvis at the University of Potsdam (INST 336/173-1; project number: 502227537).

MB acknowledges the Department of Physics and Earth Science of the University of Ferrara for the financial support through the FIRD 2025 grant.

MWC\ acknowledges support from the National Science Foundation with grant numbers PHY-2308862 and PHY-2117997.

\section*{Data Availability} 
All scripts used to generate the results can be found in our publicly available \href{https://github.com/haukekoehn/mma_analyses}{github repository}.
Larger data products such as posterior samples can be made available upon request to the first author.
The algorithm for the hierarchical inference was implemented within a version of \textsc{jester} and can be accessed on \href{https://github.com/nuclear-multimessenger-astronomy/jester}{github} under the \texttt{GWEM\_Likelihood} branch.

\bibliographystyle{mnras_custom}
\bibliography{bilbiography}

@article{Abac:2023ujg,
    author = "Abac, Adrian and Dietrich, Tim and Buonanno, Alessandra and Steinhoff, Jan and Ujevic, Maximiliano",
    title = "{New and robust gravitational-waveform model for high-mass-ratio binary neutron star systems with dynamical tidal effects}",
    eprint = "2311.07456",
    archivePrefix = "arXiv",
    primaryClass = "gr-qc",
    doi = "10.1103/PhysRevD.109.024062",
    journal = "Phys. Rev. D",
    volume = "109",
    number = "2",
    pages = "024062",
    year = "2024"
}

@article{Aksulu:2021crt,
    author = "Aksulu, M. D. and Wijers, R. A. M. J. and van Eerten, H. J. and van der Horst, A. J.",
    title = "{Exploring the GRB population: robust afterglow modelling}",
    eprint = "2106.14921",
    archivePrefix = "arXiv",
    primaryClass = "astro-ph.HE",
    doi = "10.1093/mnras/stac246",
    journal = "Mon. Not. Roy. Astron. Soc.",
    volume = "511",
    number = "2",
    pages = "2848--2867",
    year = "2022"
}

@article{Alexander:2017aly,
    author = "Alexander, K. D. and others",
    title = "{The Electromagnetic Counterpart of the Binary Neutron Star Merger LIGO/VIRGO GW170817. VI. Radio Constraints on a Relativistic Jet and Predictions for Late-Time Emission from the Kilonova Ejecta}",
    eprint = "1710.05457",
    archivePrefix = "arXiv",
    primaryClass = "astro-ph.HE",
    reportNumber = "FERMILAB-PUB-17-474-A-AE-CD",
    doi = "10.3847/2041-8213/aa905d",
    journal = "Astrophys. J. Lett.",
    volume = "848",
    number = "2",
    pages = "L21",
    year = "2017"
}

@article{Alford:2022bpp,
    author = "Alford, Mark G. and Brodie, Liam and Haber, Alexander and Tews, Ingo",
    title = "{Relativistic mean-field theories for neutron-star physics based on chiral effective field theory}",
    eprint = "2205.10283",
    archivePrefix = "arXiv",
    primaryClass = "nucl-th",
    reportNumber = "LA-UR-22-23972",
    doi = "10.1103/PhysRevC.106.055804",
    journal = "Phys. Rev. C",
    volume = "106",
    number = "5",
    pages = "055804",
    year = "2022"
}

@article{Alford:2023rgp,
    author = "Alford, Mark G. and Brodie, Liam and Haber, Alexander and Tews, Ingo",
    title = "{Tabulated equations of state from models informed by chiral effective field theory}",
    eprint = "2304.07836",
    archivePrefix = "arXiv",
    primaryClass = "nucl-th",
    reportNumber = "LA-UR-23-23837",
    doi = "10.1088/1402-4896/ad03c8",
    journal = "Phys. Scripta",
    volume = "98",
    number = "12",
    pages = "125302",
    year = "2023"
}

@article{Almualla:2020hbs,
    author = "Almualla, Mouza and Coughlin, Michael W. and Anand, Shreya and Alqassimi, Khalid and Guessoum, Nidhal and Singer, Leo P.",
    title = "{Dynamic scheduling: target of opportunity observations of gravitational wave events}",
    eprint = "2003.09718",
    archivePrefix = "arXiv",
    primaryClass = "astro-ph.HE",
    doi = "10.1093/mnras/staa1498",
    journal = "Mon. Not. Roy. Astron. Soc.",
    volume = "495",
    number = "4",
    pages = "4366--4371",
    year = "2020"
}

@article{Andreoni:2021epw,
    author = "Andreoni, Igor and others",
    title = "{Target-of-opportunity Observations of Gravitational-wave Events with Vera C. Rubin Observatory}",
    eprint = "2111.01945",
    archivePrefix = "arXiv",
    primaryClass = "astro-ph.HE",
    doi = "10.3847/1538-4365/ac617c",
    journal = "Astrophys. J. Supp.",
    volume = "260",
    number = "1",
    pages = "18",
    year = "2022"
}

@article{Andreoni:2023xlv,
    author = "Andreoni, Igor and others",
    title = "{Enabling kilonova science with Nancy Grace Roman Space Telescope}",
    eprint = "2307.09511",
    archivePrefix = "arXiv",
    primaryClass = "astro-ph.HE",
    doi = "10.1016/j.astropartphys.2023.102904",
    journal = "Astropart. Phys.",
    volume = "155",
    pages = "102904",
    year = "2024"
}

@article{Andreoni:2024pkp,
    author = "Andreoni, Igor and others",
    title = "{Rubin ToO 2024: Envisioning the Vera C. Rubin Observatory LSST Target of Opportunity program}",
    eprint = "2411.04793",
    archivePrefix = "arXiv",
    primaryClass = "astro-ph.IM",
    month = "11",
    year = "2024"
}

@article{Arcavi:2017vbi,
    author = "Arcavi, Iair and others",
    title = "{Optical Follow-up of Gravitational-wave Events with Las Cumbres Observatory}",
    eprint = "1710.05842",
    archivePrefix = "arXiv",
    primaryClass = "astro-ph.HE",
    doi = "10.3847/2041-8213/aa910f",
    journal = "Astrophys. J. Lett.",
    volume = "848",
    number = "2",
    pages = "L33",
    year = "2017"
}

@online{AXIS:2026,
    title = "{NASA kills future x-ray mission}",
    author = {Daniel Clery},
    year = 2026,
    url = {https://www.science.org/content/article/nasa-kills-future-x-ray-mission},
    urldate = {2026-10-29}
}

@article{Baka:2025yqx,
    author = "Baka, Tomasz and Narola, Harsh and Janquart, Justin and Samajdar, Anuradha and Dietrich, Tim and Van Den Broeck, Chris",
    title = "{Overlapping signals in next-generation gravitational wave observatories: A recipe for selecting the best parameter estimation technique}",
    eprint = "2507.10304",
    archivePrefix = "arXiv",
    primaryClass = "gr-qc",
    doi = "10.1103/8cwp-mxcd",
    journal = "Phys. Rev. D",
    volume = "112",
    number = "8",
    pages = "082001",
    year = "2025"
}

@article{Baker:2025taj,
    author = "Baker, A. Makai and Lasky, Paul D. and Thrane, Eric and Golomb, Jacob",
    title = "{Significant challenges for astrophysical inference with next-generation gravitational-wave observatories}",
    eprint = "2503.04073",
    archivePrefix = "arXiv",
    primaryClass = "gr-qc",
    doi = "10.1103/n6t6-5wn3",
    journal = "Phys. Rev. D",
    volume = "112",
    number = "10",
    pages = "102004",
    year = "2025"
}

@article{Banerjee:2022gkv,
    author = "Banerjee, Biswajit and others",
    title = "{Pre-merger alert to detect prompt emission in very-high-energy gamma-rays from binary neutron star mergers: Einstein Telescope and Cherenkov Telescope Array synergy}",
    eprint = "2212.14007",
    archivePrefix = "arXiv",
    primaryClass = "astro-ph.HE",
    doi = "10.1051/0004-6361/202345850",
    journal = "Astron. Astrophys.",
    volume = "678",
    pages = "A126",
    year = "2023"
}

@article{Baral:2023xst,
    author = "Baral, Pratyusava and Morisaki, Soichiro and Maga{\~n}a Hernandez, Ignacio and Creighton, Jolien",
    title = "{Localization of binary neutron star mergers with a single cosmic explorer}",
    eprint = "2304.09889",
    archivePrefix = "arXiv",
    primaryClass = "astro-ph.HE",
    reportNumber = "LIGO-P2300057",
    doi = "10.1103/PhysRevD.108.043010",
    journal = "Phys. Rev. D",
    volume = "108",
    number = "4",
    pages = "043010",
    year = "2023"
}

@article{Baral:2025geo,
    author = "Baral, Pratyusava and Morisaki, Soichiro and Gupta, Ish and Creighton, Jolien",
    title = "{Parameter estimation of gravitational-wave signals with frequency-dependent antenna responses and higher modes}",
    eprint = "2503.09627",
    archivePrefix = "arXiv",
    primaryClass = "astro-ph.IM",
    reportNumber = "LIGO DCC P2500078",
    doi = "10.1088/1361-6382/ae128b",
    journal = "Class. Quant. Grav.",
    volume = "42",
    number = "21",
    pages = "215007",
    year = "2025"
}

@article{Barna:2024bcd,
    author = "Barna, Tyler and Reed, Brandon and Andreoni, Igor and Coughlin, Michael W. and Dietrich, Tim and Groom, Steven L. and Laz, Theophile Jegou du and Pang, Peter T. H. and Purdum, Josiah N. and Rusholme, Ben",
    title = "{An online framework for fitting fast transient light curves}",
    eprint = "2404.17515",
    archivePrefix = "arXiv",
    primaryClass = "astro-ph.HE",
    doi = "10.1093/mnras/stae1164",
    journal = "Mon. Not. Roy. Astron. Soc.",
    volume = "531",
    number = "1",
    pages = "1084--1094",
    year = "2024"
}

@article{Barna:2025xxn,
    author = "Barna, Tyler and others",
    title = "{IIb or not IIb: A Catalog of ZTF Kilonova Imposters}",
    eprint = "2506.15900",
    archivePrefix = "arXiv",
    primaryClass = "astro-ph.HE",
    doi = "10.1088/1538-3873/adf578",
    journal = "Publ. Astron. Soc. Pac.",
    volume = "137",
    number = "8",
    pages = "084105",
    year = "2025"
}

@article{Barthelmy:2005hs,
    author = "Barthelmy, S. D. and others",
    title = "{The Burst Alert Telescope (BAT) on the Swift MIDEX mission}",
    eprint = "astro-ph/0507410",
    archivePrefix = "arXiv",
    doi = "10.1007/s11214-005-5096-3",
    journal = "Space Sci. Rev.",
    volume = "120",
    pages = "143",
    year = "2005"
}

@article{Bauswein:2013jpa,
    author = "Bauswein, A. and Baumgarte, T. W. and Janka, H. -T.",
    title = "{Prompt merger collapse and the maximum mass of neutron stars}",
    eprint = "1307.5191",
    archivePrefix = "arXiv",
    primaryClass = "astro-ph.SR",
    doi = "10.1103/PhysRevLett.111.131101",
    journal = "Phys. Rev. Lett.",
    volume = "111",
    number = "13",
    pages = "131101",
    year = "2013"
}

@article{Biscoveanu:2021eht,
    author = "Biscoveanu, Sylvia and Talbot, Colm and Vitale, Salvatore",
    title = "{The effect of spin mismodelling on gravitational-wave measurements of the binary neutron star mass distribution}",
    eprint = "2111.13619",
    archivePrefix = "arXiv",
    primaryClass = "astro-ph.HE",
    reportNumber = "LIGO document number LIGO-P2100426",
    doi = "10.1093/mnras/stac347",
    journal = "Mon. Not. Roy. Astron. Soc.",
    volume = "511",
    number = "3",
    pages = "4350--4359",
    year = "2022"
}

@article{Biswas:2024hja,
    author = "Biswas, Bhaskar and Rosswog, Stephan",
    title = "{Simultaneously constraining the neutron star equation of state and mass distribution through multimessenger observations and nuclear benchmarks}",
    eprint = "2408.15192",
    archivePrefix = "arXiv",
    primaryClass = "astro-ph.HE",
    doi = "10.1103/8lv3-1ywb",
    journal = "Phys. Rev. D",
    volume = "112",
    number = "2",
    pages = "023045",
    year = "2025"
}

@article{Blake:2025etn,
    author = "Blake, Chris and Turner, Ryan J.",
    title = "{The role of peculiar velocity uncertainties in standard siren cosmology}",
    eprint = "2509.03101",
    archivePrefix = "arXiv",
    primaryClass = "astro-ph.CO",
    doi = "10.33232/001c.154055",
    journal = "The Open Journal of Astrophysics",
    volume = "8",
    month = "12",
    year = "2025"
}

@article{Blandford:1977ds,
    author = "Blandford, R. D. and Znajek, R. L.",
    title = "{Electromagnetic extractions of energy from Kerr black holes}",
    doi = "10.1093/mnras/179.3.433",
    journal = "Mon. Not. Roy. Astron. Soc.",
    volume = "179",
    pages = "433--456",
    year = "1977"
}

@ARTICLE{Bellm:2019,
       author = "Bellm, E. C. and others",
        title = "{The Zwicky Transient Facility: System Overview, Performance, and First Results}",
      journal = "{Publ. Astron. Soc. Pac.}",
         year = 2019,
        month = jan,
       volume = {131},
       number = {995},
        pages = {018002},
          doi = {10.1088/1538-3873/aaecbe},
archivePrefix = {arXiv},
       eprint = {1902.01932},
}

@article{Beniamini:2015eaa,
    author = "Beniamini, Paz and Nava, Lara and Barniol Duran, Rodolfo and Piran, Tsvi",
    title = "{Energies of GRB blast waves and prompt efficiencies as implied by modelling of X-ray and GeV afterglows}",
    eprint = "1504.04833",
    archivePrefix = "arXiv",
    primaryClass = "astro-ph.HE",
    doi = "10.1093/mnras/stv2033",
    journal = "Mon. Not. Roy. Astron. Soc.",
    volume = "454",
    number = "1",
    pages = "1073--1085",
    year = "2015"
}

@article{Beniamini:2016hzc,
    author = "Beniamini, Paz and Nava, Lara and Piran, Tsvi",
    title = "{A revised analysis of gamma-ray bursts{\textquoteright} prompt efficiencies}",
    eprint = "1606.00311",
    archivePrefix = "arXiv",
    primaryClass = "astro-ph.HE",
    reportNumber = "MN-16-1239-MJ.R2",
    doi = "10.1093/mnras/stw1331",
    journal = "Mon. Not. Roy. Astron. Soc.",
    volume = "461",
    number = "1",
    pages = "51--59",
    year = "2016"
}

@article{Beniamini:2020adb,
    author = "Beniamini, Paz and Duran, Rodolfo Barniol and Petropoulou, Maria and Giannios, Dimitrios",
    title = "{Ready, Set, Launch: Time Interval between a Binary Neutron Star Merger and Short Gamma-Ray Burst Jet Formation}",
    eprint = "2001.00950",
    archivePrefix = "arXiv",
    primaryClass = "astro-ph.HE",
    doi = "10.3847/2041-8213/ab9223",
    journal = "Astrophys. J. Lett.",
    volume = "895",
    number = "2",
    pages = "L33",
    year = "2020"
}

@article{Bernuzzi:2024mfx,
    author = "Bernuzzi, Sebastiano and Magistrelli, Fabio and Jacobi, Maximilian and Logoteta, Domenico and Perego, Albino and Radice, David",
    title = "{Long-lived neutron-star remnants from asymmetric binary neutron star mergers: element formation, kilonova signals and gravitational waves}",
    eprint = "2409.18185",
    archivePrefix = "arXiv",
    primaryClass = "astro-ph.HE",
    doi = "10.1093/mnras/staf1147",
    journal = "Mon. Not. Roy. Astron. Soc.",
    volume = "256",
    pages = "271",
    year = "2025"
}

@article{Bisero:2025tkw,
    author = "Bisero, S. and Vergani, S. D. and Loffredo, E. and Branchesi, M. and Hazra, N. and Dupletsa, U. and Anderson, R. I.",
    title = "{Multi-messenger observations of binary neutron star mergers: Synergies between next-generation gravitational wave interferometers and wide-field, high-multiplex spectroscopic facilities}",
    eprint = "2507.02055",
    archivePrefix = "arXiv",
    primaryClass = "astro-ph.HE",
    doi = "10.1051/0004-6361/202556217",
    journal = "Astron. Astrophys.",
    volume = "705",
    pages = "A54",
    year = "2026"
}

@article{Bonaldi:2020ukl,
    author = "Bonaldi, A. and others",
    title = "{Square Kilometre Array Science Data Challenge 1: analysis and results}",
    eprint = "2009.13346",
    archivePrefix = "arXiv",
    primaryClass = "astro-ph.IM",
    doi = "10.1093/mnras/staa3023",
    journal = "Mon. Not. Roy. Astron. Soc.",
    volume = "500",
    number = "3",
    pages = "3821--3837",
    year = "2020"
}

@article{Branchesi:2023mws,
    author = "Branchesi, Marica and others",
    title = "{Science with the Einstein Telescope: a comparison of different designs}",
    eprint = "2303.15923",
    archivePrefix = "arXiv",
    primaryClass = "gr-qc",
    reportNumber = "ET-0084A-23",
    doi = "10.1088/1475-7516/2023/07/068",
    journal = "JCAP",
    volume = "07",
    pages = "068",
    year = "2023"
}

@article{Braun:2019gdo,
    author = "Braun, Robert and Bonaldi, Anna and Bourke, Tyler and Keane, Evan and Wagg, Jeff",
    title = "{Anticipated Performance of the Square Kilometre Array -- Phase 1 (SKA1)}",
    eprint = "1912.12699",
    archivePrefix = "arXiv",
    primaryClass = "astro-ph.IM",
    month = "12",
    year = "2019"
}

@article{Breschi:2021xrx,
    author = "Breschi, Matteo and Bernuzzi, Sebastiano and Godzieba, Daniel and Perego, Albino and Radice, David",
    title = "{Constraints on the Maximum Densities of Neutron Stars from Postmerger Gravitational Waves with Third-Generation Observations}",
    eprint = "2110.06957",
    archivePrefix = "arXiv",
    primaryClass = "gr-qc",
    doi = "10.1103/PhysRevLett.128.161102",
    journal = "Phys. Rev. Lett.",
    volume = "128",
    number = "16",
    pages = "161102",
    year = "2022"
}

@article{Brethauer:2024zxg,
    author = "Brethauer, Daniel and Kasen, Daniel and Margutti, Raffaella and Chornock, Ryan",
    title = "{Impact of Systematic Modeling Uncertainties on Kilonova Property Estimation}",
    eprint = "2408.02731",
    archivePrefix = "arXiv",
    primaryClass = "astro-ph.HE",
    doi = "10.3847/1538-4357/ad7d83",
    journal = "Astrophys. J.",
    volume = "975",
    number = "2",
    pages = "213",
    year = "2024"
}

@article{Bulla:2019muo,
    author = "Bulla, Mattia",
    title = "{POSSIS: predicting spectra, light curves and polarization for multi-dimensional models of supernovae and kilonovae}",
    eprint = "1906.04205",
    archivePrefix = "arXiv",
    primaryClass = "astro-ph.HE",
    doi = "10.1093/mnras/stz2495",
    journal = "Mon. Not. Roy. Astron. Soc.",
    volume = "489",
    number = "4",
    pages = "5037--5045",
    year = "2019"
}

@article{Bulla:2022mwo,
    author = "Bulla, Mattia",
    title = "{The critical role of nuclear heating rates, thermalization efficiencies, and opacities for kilonova modelling and parameter inference}",
    eprint = "2211.14348",
    archivePrefix = "arXiv",
    primaryClass = "astro-ph.HE",
    doi = "10.1093/mnras/stad232",
    journal = "Mon. Not. Roy. Astron. Soc.",
    volume = "520",
    number = "2",
    pages = "2558--2570",
    year = "2023"
}

@article{Burns:2015fol,
    author = "Burns, Eric and Connaughton, Valerie and Zhang, Bin-Bin and Lien, Amy and Briggs, Michael S. and Goldstein, Adam and Pelassa, Veronique and Troja, Eleonora",
    title = "{Do the Fermi Gamma-Ray Burst Monitor and Swift Burst Alert Telescope see the Same Short Gamma-Ray Bursts?}",
    eprint = "1512.00923",
    archivePrefix = "arXiv",
    primaryClass = "astro-ph.HE",
    doi = "10.3847/0004-637X/818/2/110",
    journal = "Astrophys. J.",
    volume = "818",
    number = "2",
    pages = "110",
    year = "2016"
}

@article{Cabezas:2024blackjax,
  author = {Cabezas, Alberto and Corenflos, Adrien and Lao, Junpeng and Louf, Rémi and others},
  title = "{BlackJAX: Composable Bayesian inference in JAX}",
  eprint = "2402.10797",
  archivePrefix = "arXiv",
  primaryClass = "cs.MS",
  year = "2024",
}

@article{Califano:2022syd,
    author = "Califano, Matteo and de Martino, Ivan and Vernieri, Daniele and Capozziello, Salvatore",
    title = "{Exploiting the Einstein Telescope to solve the Hubble tension}",
    eprint = "2208.13999",
    archivePrefix = "arXiv",
    primaryClass = "astro-ph.CO",
    reportNumber = "ET-0188A-22",
    doi = "10.1103/PhysRevD.107.123519",
    journal = "Phys. Rev. D",
    volume = "107",
    number = "12",
    pages = "123519",
    year = "2023"
}

@article{Camisasca:2023dqx,
    author = "Camisasca, A. E. and Steele, I. A. and Bulla, M. and Guidorzi, C. and Shrestha, M.",
    title = "{Optimizing the observation of optical kilonovae with medium-size telescopes}",
    eprint = "2307.04031",
    archivePrefix = "arXiv",
    primaryClass = "astro-ph.HE",
    doi = "10.1093/mnras/stad1102",
    journal = "Mon. Not. Roy. Astron. Soc.",
    volume = "522",
    number = "2",
    pages = "2516--2524",
    year = "2023"
}

@article{Canevarolo:2023dkh,
    author = "Canevarolo, Sofia and Chisari, Nora Elisa",
    title = "{Lensing bias on cosmological parameters from bright standard sirens}",
    eprint = "2310.12764",
    archivePrefix = "arXiv",
    primaryClass = "astro-ph.CO",
    doi = "10.1093/mnras/stae1713",
    journal = "Mon. Not. Roy. Astron. Soc.",
    volume = "533",
    number = "1",
    pages = "36--51",
    year = "2024"
}

@article{Capote:2024rmo,
    author = "Capote, E. and others",
    title = "{Advanced LIGO detector performance in the fourth observing run}",
    eprint = "2411.14607",
    archivePrefix = "arXiv",
    primaryClass = "gr-qc",
    reportNumber = "LIGO-P2400256",
    doi = "10.1103/PhysRevD.111.062002",
    journal = "Phys. Rev. D",
    volume = "111",
    number = "6",
    pages = "062002",
    year = "2025"
}

@online{CE_asd_files,
     author="Kuns, Kevin and Fulda, Paul and Barsotti, Lisa and Evans, Matthew",
     title = "CE DCC document T2000017-v9",
     year="2025",
     url = "https://dcc.cosmicexplorer.org/cgi-bin/DocDB/ShowDocument?.submit=Identifier&docid=T2000017&version=",
     urldate = "2026-10-29"}

@article{Chatterjee:2021xrm,
    author = "Chatterjee, Deep and Hegade K R, Abhishek and Holder, Gilbert and Holz, Daniel E. and Perkins, Scott and Yagi, Kent and Yunes, Nicol{\'a}s",
    title = "{Cosmology with Love: Measuring the Hubble constant using neutron star universal relations}",
    eprint = "2106.06589",
    archivePrefix = "arXiv",
    primaryClass = "gr-qc",
    doi = "10.1103/PhysRevD.104.083528",
    journal = "Phys. Rev. D",
    volume = "104",
    number = "8",
    pages = "083528",
    year = "2021"
}

@dataset{Chatterjee:2025_eos,
  author       = {Chatterjee, Debarati and
                  Davis, Philip and
                  Dexheimer, Veronica and
                  Ishizuka, Chikako and
                  Klähn, Thomas and
                  Mancini, Marco and
                  Oertel, Micaela and
                  Novak, Jêrome and
                  Pais, Helena and
                  Providência, Constança and
                  Raduta, Adriana R and
                  Servillat, Mathieu and
                  Tolos, Laura and
                  Typel, Stefan},
  title        = {Data table for EoS ABHT(QMC-RMF3) nparam=3},
  month        = feb,
  year         = 2025,
  publisher    = {Zenodo},
  doi          = {10.5281/zenodo.14809193},
}

@article{Chen:2017rfc,
    author = "Chen, Hsin-Yu and Fishbach, Maya and Holz, Daniel E.",
    title = "{A two per cent Hubble constant measurement from standard sirens within five years}",
    eprint = "1712.06531",
    archivePrefix = "arXiv",
    primaryClass = "astro-ph.CO",
    doi = "10.1038/s41586-018-0606-0",
    journal = "Nature",
    volume = "562",
    number = "7728",
    pages = "545--547",
    year = "2018"
}

@article{Chen:2020dyt,
    author = "Chen, Hsin-Yu",
    title = "{Systematic Uncertainty of Standard Sirens from the Viewing Angle of Binary Neutron Star Inspirals}",
    eprint = "2006.02779",
    archivePrefix = "arXiv",
    primaryClass = "astro-ph.HE",
    doi = "10.1103/PhysRevLett.125.201301",
    journal = "Phys. Rev. Lett.",
    volume = "125",
    number = "20",
    pages = "201301",
    year = "2020"
}

@article{Chen:2023dgw,
    author = "Chen, Hsin-Yu and Talbot, Colm and Chase, Eve A.",
    title = "{Mitigating the Counterpart Selection Effect for Standard Sirens}",
    eprint = "2307.10402",
    archivePrefix = "arXiv",
    primaryClass = "astro-ph.CO",
    doi = "10.1103/PhysRevLett.132.191003",
    journal = "Phys. Rev. Lett.",
    volume = "132",
    number = "19",
    pages = "191003",
    year = "2024"
}

@article{Ciolfi:2020wfx,
    author = "Ciolfi, Riccardo and Kalinani, Jay Vijay",
    title = "{Magnetically Driven Baryon Winds from Binary Neutron Star Merger Remnants and the Blue Kilonova of 2017 August}",
    eprint = "2004.11298",
    archivePrefix = "arXiv",
    primaryClass = "astro-ph.HE",
    doi = "10.3847/2041-8213/abb240",
    journal = "Astrophys. J. Lett.",
    volume = "900",
    number = "2",
    pages = "L35",
    year = "2020"
}

@article{Cokluk:2023xio,
    author = "{\c{C}}okluk, Kutay A. and Yakut, Kadri and Giacomazzo, Bruno",
    title = "{General relativistic simulations of high-mass binary neutron star mergers: rapid formation of low-mass stellar black holes}",
    eprint = "2301.09635",
    archivePrefix = "arXiv",
    primaryClass = "astro-ph.HE",
    doi = "10.1093/mnras/stad3752",
    journal = "Mon. Not. Roy. Astron. Soc.",
    volume = "527",
    number = "3",
    pages = "8043--8053",
    year = "2023"
}

@article{Colombo:2022zzp,
    author = "Colombo, Alberto and Salafia, Om Sharan and Gabrielli, Francesco and Ghirlanda, Giancarlo and Giacomazzo, Bruno and Perego, Albino and Colpi, Monica",
    title = "{Multi-messenger Observations of Binary Neutron Star Mergers in the O4 Run}",
    eprint = "2204.07592",
    archivePrefix = "arXiv",
    primaryClass = "astro-ph.HE",
    doi = "10.3847/1538-4357/ac8d00",
    journal = "Astrophys. J.",
    volume = "937",
    number = "2",
    pages = "79",
    year = "2022"
}

@article{Colombo:2025sdm,
    author = "Colombo, Alberto and Salafia, Om Sharan and Ghirlanda, Giancarlo and Iacovelli, Francesco and Mancarella, Michele and Broekgaarden, Floor S. and Nava, Lara and Giacomazzo, Bruno and Colpi, Monica",
    title = "{Multi-messenger observations in the Einstein Telescope era: Binary neutron star and black hole{\textendash}neutron star mergers}",
    eprint = "2503.00116",
    archivePrefix = "arXiv",
    primaryClass = "astro-ph.HE",
    doi = "10.1051/0004-6361/202554326",
    journal = "Astron. Astrophys.",
    volume = "704",
    pages = "A260",
    year = "2025"
}

@article{Corsi:2019xxn,
    author = "Corsi, Alessandra and others",
    title = "{Astro2020 Science White Paper: Radio Counterparts of Compact Object Mergers in the Era of Gravitational-Wave Astronomy}",
    eprint = "1903.10589",
    archivePrefix = "arXiv",
    primaryClass = "astro-ph.IM",
    month = "3",
    year = "2019"
}

@article{Coughlin:2018lta,
    author = "Coughlin, Michael W. and others",
    title = "{Optimizing searches for electromagnetic counterparts of gravitational wave triggers}",
    eprint = "1803.02255",
    archivePrefix = "arXiv",
    primaryClass = "astro-ph.IM",
    doi = "10.1093/mnras/sty1066",
    journal = "Mon. Not. Roy. Astron. Soc.",
    volume = "478",
    number = "1",
    pages = "692--702",
    year = "2018"
}

@article{Coughlin:2018fis,
    author = "Coughlin, Michael W. and Dietrich, Tim and Margalit, Ben and Metzger, Brian D.",
    title = "{Multimessenger Bayesian parameter inference of a binary neutron star merger}",
    eprint = "1812.04803",
    archivePrefix = "arXiv",
    primaryClass = "astro-ph.HE",
    doi = "10.1093/mnrasl/slz133",
    journal = "Mon. Not. Roy. Astron. Soc.",
    volume = "489",
    number = "1",
    pages = "L91--L96",
    year = "2019"
}

@article{Coughlin:2019qkn,
    author = "Coughlin, Michael W. and others",
    title = "{Optimizing multitelescope observations of gravitational-wave counterparts}",
    eprint = "1909.01244",
    archivePrefix = "arXiv",
    primaryClass = "astro-ph.IM",
    doi = "10.1093/mnras/stz2485",
    journal = "Mon. Not. Roy. Astron. Soc.",
    volume = "489",
    number = "4",
    pages = "5775--5783",
    year = "2019"
}

@article{Coulter:2017wya,
    author = "Coulter, D. A. and others",
    title = "{Swope Supernova Survey 2017a (SSS17a), the Optical Counterpart to a Gravitational Wave Source}",
    eprint = "1710.05452",
    archivePrefix = "arXiv",
    primaryClass = "astro-ph.HE",
    doi = "10.1126/science.aap9811",
    journal = "Science",
    volume = "358",
    pages = "1556",
    year = "2017"
}

@article{Cozzumbo:2024vxw,
    author = "Cozzumbo, Andrea and Dupletsa, Ulyana and Calder{\'o}n, Rodrigo and Murgia, Riccardo and Oganesyan, Gor and Branchesi, Marica",
    title = "{Model-independent cosmology with joint observations of gravitational waves and {\ensuremath{\gamma}}-ray bursts}",
    eprint = "2411.02490",
    archivePrefix = "arXiv",
    primaryClass = "astro-ph.CO",
    doi = "10.1088/1475-7516/2025/05/021",
    journal = "JCAP",
    volume = "05",
    pages = "021",
    year = "2025"
}

@article{DallOsso:2023gdk,
    author = "Dall'Osso, Simone and Stratta, Giulia and Perna, Rosalba and de Cesare, Giovanni and Stella, Luigi",
    title = "{Magnetar Central Engines in Gamma-Ray Bursts Follow the Universal Relation of Accreting Magnetic Stars}",
    eprint = "2305.00029",
    archivePrefix = "arXiv",
    primaryClass = "astro-ph.HE",
    doi = "10.3847/2041-8213/acccec",
    journal = "Astrophys. J. Lett.",
    volume = "949",
    number = "2",
    pages = "L32",
    year = "2023"
}

@article{Dalya:2018cnd,
    author = "D{\'a}lya, Gergely and Galg{\'o}czi, G{\'a}bor and Dobos, L{\'a}szl{\'o} and Frei, Zsolt and Heng, Ik Siong and Macas, Ronaldas and Messenger, Christopher and Raffai, P{\'e}ter and de Souza, Rafael S.",
    title = "{GLADE: A galaxy catalogue for multimessenger searches in the advanced gravitational-wave detector era}",
    eprint = "1804.05709",
    archivePrefix = "arXiv",
    primaryClass = "astro-ph.HE",
    doi = "10.1093/mnras/sty1703",
    journal = "Mon. Not. Roy. Astron. Soc.",
    volume = "479",
    number = "2",
    pages = "2374--2381",
    year = "2018"
}

@article{Dalya:2021ewn,
    author = "D{\'a}lya, G. and others",
    title = "{GLADE+~: an extended galaxy catalogue for multimessenger searches with advanced gravitational-wave detectors}",
    eprint = "2110.06184",
    archivePrefix = "arXiv",
    primaryClass = "astro-ph.CO",
    doi = "10.1093/mnras/stac1443",
    journal = "Mon. Not. Roy. Astron. Soc.",
    volume = "514",
    number = "1",
    pages = "1403--1411",
    year = "2022"
}

@article{DES:2017kbs,
    author = "Soares-Santos, M. and others",
    collaboration = "DES, Dark Energy Camera GW-EM",
    title = "{The Electromagnetic Counterpart of the Binary Neutron Star Merger LIGO/Virgo GW170817. I. Discovery of the Optical Counterpart Using the Dark Energy Camera}",
    eprint = "1710.05459",
    archivePrefix = "arXiv",
    primaryClass = "astro-ph.HE",
    reportNumber = "FERMILAB-PUB-17-454-AE-CD-PPD",
    doi = "10.3847/2041-8213/aa9059",
    journal = "Astrophys. J. Lett.",
    volume = "848",
    number = "2",
    pages = "L16",
    year = "2017"
}

@article{DES:2017dgt,
    author = "Scolnic, D. and others",
    collaboration = "DES",
    title = "{How Many Kilonovae Can Be Found in Past, Present, and Future Survey Data Sets?}",
    eprint = "1710.05845",
    archivePrefix = "arXiv",
    primaryClass = "astro-ph.IM",
    reportNumber = "FERMILAB-PUB-17-452-AE",
    doi = "10.3847/2041-8213/aa9d82",
    journal = "Astrophys. J. Lett.",
    volume = "852",
    number = "1",
    pages = "L3",
    year = "2018"
}

@article{DES:2023hft,
    author = "Bom, Clecio R. and others",
    collaboration = "DES",
    title = "{Designing an Optimal Kilonova Search Using DECam for Gravitational-wave Events}",
    eprint = "2302.04878",
    archivePrefix = "arXiv",
    primaryClass = "astro-ph.HE",
    reportNumber = "DES-2022-0714, FERMILAB-PUB-23-048-PPD",
    doi = "10.3847/1538-4357/ad0462",
    journal = "Astrophys. J.",
    volume = "960",
    number = "2",
    pages = "122",
    year = "2024"
}

@article{DES:2025key,
    author = "Bechtol, K. and others",
    collaboration = "DES",
    title = "{Dark Energy Survey Year 6 Results: Photometric Dataset for Cosmology}",
    eprint = "2501.05739",
    archivePrefix = "arXiv",
    primaryClass = "astro-ph.CO",
    reportNumber = "DES-2023-0761, FERMILAB-PUB-24-0932-PPD",
    doi = "10.3847/1538-4365/ae18d3",
    journal = "Astrophys. J. Suppl.",
    volume = "282",
    number = "2",
    pages = "62",
    year = "2026"
}

@article{DeSantis:2026wfw,
    author = "De Santis, Alessio Ludovico and Ronchini, Samuele and Santoliquido, Filippo and Branchesi, Marica",
    title = "{Constraining binary neutron star population synthesis models using short gamma-ray burst data}",
    eprint = "2602.13391",
    archivePrefix = "arXiv",
    primaryClass = "astro-ph.HE",
    doi = "10.1051/0004-6361/202659597",
    journal = "Astron. Astrophys.",
    volume = "710",
    pages = "A388",
    year = "2026"
}

@article{Dessart:2008zd,
    author = "Dessart, L. and Ott, C. D. and Burrows, A. and Rosswog, S. and Livne, E.",
    title = "{Neutrino signatures and the neutrino-driven wind in Binary Neutron Star Mergers}",
    eprint = "0806.4380",
    archivePrefix = "arXiv",
    primaryClass = "astro-ph",
    doi = "10.1088/0004-637X/690/2/1681",
    journal = "Astrophys. J.",
    volume = "690",
    pages = "1681",
    year = "2009"
}

@article{Dhani:2025yxf,
    author = "Dhani, Arnab and Camilletti, Alessandro and Radice, David and Kashyap, Rahul and Sathyaprakash, Bangalore and Logoteta, Domenico and Perego, Albino",
    title = "{Distinguishing prompt-collapse binary neutron star mergers from binary black Holes: Tidal effects and remnant properties}",
    doi = "10.1103/2yng-9l9s",
    journal = "Phys. Rev. D",
    volume = "112",
    number = "12",
    pages = "124003",
    year = "2025"
}

@article{Dietrich:2016lyp,
    author = "Dietrich, Tim and Bernuzzi, Sebastiano and Ujevic, Maximiliano and Tichy, Wolfgang",
    title = "{Gravitational waves and mass ejecta from binary neutron star mergers: Effect of the stars' rotation}",
    eprint = "1611.07367",
    archivePrefix = "arXiv",
    primaryClass = "gr-qc",
    doi = "10.1103/PhysRevD.95.044045",
    journal = "Phys. Rev. D",
    volume = "95",
    number = "4",
    pages = "044045",
    year = "2017"
}

@article{Dietrich:2020efo,
    author = "Dietrich, Tim and Coughlin, Michael W. and Pang, Peter T. H. and Bulla, Mattia and Heinzel, Jack and Issa, Lina and Tews, Ingo and Antier, Sarah",
    title = "{Multimessenger constraints on the neutron-star equation of state and the Hubble constant}",
    eprint = "2002.11355",
    archivePrefix = "arXiv",
    primaryClass = "astro-ph.HE",
    reportNumber = "LA-UR-20-21470",
    doi = "10.1126/science.abb4317",
    journal = "Science",
    volume = "370",
    number = "6523",
    pages = "1450--1453",
    year = "2020"
}

@article{Dobie:2019eeg,
    author = "Dobie, D. and Murphy, T. and Kaplan, D. L. and Ghosh, S. and Bannister, K. W. and Hunstead, R. W.",
    title = "{An optimised gravitational wave follow-up strategy with the Australian Square Kilometre Array Pathfinder}",
    eprint = "1903.01481",
    archivePrefix = "arXiv",
    primaryClass = "astro-ph.IM",
    doi = "10.1017/pasa.2019.9",
    journal = "Publ. Astron. Soc. Austral.",
    volume = "36",
    pages = "e019",
    year = "2019"
}

@article{Dobie:2021qya,
    author = "Dobie, Dougal and Murphy, Tara and Kaplan, David L. and Hotokezaka, Kenta and Ataides, Juan Pablo Bonilla and Mahony, Elizabeth K. and Sadler, Elaine M.",
    title = "{Radio afterglows from compact binary coalescences: prospects for next-generation telescopes}",
    eprint = "2105.08933",
    archivePrefix = "arXiv",
    primaryClass = "astro-ph.HE",
    doi = "10.1093/mnras/stab1468",
    journal = "Mon. Not. Roy. Astron. Soc.",
    volume = "505",
    number = "2",
    pages = "2647--2661",
    year = "2021"
}

@article{Dreas:2026jte,
    author = "Dreas, Emma and Salafia, Om Sharan and Pavan, Andrea and Ciolfi, Riccardo and Celotti, Annalisa",
    title = "{Evolution and afterglow emission of gamma-ray burst jets from binary neutron star mergers}",
    eprint = "2601.08714",
    archivePrefix = "arXiv",
    primaryClass = "astro-ph.HE",
    month = "1",
    year = "2026"
}

@article{Du:2025wto,
    author = "Du, Yun-Fei and Yorgancioglu, Emre Seyit and Yi, Shu-Xu and Cao, Tian-Yong and Zhang, Shuang-Nan",
    title = "{A systematic study of binary neutron star merger rate density history using simulated gravitational wave and short gamma-ray burst observations}",
    eprint = "2507.04019",
    archivePrefix = "arXiv",
    primaryClass = "astro-ph.HE",
    doi = "10.1093/mnras/staf1043",
    journal = "Mon. Not. Roy. Astron. Soc.",
    volume = "541",
    number = "2",
    pages = "798--805",
    year = "2025"
}

@article{Duffell:2018iig,
    author = "Duffell, Paul C. and Quataert, Eliot and Kasen, Daniel and Klion, Hannah",
    title = "{Jet Dynamics in Compact Object Mergers: GW170817 Likely had a Successful Jet}",
    eprint = "1806.10616",
    archivePrefix = "arXiv",
    primaryClass = "astro-ph.HE",
    doi = "10.3847/1538-4357/aae084",
    journal = "Astrophys. J.",
    volume = "866",
    number = "1",
    pages = "3",
    year = "2018"
}

@article{Dupletsa:2022scg,
    author = "Dupletsa, Ulyana and Harms, Jan and Banerjee, Biswajit and Branchesi, Marica and Goncharov, Boris and Maselli, Andrea and Oliveira, Ana Carolina Silva and Ronchini, Samuele and Tissino, Jacopo",
    title = "{gwfish: A simulation software to evaluate parameter-estimation capabilities of gravitational-wave detector networks}",
    eprint = "2205.02499",
    archivePrefix = "arXiv",
    primaryClass = "gr-qc",
    doi = "10.1016/j.ascom.2022.100671",
    journal = "Astron. Comput.",
    volume = "42",
    pages = "100671",
    year = "2023"
}

@article{Ecker:2024kzs,
    author = {Ecker, Christian and Topolski, Konrad and J{\"a}rvinen, Matti and Stehr, Alina},
    title = "{Prompt black hole formation in binary neutron star mergers}",
    eprint = "2402.11013",
    archivePrefix = "arXiv",
    primaryClass = "astro-ph.HE",
    doi = "10.1103/PhysRevD.111.023001",
    journal = "Phys. Rev. D",
    volume = "111",
    number = "2",
    pages = "023001",
    year = "2025"
}

@article{ELT_MICADO,
    author="Davies, R. and others",
    collaboration="The MICADO Consortium",
    title = "{MICADO: The Multi-Adaptive Optics Camera for Deep Observations.}",
    year = "2021", 
    journal = "The Messenger",
    pages = "17-21",
    doi = "10.18727/0722-6691/5217",
    volume = "182",
}

@article{Escorial:2022nvp,
    author = "Escorial, Alicia Rouco and others",
    title = "{The Jet Opening Angle and Event Rate Distributions of Short Gamma-Ray Bursts from Late-time X-Ray Afterglows}",
    eprint = "2210.05695",
    archivePrefix = "arXiv",
    primaryClass = "astro-ph.HE",
    doi = "10.3847/1538-4357/acf830",
    journal = "Astrophys. J.",
    volume = "959",
    number = "1",
    pages = "13",
    year = "2023"
}

@online{ET_asd_files,
     author="Danilishin, Stefan and Zhang, Teng",
     title = "Einstein Telescope sensitivity curves used for CoBA Science
study (ET-0291A-22)",
     year="2023",
     url = "https://apps.et-gw.eu/tds/?r=18213",
     urldate ="2026-10-29"
}

@article{Evans:2021gyd,
    author = "Evans, Matthew and others",
    title = "{A Horizon Study for Cosmic Explorer: Science, Observatories, and Community}",
    eprint = "2109.09882",
    archivePrefix = "arXiv",
    primaryClass = "astro-ph.IM",
    reportNumber = "CE-P2100003-v7, Cosmic Explorer technical report CE-P2100003-v6",
    month = "9",
    year = "2021"
}

@article{Fahlman:2022jkh,
    author = "Fahlman, Steven and Fern{\'a}ndez, Rodrigo",
    title = "{Long-term 3D MHD simulations of black hole accretion discs formed in neutron star mergers}",
    eprint = "2204.03005",
    archivePrefix = "arXiv",
    primaryClass = "astro-ph.HE",
    doi = "10.1093/mnras/stac948",
    journal = "Mon. Not. Roy. Astron. Soc.",
    volume = "513",
    number = "2",
    pages = "2689--2707",
    year = "2022"
}

@article{Farr:2019rap,
    author = "Farr, Will M.",
    title = "{Accuracy Requirements for Empirically-Measured Selection Functions}",
    eprint = "1904.10879",
    archivePrefix = "arXiv",
    primaryClass = "astro-ph.IM",
    doi = "10.3847/2515-5172/ab1d5f",
    journal = "Research Notes of the AAS",
    volume = "3",
    number = "5",
    pages = "66",
    year = "2019"
}

@article{Farrow:2019xnc,
    author = "Farrow, Nicholas and Zhu, Xing-Jiang and Thrane, Eric",
    title = "{The mass distribution of Galactic double neutron stars}",
    eprint = "1902.03300",
    archivePrefix = "arXiv",
    primaryClass = "astro-ph.HE",
    doi = "10.3847/1538-4357/ab12e3",
    journal = "Astrophys. J.",
    volume = "876",
    number = "1",
    pages = "18",
    year = "2019"
}

@article{Feeney:2018mkj,
    author = "Feeney, Stephen M. and Peiris, Hiranya V. and Williamson, Andrew R. and Nissanke, Samaya M. and Mortlock, Daniel J. and Alsing, Justin and Scolnic, Dan",
    title = "{Prospects for resolving the Hubble constant tension with standard sirens}",
    eprint = "1802.03404",
    archivePrefix = "arXiv",
    primaryClass = "astro-ph.CO",
    doi = "10.1103/PhysRevLett.122.061105",
    journal = "Phys. Rev. Lett.",
    volume = "122",
    number = "6",
    pages = "061105",
    year = "2019"
}

@article{Ferdman:2013xia,
    author = "Ferdman, R. D. and others",
    title = "{The double pulsar: evidence for neutron star formation without an iron core-collapse supernova}",
    eprint = "1302.2914",
    archivePrefix = "arXiv",
    primaryClass = "astro-ph.SR",
    doi = "10.1088/0004-637X/767/1/85",
    journal = "Astrophys. J.",
    volume = "767",
    pages = "85",
    year = "2013"
}

@online{FERMI-GBM-catalog,
    title = "{FERMIGBRST - Fermi GBM Burst Catalog}",
    author = {NASA},
    year = 2026,
    url = {https://heasarc.gsfc.nasa.gov/cgi-bin/W3Browse/w3query.pl?&tablehead=name%3Dheasarc%5Ffermigbrst%26description%3DFermi+GBM+Burst+Catalog%26url%3Dhttps%3A%2F%2Fheasarc%2Egsfc%2Enasa%2Egov%2FW3Browse%2Ffermi%2Ffermigbrst%2Ehtml%26archive%3DY%26radius%3D180%26mission%3DFERMI%26priority%3D1&mission=FERMI&Action=More+Options&Action=Parameter+Search&ConeAdd=1},
    urldate = {2026-10-30}
}

@article{Fernandez:2015use,
    author = "Fern{\'a}ndez, Rodrigo and Metzger, Brian D.",
    title = "{Electromagnetic Signatures of Neutron Star Mergers in the Advanced LIGO Era}",
    eprint = "1512.05435",
    archivePrefix = "arXiv",
    primaryClass = "astro-ph.HE",
    doi = "10.1146/annurev-nucl-102115-044819",
    journal = "Ann. Rev. Nucl. Part. Sci.",
    volume = "66",
    pages = "23--45",
    year = "2016"
}

@article{Fernandez:2018kax,
    author = "Fern{\'a}ndez, Rodrigo and Tchekhovskoy, Alexander and Quataert, Eliot and Foucart, Francois and Kasen, Daniel",
    title = "{Long-term GRMHD simulations of neutron star merger accretion discs: implications for electromagnetic counterparts}",
    eprint = "1808.00461",
    archivePrefix = "arXiv",
    primaryClass = "astro-ph.HE",
    doi = "10.1093/mnras/sty2932",
    journal = "Mon. Not. Roy. Astron. Soc.",
    volume = "482",
    number = "3",
    pages = "3373--3393",
    year = "2019"
}

@article{Finstad:2022oni,
    author = "Finstad, Daniel and White, Laurel V. and Brown, Duncan A.",
    title = "{Prospects for a Precise Equation of State Measurement from Advanced LIGO and Cosmic Explorer}",
    eprint = "2211.01396",
    archivePrefix = "arXiv",
    primaryClass = "astro-ph.HE",
    doi = "10.3847/1538-4357/acf12f",
    journal = "Astrophys. J.",
    volume = "955",
    number = "1",
    pages = "45",
    year = "2023"
}

@article{Fong:2015oha,
    author = "Fong, Wen-fai and Berger, Edo and Margutti, Raffaella and Zauderer, B. Ashley",
    title = "{A Decade of Short-duration Gamma-ray Burst Broadband Afterglows: Energetics, Circumburst Densities, and jet Opening Angles}",
    eprint = "1509.02922",
    archivePrefix = "arXiv",
    primaryClass = "astro-ph.HE",
    doi = "10.1088/0004-637X/815/2/102",
    journal = "Astrophys. J.",
    volume = "815",
    number = "2",
    pages = "102",
    year = "2015"
}

@article{Fulton:2025hsx,
    author = "Fulton, M. D. and others",
    title = "{Results from the Pan-STARRS search for kilonovae: contamination by massive stellar outbursts}",
    eprint = "2506.07082",
    archivePrefix = "arXiv",
    primaryClass = "astro-ph.HE",
    doi = "10.1093/mnras/staf1165",
    journal = "Mon. Not. Roy. Astron. Soc.",
    volume = "542",
    number = "2",
    pages = "541--559",
    year = "2025"
}

@article{Gamba:2020wgg,
    author = "Gamba, Rossella and Breschi, Matteo and Bernuzzi, Sebastiano and Agathos, Michalis and Nagar, Alessandro",
    title = "{Waveform systematics in the gravitational-wave inference of tidal parameters and equation of state from binary neutron star signals}",
    eprint = "2009.08467",
    archivePrefix = "arXiv",
    primaryClass = "gr-qc",
    doi = "10.1103/PhysRevD.103.124015",
    journal = "Phys. Rev. D",
    volume = "103",
    number = "12",
    pages = "124015",
    year = "2021"
}

@ARTICLE{Gardner:2023,
       author = {{Gardner}, Jonathan P. and {Mather}, John C. and {Abbott}, Randy and {Abell}, James S. and {Abernathy}, Mark and {Abney}, Faith E. and {Abraham}, John G. and {Abraham}, Roberto and {Abul-Huda}, Yasin M. and {Acton}, Scott and {Adams}, Cynthia K. and {Adams}, Evan and {Adler}, David S. and {Adriaensen}, Maarten and {Aguilar}, Jonathan Albert and {Ahmed}, Mansoor and {Ahmed}, Nasif S. and {Ahmed}, Tanjira and {Albat}, R{\"u}deger and {Albert}, Lo{\"\i}c and {Alberts}, Stacey and {Aldridge}, David and {Allen}, Mary Marsha and {Allen}, Shaune S. and {Altenburg}, Martin and {Altunc}, Serhat and {Alvarez}, Jose Lorenzo and {{\'A}lvarez-M{\'a}rquez}, Javier and {Alves de Oliveira}, Catarina and {Ambrose}, Leslie L. and {Anandakrishnan}, Satya M. and {Andersen}, Gregory C. and {Anderson}, Harry James and {Anderson}, Jay and {Anderson}, Kristen and {Anderson}, Sara M. and {Aprea}, Julio and {Archer}, Benita J. and {Arenberg}, Jonathan W. and {Argyriou}, Ioannis and {Arribas}, Santiago and {Artigau}, {\'E}tienne and {Arvai}, Amanda Rose and {Atcheson}, Paul and {Atkinson}, Charles B. and {Averbukh}, Jesse and {Aymergen}, Cagatay and {Bacinski}, John J. and {Baggett}, Wayne E. and {Bagnasco}, Giorgio and {Baker}, Lynn L. and {Balzano}, Vicki Ann and {Banks}, Kimberly A. and {Baran}, David A. and {Barker}, Elizabeth A. and {Barrett}, Larry K. and {Barringer}, Bruce O. and {Barto}, Allison and {Bast}, William and {Baudoz}, Pierre and {Baum}, Stefi and {Beatty}, Thomas G. and {Beaulieu}, Mathilde and {Bechtold}, Kathryn and {Beck}, Tracy and {Beddard}, Megan M. and {Beichman}, Charles and {Bellagama}, Larry and {Bely}, Pierre and {Berger}, Timothy W. and {Bergeron}, Louis E. and {Bernier}, Antoine-Darveau and {Bertch}, Maria D. and {Beskow}, Charlotte and {Betz}, Laura E. and {Biagetti}, Carl P. and {Birkmann}, Stephan and {Bjorklund}, Kurt F. and {Blackwood}, James D. and {Blazek}, Ronald Paul and {Blossfeld}, Stephen and {Bluth}, Marcel and {Boccaletti}, Anthony and {Boegner}, Jr., Martin E. and {Bohlin}, Ralph C. and {Boia}, John Joseph and {B{\"o}ker}, Torsten and {Bonaventura}, N. and {Bond}, Nicholas A. and {Bosley}, Kari Ann and {Boucarut}, Rene A. and {Bouchet}, Patrice and {Bouwman}, Jeroen and {Bower}, Gary and {Bowers}, Ariel S. and {Bowers}, Charles W. and {Boyce}, Leslye A. and {Boyer}, Christine T. and {Boyer}, Martha L. and {Boyer}, Michael and {Boyer}, Robert and {Bradley}, Larry D. and {Brady}, Gregory R. and {Brandl}, Bernhard R. and {Brannen}, Judith L. and {Breda}, David and {Bremmer}, Harold G. and {Brennan}, David and {Bresnahan}, Pamela A. and {Bright}, Stacey N. and {Broiles}, Brian J. and {Bromenschenkel}, Asa and {Brooks}, Brian H. and {Brooks}, Keira J. and {Brown}, Bob and {Brown}, Bruce and {Brown}, Thomas M. and {Bruce}, Barry W. and {Bryson}, Jonathan G. and {Bujanda}, Edwin D. and {Bullock}, Blake M. and {Bunker}, A.~J. and {Bureo}, Rafael and {Burt}, Irving J. and {Bush}, James Aaron and {Bushouse}, Howard A. and {Bussman}, Marie C. and {Cabaud}, Olivier and {Cale}, Steven and {Calhoon}, Charles D. and {Calvani}, Humberto and {Canipe}, Alicia M. and {Caputo}, Francis M. and {Cara}, Mihai and {Carey}, Larkin and {Case}, Michael Eli and {Cesari}, Thaddeus and {Cetorelli}, Lee D. and {Chance}, Don R. and {Chandler}, Lynn and {Chaney}, Dave and {Chapman}, George N. and {Charlot}, S. and {Chayer}, Pierre and {Cheezum}, Jeffrey I. and {Chen}, Bin and {Chen}, Christine H. and {Cherinka}, Brian and {Chichester}, Sarah C. and {Chilton}, Zachary S. and {Chittiraibalan}, Dharini and {Clampin}, Mark and {Clark}, Charles R. and {Clark}, Kerry W. and {Clark}, Stephanie M. and {Claybrooks}, Edward E. and {Cleveland}, Keith A. and {Cohen}, Andrew L. and {Cohen}, Lester M. and {Col{\'o}n}, Knicole D. and {Coleman}, Benee L. and {Colina}, Luis and {Comber}, Brian J. and {Comeau}, Thomas M. and {Comer}, Thomas and {Conde Reis}, Alain and {Connolly}, Dennis C. and {Conroy}, Kyle E. and {Contos}, Adam R. and {Contreras}, James and {Cook}, Neil J. and {Cooper}, James L. and {Cooper}, Rachel Aviva and {Correia}, Michael F. and {Correnti}, Matteo and {Cossou}, Christophe and {Costanza}, Brian F. and {Coulais}, Alain and {Cox}, Colin R. and {Coyle}, Ray T. and {Cracraft}, Misty M. and {Crew}, Keith A. and {Curtis}, Gary J. and {Cusveller}, Bianca and {Da Costa Maciel}, Cleyciane and {Dailey}, Christopher T. and {Daugeron}, Fr{\'e}d{\'e}ric and {Davidson}, Greg S. and {Davies}, James E. and {Davis}, Katherine Anne and {Davis}, Michael S. and {Day}, Ratna and {de Chambure}, Daniel and {de Jong}, Pauline and {De Marchi}, Guido and {Dean}, Bruce H. and {Decker}, John E. and {Delisa}, Amy S. and {Dell}, Lawrence C. and {Dellagatta}, Gail},
        title = "{The James Webb Space Telescope Mission}",
      journal = {\pasp},
         year = 2023,
        month = jun,
       volume = {135},
       number = {1048},
          eid = {068001},
        pages = {068001},
          doi = {10.1088/1538-3873/acd1b5},
archivePrefix = {arXiv},
       eprint = {2304.04869},
 primaryClass = {astro-ph.IM},
       adsurl = {https://ui.adsabs.harvard.edu/abs/2023PASP..135f8001G}
}

@article{Gehrels:2015uga,
    author = "Gehrels, Neil and Cannizzo, John K. and Kanner, Jonah and Kasliwal, Mansi M. and Nissanke, Samaya and Singer, Leo P.",
    title = "{Galaxy Strategy for LIGO-Virgo Gravitational Wave Counterpart Searches}",
    eprint = "1508.03608",
    archivePrefix = "arXiv",
    primaryClass = "astro-ph.HE",
    doi = "10.3847/0004-637X/820/2/136",
    journal = "Astrophys. J.",
    volume = "820",
    number = "2",
    pages = "136",
    year = "2016"
}

@article{Gerosa:2020pgy,
    author = "Gerosa, Davide and Pratten, Geraint and Vecchio, Alberto",
    title = "{Gravitational-wave selection effects using neural-network classifiers}",
    eprint = "2007.06585",
    archivePrefix = "arXiv",
    primaryClass = "astro-ph.HE",
    doi = "10.1103/PhysRevD.102.103020",
    journal = "Phys. Rev. D",
    volume = "102",
    number = "10",
    pages = "103020",
    year = "2020"
}

@article{Gerosa:2024isl,
    author = "Gerosa, Davide and Bellotti, Malvina",
    title = "{Quick recipes for gravitational-wave selection effects}",
    eprint = "2404.16930",
    archivePrefix = "arXiv",
    primaryClass = "astro-ph.HE",
    doi = "10.1088/1361-6382/ad4509",
    journal = "Class. Quant. Grav.",
    volume = "41",
    number = "12",
    pages = "125002",
    year = "2024"
}

@article{Ghirlanda:2018uyx,
    author = "Ghirlanda, G. and others",
    title = "{Compact radio emission indicates a structured jet was produced by a binary neutron star merger}",
    eprint = "1808.00469",
    archivePrefix = "arXiv",
    primaryClass = "astro-ph.HE",
    doi = "10.1126/science.aau8815",
    journal = "Science",
    volume = "363",
    pages = "968",
    year = "2019"
}

@article{Ghosh:2024cwc,
    author = "Ghosh, Tathagata and Biswas, Bhaskar and Bose, Sukanta and Kapadia, Shasvath J.",
    title = "{Joint Inference of Population, Cosmology, and Neutron Star Equation of State from Gravitational Waves of Dark Binary Neutron Stars}",
    eprint = "2407.16669",
    archivePrefix = "arXiv",
    primaryClass = "gr-qc",
    reportNumber = "LIGO-P2400303",
    doi = "10.3847/1538-4365/ae0472",
    journal = "Astrophys. J. Suppl.",
    volume = "281",
    number = "1",
    pages = "11",
    year = "2025"
}

@article{Gillanders:2025fwf,
    author = "Gillanders, J. H. and others",
    title = "{Pan-STARRS Follow-up of the Gravitational-wave Event S250818k and the Light Curve of SN2025ulz}",
    eprint = "2510.01142",
    archivePrefix = "arXiv",
    primaryClass = "astro-ph.HE",
    doi = "10.3847/2041-8213/ae2125",
    journal = "Astrophys. J. Lett.",
    volume = "995",
    number = "1",
    pages = "L27",
    year = "2025"
}

@article{Gittins:2026ntx,
    author = "Gittins, Fabian and Narola, Harsh and Wouters, Thibeau and Pang, Peter T. H. and Hinderer, Tanja and Van Den Broeck, Chris",
    title = "{Detecting Tidal Resonances in Binary Neutron Stars}",
    eprint = "2606.06376",
    archivePrefix = "arXiv",
    primaryClass = "gr-qc",
    reportNumber = "INT-PUB-26-022",
    month = "6",
    year = "2026"
}

@article{Goldstein:2017mmi,
    author = "Goldstein, A. and others",
    title = "{An Ordinary Short Gamma-Ray Burst with Extraordinary Implications: Fermi-GBM Detection of GRB 170817A}",
    eprint = "1710.05446",
    archivePrefix = "arXiv",
    primaryClass = "astro-ph.HE",
    doi = "10.3847/2041-8213/aa8f41",
    journal = "Astrophys. J. Lett.",
    volume = "848",
    number = "2",
    pages = "L14",
    year = "2017"
}

@article{Golomb:2021tll,
    author = "Golomb, Jacob and Talbot, Colm",
    title = "{Hierarchical Inference of Binary Neutron Star Mass Distribution and Equation of State with Gravitational Waves}",
    eprint = "2106.15745",
    archivePrefix = "arXiv",
    primaryClass = "astro-ph.HE",
    doi = "10.3847/1538-4357/ac43bc",
    journal = "Astrophys. J.",
    volume = "926",
    number = "1",
    pages = "79",
    year = "2022"
}

@article{Golomb:2024lds,
    author = "Golomb, Jacob and Legred, Isaac and Chatziioannou, Katerina and Landry, Philippe",
    title = "{Interplay of astrophysics and nuclear physics in determining the properties of neutron stars}",
    eprint = "2410.14597",
    archivePrefix = "arXiv",
    primaryClass = "astro-ph.HE",
    doi = "10.1103/PhysRevD.111.023029",
    journal = "Phys. Rev. D",
    volume = "111",
    number = "2",
    pages = "023029",
    year = "2025"
}

@article{Gottlieb:2024mwu,
    author = "Gottlieb, Ore and Metzger, Brian D. and Foucart, Francois and Ramirez-Ruiz, Enrico",
    title = "{A Unified Model of Kilonovae and Gamma-Ray Bursts in Binary Mergers Establishes Neutron Stars as the Central Engines of Short GRBs}",
    eprint = "2411.13657",
    archivePrefix = "arXiv",
    primaryClass = "astro-ph.HE",
    doi = "10.3847/1538-4357/adc577",
    journal = "Astrophys. J.",
    volume = "984",
    number = "1",
    pages = "77",
    year = "2025"
}

@article{Grunthal:2021kqg,
    author = "Grunthal, Kathrin and Kramer, Michael and Desvignes, Gregory",
    title = "{Revisiting the Galactic Double Neutron Star merger and LIGO detection rates}",
    eprint = "2107.13307",
    archivePrefix = "arXiv",
    primaryClass = "astro-ph.HE",
    doi = "10.1093/mnras/stab2198",
    journal = "Mon. Not. Roy. Astron. Soc.",
    volume = "507",
    number = "4",
    pages = "5658--5670",
    year = "2021"
}

@article{Gupta:2019okl,
    author = "Gupta, Anuradha and Fox, Derek and Sathyaprakash, B. S. and Schutz, B. F.",
    title = "{Calibrating the cosmic distance ladder using gravitational-wave observations}",
    eprint = "1907.09897",
    archivePrefix = "arXiv",
    primaryClass = "astro-ph.CO",
    doi = "10.3847/1538-4357/ab4c92",
    journal = "Astrophys. J.",
    volume = "886",
    number = "1",
    pages = "71",
    reportNumber = "LIGO-P1900172",
    month = "11",
    year = "2019"
}

@article{Gupta:2022qgg,
    author = "Gupta, Pawan Kumar and Puecher, Anna and Pang, Peter T. H. and Janquart, Justin and Koekoek, Gideon and Broeck Van Den, Chris",
    title = "{Determining the equation of state of neutron stars with Einstein Telescope using tidal effects and r-mode excitations from a population of binary inspirals}",
    eprint = "2205.01182",
    archivePrefix = "arXiv",
    primaryClass = "gr-qc",
    month = "5",
    year = "2022"
}

@article{Gupte:2020mfp,
    author = "Gupte, Nihar and Bartos, Imre",
    title = "{Optimal Gravitational-wave Follow-up Tiling Strategies Using a Genetic Algorithm}",
    eprint = "2003.04839",
    archivePrefix = "arXiv",
    primaryClass = "astro-ph.IM",
    doi = "10.1103/PhysRevD.101.123008",
    journal = "Phys. Rev. D",
    volume = "101",
    number = "12",
    pages = "123008",
    year = "2020"
}

@article{Hall:2025qsm,
    author = "Hall, Xander J. and others",
    title = "{AT2025ulz and S250818k: Investigating early time observations of a subsolar mass gravitational-wave binary neutron star merger candidate}",
    eprint = "2510.24620",
    archivePrefix = "arXiv",
    primaryClass = "astro-ph.HE",
    month = "10",
    year = "2025"
}

@article{Hallinan:2017woc,
    author = "Hallinan, G. and others",
    title = "{A Radio Counterpart to a Neutron Star Merger}",
    eprint = "1710.05435",
    archivePrefix = "arXiv",
    primaryClass = "astro-ph.HE",
    doi = "10.1126/science.aap9855",
    journal = "Science",
    volume = "358",
    pages = "1579",
    year = "2017"
}

@article{Hallinan:2019qyo,
    author = "Hallinan, G. and others",
    title = "{The DSA-2000 -- A Radio Survey Camera}",
    eprint = "1907.07648",
    archivePrefix = "arXiv",
    primaryClass = "astro-ph.IM",
    month = "7",
    year = "2019"
}

@article{Han:2025fii,
    author = "Han, Tao and Zhang, Jing-Fei and Zhang, Xin",
    title = "{Multi-messenger standard-siren cosmology for third-generation gravitational-wave detectors: forecasts considering observations of gamma-ray bursts and kilonovae}",
    eprint = "2504.17741",
    archivePrefix = "arXiv",
    primaryClass = "astro-ph.CO",
    doi = "10.1140/epjc/s10052-025-15114-9",
    journal = "Eur. Phys. J. C",
    volume = "86",
    number = "1",
    pages = "8",
    year = "2026"
}

@article{Haggard:2017qne,
    author = "Haggard, Daryl and Nynka, Melania and Ruan, John J. and Kalogera, Vicky and Bradley Cenko, S. and Evans, Phil and Kennea, Jamie A.",
    title = "{A Deep Chandra X-ray Study of Neutron Star Coalescence GW170817}",
    eprint = "1710.05852",
    archivePrefix = "arXiv",
    primaryClass = "astro-ph.HE",
    doi = "10.3847/2041-8213/aa8ede",
    journal = "Astrophys. J. Lett.",
    volume = "848",
    number = "2",
    pages = "L25",
    year = "2017"
}

@article{Hayashi:2024jwt,
    author = "Hayashi, Kota and Kiuchi, Kenta and Kyutoku, Koutarou and Sekiguchi, Yuichiro and Shibata, Masaru",
    title = "{Jet from Binary Neutron Star Merger with Prompt Black Hole Formation}",
    eprint = "2410.10958",
    archivePrefix = "arXiv",
    primaryClass = "astro-ph.HE",
    doi = "10.1103/PhysRevLett.134.211407",
    journal = "Phys. Rev. Lett.",
    volume = "134",
    number = "21",
    pages = "211407",
    year = "2025"
}

@article{Henkel:2022naw,
    author = "Henkel, Amelia and Foucart, Francois and Raaijmakers, Geert and Nissanke, Samaya",
    title = "{Study of the agreement between binary neutron star ejecta models derived from numerical relativity simulations}",
    eprint = "2207.07658",
    archivePrefix = "arXiv",
    primaryClass = "astro-ph.HE",
    doi = "10.1103/PhysRevD.107.063028",
    journal = "Phys. Rev. D",
    volume = "107",
    number = "6",
    pages = "063028",
    year = "2023"
}

@article{Henkel:2025ayq,
    author = "Henkel, Amelia and Foucart, Francois and Melfor, Selah and Nissanke, Samaya and Wernersson, Alexandra and Bhardwaj, Uddipta",
    title = "{The impact of binary neutron star outflow mass models on kilonova parameter estimation}",
    eprint = "2504.03900",
    archivePrefix = "arXiv",
    primaryClass = "astro-ph.HE",
    month = "4",
    year = "2025"
}

@article{Hjorth:2017yza,
    author = "Hjorth, Jens and Levan, Andrew J. and Tanvir, Nial R. and Lyman, Joe D. and Wojtak, Rados and Schr{\o}der, Sophie L. and Mandel, Ilya and Gall, Christa and Bruun, Sofie H.",
    title = "{The Distance to NGC 4993: The Host Galaxy of the Gravitational-wave Event GW170817}",
    eprint = "1710.05856",
    archivePrefix = "arXiv",
    primaryClass = "astro-ph.GA",
    doi = "10.3847/2041-8213/aa9110",
    journal = "Astrophys. J. Lett.",
    volume = "848",
    number = "2",
    pages = "L31",
    year = "2017"
}

@ARTICLE{Hodapp:2004,
       author = {{Hodapp}, K.~W. and {Kaiser}, N. and {Aussel}, H. and {Burgett}, W. and {Chambers}, K.~C. and {Chun}, M. and {Dombeck}, T. and {Douglas}, A. and {Hafner}, D. and {Heasley}, J. and {Hoblitt}, J. and {Hude}, C. and {Isani}, S. and {Jedicke}, R. and {Jewitt}, D. and {Laux}, U. and {Luppino}, G.~A. and {Lupton}, R. and {Maberry}, M. and {Magnier}, E. and {Mannery}, E. and {Monet}, D. and {Morgan}, J. and {Onaka}, P. and {Price}, P. and {Ryan}, A. and {Siegmund}, W. and {Szapudi}, I. and {Tonry}, J. and {Wainscoat}, R. and {Waterson}, M.},
        title = "{Design of the Pan-STARRS telescopes}",
      journal = {Astronomische Nachrichten},
         year = 2004,
        month = oct,
       volume = {325},
       number = {6},
        pages = {636-642},
          doi = {10.1002/asna.200410300},
       adsurl = {https://ui.adsabs.harvard.edu/abs/2004AN....325..636H}
}

@article{Hotokezaka:2018dfi,
    author = "Hotokezaka, Kenta and Nakar, Ehud and Gottlieb, Ore and Nissanke, Samaya and Masuda, Kento and Hallinan, Gregg and Mooley, Kunal P. and Deller, Adam. T.",
    title = "{A Hubble constant measurement from superluminal motion of the jet in GW170817}",
    eprint = "1806.10596",
    archivePrefix = "arXiv",
    primaryClass = "astro-ph.CO",
    doi = "10.1038/s41550-019-0820-1",
    journal = "Nature Astron.",
    volume = "3",
    number = "10",
    pages = "940--944",
    year = "2019"
}

@article{Howlett:2019mdh,
    author = "Howlett, Cullan and Davis, Tamara M.",
    title = "{Standard siren speeds: improving velocities in gravitational-wave measurements of $H_0$}",
    eprint = "1909.00587",
    archivePrefix = "arXiv",
    primaryClass = "astro-ph.CO",
    doi = "10.1093/mnras/staa049",
    journal = "Mon. Not. Roy. Astron. Soc.",
    volume = "492",
    number = "3",
    pages = "3803--3815",
    year = "2020"
}

@article{Hussenot-Desenonges:2025gik,
    author = "Hussenot-Desenonges, Thomas and Pillas, Marion and Antier, Sarah and Hello, Patrice and Pang, Peter T. H.",
    title = "{Kilonova modelling and parameter inference: Understanding uncertainties and evaluating compatibility between observations and models}",
    eprint = "2505.21392",
    archivePrefix = "arXiv",
    primaryClass = "astro-ph.HE",
    month = "5",
    year = "2025"
}

@article{Jhawar:2024ezm,
    author = "Jhawar, Sahil and Wouters, Thibeau and Pang, Peter T. H. and Bulla, Mattia and Coughlin, Michael W. and Dietrich, Tim",
    title = "{Data-driven approach for modeling the temporal and spectral evolution of kilonova systematic uncertainties}",
    eprint = "2410.21978",
    archivePrefix = "arXiv",
    primaryClass = "astro-ph.HE",
    doi = "10.1103/PhysRevD.111.043046",
    journal = "Phys. Rev. D",
    volume = "111",
    number = "4",
    pages = "043046",
    year = "2025"
}

@article{Kalinani:2025itu,
    author = "Kalinani, Jay V. and Ciolfi, Riccardo and Campanelli, Manuela and Giacomazzo, Bruno and Pavan, Andrea and Wen, Allen and Zlochower, Yosef",
    title = "{Jet{\textendash}Environment Interaction after Delayed Collapse in Binary Neutron Star Mergers}",
    eprint = "2505.09426",
    archivePrefix = "arXiv",
    primaryClass = "astro-ph.HE",
    doi = "10.3847/2041-8213/ae402a",
    journal = "Astrophys. J. Lett.",
    volume = "1000",
    number = "1",
    pages = "L35",
    year = "2026"
}

@article{Kashyap:2021wzs,
    author = "Kashyap, Rahul and others",
    title = "{Numerical relativity simulations of prompt collapse mergers: Threshold mass and phenomenological constraints on neutron star properties after GW170817}",
    eprint = "2111.05183",
    archivePrefix = "arXiv",
    primaryClass = "astro-ph.HE",
    doi = "10.1103/PhysRevD.105.103022",
    journal = "Phys. Rev. D",
    volume = "105",
    number = "10",
    pages = "103022",
    year = "2022"
}

@article{Kasliwal:2025keb,
    author = "Kasliwal, Mansi M. and others",
    title = "{ZTF25abjmnps (AT2025ulz) and S250818k: A Candidate Superkilonova from a Subthreshold Subsolar Gravitational-wave Trigger}",
    eprint = "2510.23732",
    archivePrefix = "arXiv",
    primaryClass = "astro-ph.HE",
    doi = "10.3847/2041-8213/ae2000",
    journal = "Astrophys. J. Lett.",
    volume = "995",
    number = "2",
    pages = "L59",
    year = "2025"
}

@article{Kathirgamaraju:2019xwu,
    author = "Kathirgamaraju, Adithan and Giannios, Dimitrios and Beniamini, Paz",
    title = "{Observable Features of GW170817 Kilonova Afterglow}",
    eprint = "1901.00868",
    archivePrefix = "arXiv",
    primaryClass = "astro-ph.HE",
    doi = "10.1093/mnras/stz1564",
    journal = "Mon. Not. Roy. Astron. Soc.",
    volume = "487",
    number = "3",
    pages = "3914--3921",
    year = "2019"
}

@article{Kaur:2024yag,
    author = "Kaur, Ravjit and O'Connor, Brendan and Palmese, Antonella and Kunnumkai, Keerthi",
    title = "{Detecting Prompt and Afterglow Jet Emission of Gravitational-wave Events from LIGO/Virgo/KAGRA and Next-generation Detectors}",
    eprint = "2410.10579",
    archivePrefix = "arXiv",
    primaryClass = "astro-ph.HE",
    doi = "10.3847/1538-4357/ae4341",
    journal = "Astrophys. J.",
    volume = "1000",
    number = "1",
    pages = "74",
    year = "2026"
}

@article{Kawaguchi:2019nju,
    author = "Kawaguchi, Kyohei and Shibata, Masaru and Tanaka, Masaomi",
    title = "{Diversity of kilonova light curves}",
    eprint = "1908.05815",
    archivePrefix = "arXiv",
    primaryClass = "astro-ph.HE",
    doi = "10.3847/1538-4357/ab61f6",
    journal = "Astrophys. J.",
    volume = "889",
    number = "2",
    pages = "171",
    year = "2020"
}

@article{Khadkikar:2025ith,
    author = "Khadkikar, Sanika and Gupta, Ish and Kashyap, Rahul and Chandra, Koustav and Gamba, Rossella and Sathyaprakash, Bangalore S.",
    title = "{Precise and accurate neutron star radius measurements with next-generation gravitational wave detectors}",
    eprint = "2502.03463",
    archivePrefix = "arXiv",
    primaryClass = "astro-ph.HE",
    reportNumber = "LIGO DCC Number: LIGO-P2400544",
    doi = "10.1103/h6rl-mcjm",
    journal = "Phys. Rev. D",
    volume = "112",
    number = "6",
    pages = "063020",
    year = "2025"
}

@article{King:2025tqo,
    author = "King, Brendan L. and De, Soumi and Korobkin, Oleg and Coughlin, Michael W. and Pang, Peter T. H. and Strother, Terrance T.",
    title = "{Inferring Neutron Star Merger Ejecta Morphology with Kilonovae}",
    eprint = "2505.16876",
    archivePrefix = "arXiv",
    primaryClass = "astro-ph.HE",
    reportNumber = "LA-UR-25-24747",
    doi = "10.1088/1538-3873/ae10df",
    journal = "Publ. Astron. Soc. Pac.",
    volume = "137",
    number = "10",
    pages = "104507",
    year = "2025"
}

@article{Kiuchi:2019lls,
    author = "Kiuchi, Kenta and Kyutoku, Koutarou and Shibata, Masaru and Taniguchi, Keisuke",
    title = "{Revisiting the lower bound on tidal deformability derived by AT 2017gfo}",
    eprint = "1903.01466",
    archivePrefix = "arXiv",
    primaryClass = "astro-ph.HE",
    doi = "10.3847/2041-8213/ab1e45",
    journal = "Astrophys. J. Lett.",
    volume = "876",
    number = "2",
    pages = "L31",
    year = "2019"
}

@article{Koehn:2025zzb,
    author = "Koehn, Hauke and Wouters, Thibeau and Pang, Peter T. H. and Bulla, Mattia and Rose, Henrik and Wichern, Hannah and Dietrich, Tim",
    title = "{Efficient Bayesian analysis of kilonovae and gamma ray burst afterglows with FIESTA}",
    eprint = "2507.13807",
    archivePrefix = "arXiv",
    primaryClass = "astro-ph.HE",
    doi = "10.1051/0004-6361/202556626",
    journal = "Astron. Astrophys.",
    volume = "704",
    pages = "A55",
    year = "2025"
}

@article{Kolsch:2021lub,
    author = {K{\"o}lsch, Maximilian and Dietrich, Tim and Ujevic, Maximiliano and Bruegmann, Bernd},
    title = "{Investigating the mass-ratio dependence of the prompt-collapse threshold with numerical-relativity simulations}",
    eprint = "2112.11851",
    archivePrefix = "arXiv",
    primaryClass = "gr-qc",
    doi = "10.1103/PhysRevD.106.044026",
    journal = "Phys. Rev. D",
    volume = "106",
    number = "4",
    pages = "044026",
    year = "2022"
}

@article{Kruger:2020gig,
    author = {Kr{\"u}ger, Christian J{\"u}rgen and Foucart, Francois},
    title = "{Estimates for Disk and Ejecta Masses Produced in Compact Binary Mergers}",
    eprint = "2002.07728",
    archivePrefix = "arXiv",
    primaryClass = "astro-ph.HE",
    doi = "10.1103/PhysRevD.101.103002",
    journal = "Phys. Rev. D",
    volume = "101",
    number = "10",
    pages = "103002",
    year = "2020"
}

@article{Kumar:2022tto,
    author = "Kumar, Sumit and Nitz, Alexander H. and Forteza, Xisco Jim{\'e}nez",
    title = "{Parameter Estimation with Nonstationary Noise in Gravitational-wave Data}",
    eprint = "2202.12762",
    archivePrefix = "arXiv",
    primaryClass = "astro-ph.IM",
    doi = "10.3847/1538-4357/adb973",
    journal = "Astrophys. J.",
    volume = "982",
    number = "2",
    pages = "67",
    year = "2025"
}

@article{Kunert:2021hgm,
    author = "Kunert, Nina and Pang, Peter T. H. and Tews, Ingo and Coughlin, Michael W. and Dietrich, Tim",
    title = "{Quantifying modeling uncertainties when combining multiple gravitational-wave detections from binary neutron star sources}",
    eprint = "2110.11835",
    archivePrefix = "arXiv",
    primaryClass = "astro-ph.HE",
    reportNumber = "LA-UR-21-27560",
    doi = "10.1103/PhysRevD.105.L061301",
    journal = "Phys. Rev. D",
    volume = "105",
    number = "6",
    pages = "L061301",
    year = "2022"
}

@article{Lamb:2020ccz,
    author = "Lamb, Gavin P. and Levan, Andrew J. and Tanvir, Nial R.",
    title = "{GRB170817A as a Refreshed Shock Afterglow Viewed Off-axis}",
    eprint = "2005.12426",
    archivePrefix = "arXiv",
    primaryClass = "astro-ph.HE",
    doi = "10.3847/1538-4357/aba75a",
    journal = "Astrophys. J.",
    volume = "899",
    number = "2",
    pages = "105",
    year = "2020",
    note = "[Erratum: Astrophys.J. 910, 166 (2021)]"
}

@article{Landry:2021hvl,
    author = "Landry, Philippe and Read, Jocelyn S.",
    title = "{The Mass Distribution of Neutron Stars in Gravitational-wave Binaries}",
    eprint = "2107.04559",
    archivePrefix = "arXiv",
    primaryClass = "astro-ph.HE",
    doi = "10.3847/2041-8213/ac2f3e",
    journal = "Astrophys. J. Lett.",
    volume = "921",
    number = "2",
    pages = "L25",
    year = "2021"
}

@article{Leeney:2025ajx,
    author = "Leeney, S. A. K. and Handley, W. J. and Bevins, H. T. J. and Acedo, E. de Lera",
    title = "{Bayesian Anomaly Detection for Ia Cosmology: Automating SALT3 Data Curation}",
    eprint = "2509.13394",
    archivePrefix = "arXiv",
    primaryClass = "astro-ph.IM",
    month = "9",
    year = "2025"
}

@article{Li:2021pns,
    author = "Li, X. Q. and others",
    title = "{Inflight performance of the GECAM Gamma-ray and Charge particle Detectors}",
    eprint = "2112.04772",
    archivePrefix = "arXiv",
    primaryClass = "physics.ins-det",
    month = "12",
    year = "2021"
}

@article{LIGOScientific:2017adf,
    author = "Abbott, B. P. and others",
    collaboration = "LIGO Scientific, Virgo, 1M2H, Dark Energy Camera GW-E, DES, DLT40, Las Cumbres Observatory, VINROUGE, MASTER",
    title = "{A gravitational-wave standard siren measurement of the Hubble constant}",
    eprint = "1710.05835",
    archivePrefix = "arXiv",
    primaryClass = "astro-ph.CO",
    reportNumber = "LIGO-P1700296, FERMILAB-PUB-17-472-A-AE",
    doi = "10.1038/nature24471",
    journal = "Nature",
    volume = "551",
    number = "7678",
    pages = "85--88",
    year = "2017"
}

@article{Lipunov:2017dwd,
    author = "Lipunov, V. M. and others",
    title = "{MASTER Optical Detection of the First LIGO/Virgo Neutron Star Binary Merger GW170817}",
    eprint = "1710.05461",
    archivePrefix = "arXiv",
    primaryClass = "astro-ph.HE",
    doi = "10.3847/2041-8213/aa92c0",
    journal = "Astrophys. J. Lett.",
    volume = "850",
    number = "1",
    pages = "L1",
    year = "2017"
}

@article{LIGOScientific:2017vwq,
    author = "Abbott, B. P. and others",
    collaboration = "LIGO Scientific, Virgo",
    title = "{GW170817: Observation of Gravitational Waves from a Binary Neutron Star Inspiral}",
    eprint = "1710.05832",
    archivePrefix = "arXiv",
    primaryClass = "gr-qc",
    reportNumber = "LIGO-P170817",
    doi = "10.1103/PhysRevLett.119.161101",
    journal = "PRL",
    volume = "119",
    number = "16",
    pages = "161101",
    year = "2017"
}

@article{LIGOScientific:2017ync,
    author = "Abbott, B. P. and others",
    collaboration = "LIGO Scientific, Virgo, Fermi GBM, INTEGRAL, IceCube, AstroSat Cadmium Zinc Telluride Imager Team, IPN, Insight-Hxmt, ANTARES, Swift, AGILE Team, 1M2H Team, Dark Energy Camera GW-EM, DES, DLT40, GRAWITA, Fermi-LAT, ATCA, ASKAP, Las Cumbres Observatory Group, OzGrav, DWF (Deeper Wider Faster Program), AST3, CAASTRO, VINROUGE, MASTER, J-GEM, GROWTH, JAGWAR, CaltechNRAO, TTU-NRAO, NuSTAR, Pan-STARRS, MAXI Team, TZAC Consortium, KU, Nordic Optical Telescope, ePESSTO, GROND, Texas Tech University, SALT Group, TOROS, BOOTES, MWA, CALET, IKI-GW Follow-up, H.E.S.S., LOFAR, LWA, HAWC, Pierre Auger, ALMA, Euro VLBI Team, Pi of Sky, Chandra Team at McGill University, DFN, ATLAS Telescopes, High Time Resolution Universe Survey, RIMAS, RATIR, SKA South Africa/MeerKAT",
    title = "{Multi-messenger Observations of a Binary Neutron Star Merger}",
    eprint = "1710.05833",
    archivePrefix = "arXiv",
    primaryClass = "astro-ph.HE",
    reportNumber = "LIGO-P1700294, VIR-0802A-17, FERMILAB-PUB-17-478-A-AE-CD",
    doi = "10.3847/2041-8213/aa91c9",
    journal = "Astrophys. J. Lett.",
    volume = "848",
    number = "2",
    pages = "L12",
    year = "2017"
}

@article{LIGOScientific:2017zic,
    author = "Abbott, B. P. and others",
    collaboration = "LIGO Scientific, Virgo, Fermi-GBM, INTEGRAL",
    title = "{Gravitational Waves and Gamma-rays from a Binary Neutron Star Merger: GW170817 and GRB 170817A}",
    eprint = "1710.05834",
    archivePrefix = "arXiv",
    primaryClass = "astro-ph.HE",
    reportNumber = "LIGO-P1700308",
    doi = "10.3847/2041-8213/aa920c",
    journal = "Astrophys. J. Lett.",
    volume = "848",
    number = "2",
    pages = "L13",
    year = "2017"
}

@article{LIGOScientific:2026ctl,
    author = "Abac, None and others",
    collaboration = "LIGO Scientific, VIRGO, KAGRA",
    title = "{GWTC-5.0: Population Properties of Merging Compact Binaries}",
    eprint = "2605.27226",
    archivePrefix = "arXiv",
    primaryClass = "astro-ph.HE",
    reportNumber = "LIGO-P2600045",
    month = "5",
    year = "2026"
}

@article{Loffredo:2024gmx,
    author = "Loffredo, E. and others",
    title = "{Prospects for optical detections from binary neutron star mergers with the next-generation multi-messenger observatories}",
    eprint = "2411.02342",
    archivePrefix = "arXiv",
    primaryClass = "astro-ph.HE",
    doi = "10.1051/0004-6361/202452863",
    journal = "Astron. Astrophys.",
    volume = "697",
    pages = "A36",
    year = "2025"
}

@article{Madau:2014bja,
    author = "Madau, Piero and Dickinson, Mark",
    title = "{Cosmic Star Formation History}",
    eprint = "1403.0007",
    archivePrefix = "arXiv",
    primaryClass = "astro-ph.CO",
    doi = "10.1146/annurev-astro-081811-125615",
    journal = "Ann. Rev. Astron. Astrophys.",
    volume = "52",
    pages = "415--486",
    year = "2014"
}

@article{Mancarella:2024qle,
    author = "Mancarella, Michele and Iacovelli, Francesco and Foffa, Stefano and Muttoni, Niccol{\`o} and Maggiore, Michele",
    title = "{Accurate Standard Siren Cosmology with Joint Gravitational-Wave and {\ensuremath{\gamma}}-Ray Burst Observations}",
    eprint = "2405.02286",
    archivePrefix = "arXiv",
    primaryClass = "astro-ph.HE",
    doi = "10.1103/PhysRevLett.133.261001",
    journal = "Phys. Rev. Lett.",
    volume = "133",
    number = "26",
    pages = "261001",
    year = "2024"
}

@article{Mandel:2018mve,
    author = "Mandel, Ilya and Farr, Will M. and Gair, Jonathan R.",
    title = "{Extracting distribution parameters from multiple uncertain observations with selection biases}",
    eprint = "1809.02063",
    archivePrefix = "arXiv",
    primaryClass = "physics.data-an",
    doi = "10.1093/mnras/stz896",
    journal = "Mon. Not. Roy. Astron. Soc.",
    volume = "486",
    number = "1",
    pages = "1086--1093",
    year = "2019"
}

@article{Margueron:2017eqc,
    author = "Margueron, J{\'e}r{\^o}me and Hoffmann Casali, Rudiney and Gulminelli, Francesca",
    title = "{Equation of state for dense nucleonic matter from metamodeling. I. Foundational aspects}",
    eprint = "1708.06894",
    archivePrefix = "arXiv",
    primaryClass = "nucl-th",
    reportNumber = "INT-PUB-17-029",
    doi = "10.1103/PhysRevC.97.025805",
    journal = "Phys. Rev. C",
    volume = "97",
    number = "2",
    pages = "025805",
    year = "2018"
}

@article{Margueron:2017lup,
    author = "Margueron, J{\'e}r{\^o}me and Hoffmann Casali, Rudiney and Gulminelli, Francesca",
    title = "{Equation of state for dense nucleonic matter from metamodeling. II. Predictions for neutron star properties}",
    eprint = "1708.06895",
    archivePrefix = "arXiv",
    primaryClass = "nucl-th",
    reportNumber = "INT-PUB-17-030",
    doi = "10.1103/PhysRevC.97.025806",
    journal = "Phys. Rev. C",
    volume = "97",
    number = "2",
    pages = "025806",
    year = "2018"
}

@ARTICLE{Martin:2025a,
    author = "Martin, William and Mortlock, Daniel",
    title = "{An approach to robust Bayesian regression in astronomy}",
    eprint = {2411.02380},
    archivePrefix = {arXiv},
    primaryClass = {astro-ph.IM},
    doi = {10.1093/rasti/rzaf035},
    journal = "RAS Techniques and Instruments",
    volume = "4",
    pages = "rzaf035",
    year = "2025",
}

@article{Miceli:2022efx,
    author = "Miceli, Davide and Nava, Lara",
    title = "{Gamma-Ray Bursts Afterglow Physics and the VHE Domain}",
    eprint = "2205.12146",
    archivePrefix = "arXiv",
    primaryClass = "astro-ph.HE",
    doi = "10.3390/galaxies10030066",
    journal = "Galaxies",
    volume = "10",
    number = "3",
    pages = "66",
    year = "2022"
}

@article{Mitra:2020vzq,
    author = "Mitra, Ayan and Mifsud, Jurgen and Mota, David F. and Parkinson, David",
    title = "{Cosmology with the Einstein Telescope: No Slip Gravity Model and Redshift Specifications}",
    eprint = "2010.00189",
    archivePrefix = "arXiv",
    primaryClass = "astro-ph.CO",
    doi = "10.1093/mnras/stab165",
    journal = "Mon. Not. Roy. Astron. Soc.",
    volume = "502",
    number = "4",
    pages = "5563--5575",
    year = "2021"
}

@article{ET:2019dnz,
    author = "Maggiore, Michele and others",
    collaboration = "ET",
    title = "{Science Case for the Einstein Telescope}",
    eprint = "1912.02622",
    archivePrefix = "arXiv",
    primaryClass = "astro-ph.CO",
    doi = "10.1088/1475-7516/2020/03/050",
    journal = "JCAP",
    volume = "03",
    pages = "050",
    year = "2020"
}

@article{ET:2025xjr,
    author = "Abac, Adrian and others",
    collaboration = "ET",
    title = "{The Science of the Einstein Telescope}",
    eprint = "2503.12263",
    archivePrefix = "arXiv",
    primaryClass = "gr-qc",
    reportNumber = "ET-0036C-25",
    doi = "10.1088/1475-7516/2026/03/081",
    journal = "JCAP",
    volume = "03",
    pages = "081",
    year = "2026"
}

@article{Marchesi:2020smf,
    author = "Marchesi, Stefano and others",
    title = "{Mock catalogs for the extragalactic X-ray sky: simulating AGN surveys with Athena and with the AXIS probe}",
    eprint = "2008.09133",
    archivePrefix = "arXiv",
    primaryClass = "astro-ph.IM",
    doi = "10.1051/0004-6361/202038622",
    journal = "Astron. Astrophys.",
    volume = "642",
    pages = "A184",
    year = "2020"
}

@article{Margutti:2017cjl,
    author = "Margutti, Raffaella and others",
    title = "{The Electromagnetic Counterpart of the Binary Neutron Star Merger LIGO/VIRGO GW170817. V. Rising X-ray Emission from an Off-Axis Jet}",
    eprint = "1710.05431",
    archivePrefix = "arXiv",
    primaryClass = "astro-ph.HE",
    reportNumber = "FERMILAB-PUB-17-499-A-AE-CD",
    doi = "10.3847/2041-8213/aa9057",
    journal = "Astrophys. J. Lett.",
    volume = "848",
    number = "2",
    pages = "L20",
    year = "2017"
}

@article{Meegan:2009qu,
    author = "Meegan, Charles and others",
    title = "{The Fermi Gamma-Ray Burst Monitor}",
    eprint = "0908.0450",
    archivePrefix = "arXiv",
    primaryClass = "astro-ph.IM",
    doi = "10.1088/0004-637X/702/1/791",
    journal = "Astrophys. J.",
    volume = "702",
    pages = "791--804",
    year = "2009"
}

@article{Merfeld:2025els,
    author = "Merfeld, Kara and Corsi, Alessandra",
    title = "{Probing Binary Neutron Star Merger Ejecta and Remnants with Gravitational-wave and Radio Observations}",
    eprint = "2506.22835",
    archivePrefix = "arXiv",
    primaryClass = "astro-ph.HE",
    doi = "10.3847/1538-4357/ae80ab",
    journal = "Astrophys. J.",
    volume = "1006",
    number = "2",
    pages = "215",
    year = "2026"
}

@article{Metzger:2014yda,
    author = "Metzger, Brian D. and Bauswein, Andreas and Goriely, Stephane and Kasen, Daniel",
    title = "{Neutron-powered precursors of kilonovae}",
    eprint = "1409.0544",
    archivePrefix = "arXiv",
    primaryClass = "astro-ph.HE",
    reportNumber = "INT-PUB-14-053",
    doi = "10.1093/mnras/stu2225",
    journal = "Mon. Not. Roy. Astron. Soc.",
    volume = "446",
    pages = "1115--1120",
    year = "2015"
}

@article{Metzger:2018szx,
    author = "Metzger, Brian D. and Beniamini, Paz and Giannios, Dimitrios",
    title = "{Effects of Fallback Accretion on Protomagnetar Outflows in Gamma-Ray Bursts and Superluminous Supernovae}",
    eprint = "1802.07750",
    archivePrefix = "arXiv",
    primaryClass = "astro-ph.HE",
    doi = "10.3847/1538-4357/aab70c",
    journal = "Astrophys. J.",
    volume = "857",
    number = "2",
    pages = "95",
    year = "2018"
}

@article{Miller:2014aaa,
    author = "Miller, M. Coleman and Miller, Jon M.",
    title = "{The Masses and Spins of Neutron Stars and Stellar-Mass Black Holes}",
    eprint = "1408.4145",
    archivePrefix = "arXiv",
    primaryClass = "astro-ph.HE",
    doi = "10.1016/j.physrep.2014.09.003",
    journal = "Phys. Rept.",
    volume = "548",
    pages = "1--34",
    year = "2014"
}

@article{Morisaki:2021ngj,
    author = "Morisaki, Soichiro",
    title = "{Accelerating parameter estimation of gravitational waves from compact binary coalescence using adaptive frequency resolutions}",
    eprint = "2104.07813",
    archivePrefix = "arXiv",
    primaryClass = "gr-qc",
    doi = "10.1103/PhysRevD.104.044062",
    journal = "Phys. Rev. D",
    volume = "104",
    number = "4",
    pages = "044062",
    year = "2021"
}

@article{Morsony:2023afu,
    author = "Morsony, Brian James and Santos, Ryan De Los and Hernandez, Rubin and Bustamante, Joshua and Yassuiae, Brandon and Astorga, German and Parra, Juan and Workman, Jared C.",
    title = "{The afterglow of GW170817 from every angle: prospects for detecting the afterglows of binary neutron star mergers}",
    eprint = "2306.00076",
    archivePrefix = "arXiv",
    primaryClass = "astro-ph.HE",
    doi = "10.1093/mnras/stae1638",
    journal = "Mon. Not. Roy. Astron. Soc.",
    volume = "533",
    number = "1",
    pages = "510--524",
    year = "2024"
}

@article{Most:2019pac,
    author = "Most, Elias R. and Papenfort, L. Jens and Tsokaros, Antonios and Rezzolla, Luciano",
    title = "{Impact of high spins on the ejection of mass in GW170817}",
    eprint = "1904.04220",
    archivePrefix = "arXiv",
    primaryClass = "astro-ph.HE",
    doi = "10.3847/1538-4357/ab3ebb",
    journal = "Astrophys. J.",
    volume = "884",
    pages = "40",
    year = "2019"
}

@article{Mozzon:2021wam,
    author = "Mozzon, Simone and Ashton, Gregory and Nuttall, Laura K. and Williamson, Andrew R.",
    title = "{Does nonstationary noise in LIGO and Virgo affect the estimation of H0?}",
    eprint = "2110.11731",
    archivePrefix = "arXiv",
    primaryClass = "astro-ph.CO",
    doi = "10.1103/PhysRevD.106.043504",
    journal = "Phys. Rev. D",
    volume = "106",
    number = "4",
    pages = "043504",
    year = "2022"
}

@article{Mukherjee:2019qmm,
    author = "Mukherjee, Suvodip and Lavaux, Guilhem and Bouchet, Fran{\c{c}}ois R. and Jasche, Jens and Wandelt, Benjamin D. and Nissanke, Samaya M. and Leclercq, Florent and Hotokezaka, Kenta",
    title = "{Velocity correction for Hubble constant measurements from standard sirens}",
    eprint = "1909.08627",
    archivePrefix = "arXiv",
    primaryClass = "astro-ph.CO",
    doi = "10.1051/0004-6361/201936724",
    journal = "Astron. Astrophys.",
    volume = "646",
    pages = "A65",
    year = "2021"
}

@article{Muller:2024wzl,
    author = {M{\"u}ller, Michael and Mukherjee, Suvodip and Ryan, Geoffrey},
    title = "{Be Careful in Multimessenger Inference of the Hubble Constant: A Path Forward for Robust Inference}",
    eprint = "2406.11965",
    archivePrefix = "arXiv",
    primaryClass = "astro-ph.CO",
    doi = "10.3847/2041-8213/ad8dd1",
    journal = "Astrophys. J. Lett.",
    volume = "977",
    number = "2",
    pages = "L45",
    year = "2024"
}

@article{Murguia-Berthier:2016fys,
    author = "Murguia-Berthier, Ariadna and Ramirez-Ruiz, Enrico and Montes, Gabriela and De Colle, Fabio and Rezzolla, Luciano and Rosswog, Stephan and Takami, Kentaro and Perego, Albino and Lee, William H.",
    title = "{The Properties of Short gamma-ray burst Jets Triggered by neutron star mergers}",
    eprint = "1609.04828",
    archivePrefix = "arXiv",
    primaryClass = "astro-ph.HE",
    doi = "10.3847/2041-8213/aa5b9e",
    journal = "Astrophys. J. Lett.",
    volume = "835",
    number = "2",
    pages = "L34",
    year = "2017"
}

@article{Nakar:2011cw,
    author = "Nakar, Ehud and Piran, Tsvi",
    title = "{Radio Remnants of Compact Binary Mergers - the Electromagnetic Signal that will follow the Gravitational Waves}",
    eprint = "1102.1020",
    archivePrefix = "arXiv",
    primaryClass = "astro-ph.HE",
    doi = "10.1038/nature10365",
    journal = "Nature",
    volume = "478",
    pages = "82--84",
    year = "2011"
}

@article{NANOGrav:2023hde,
    author = "Agazie, Gabriella and others",
    collaboration = "NANOGrav",
    title = "{The NANOGrav 15 yr Data Set: Observations and Timing of 68 Millisecond Pulsars}",
    eprint = "2306.16217",
    archivePrefix = "arXiv",
    primaryClass = "astro-ph.HE",
    doi = "10.3847/2041-8213/acda9a",
    journal = "Astrophys. J. Lett.",
    volume = "951",
    number = "1",
    pages = "L9",
    year = "2023"
}

@article{Narola:2024qdh,
    author = "Narola, Harsh and others",
    title = "{Null-stream-based third-generation-ready glitch mitigation for gravitational wave measurements}",
    eprint = "2411.15506",
    archivePrefix = "arXiv",
    primaryClass = "gr-qc",
    doi = "10.1103/l6tp-ykxp",
    journal = "Phys. Rev. D",
    volume = "112",
    number = "2",
    pages = "024079",
    year = "2025"
}

@article{Nedora:2019jhl,
    author = "Nedora, Vsevolod and Bernuzzi, Sebastiano and Radice, David and Perego, Albino and Endrizzi, Andrea and Ortiz, N{\'e}stor",
    title = "{Spiral-wave wind for the blue kilonova}",
    eprint = "1907.04872",
    archivePrefix = "arXiv",
    primaryClass = "astro-ph.HE",
    doi = "10.3847/2041-8213/ab5794",
    journal = "Astrophys. J. Lett.",
    volume = "886",
    number = "2",
    pages = "L30",
    year = "2019"
}

@article{Nedora:2020qtd,
    author = "Nedora, Vsevolod and Schianchi, Federico and Bernuzzi, Sebastiano and Radice, David and Daszuta, Boris and Endrizzi, Andrea and Perego, Albino and Prakash, Aviral and Zappa, Francesco",
    title = "{Mapping dynamical ejecta and disk masses from numerical relativity simulations of neutron star mergers}",
    eprint = "2011.11110",
    archivePrefix = "arXiv",
    primaryClass = "astro-ph.HE",
    doi = "10.1088/1361-6382/ac35a8",
    journal = "Class. Quant. Grav.",
    volume = "39",
    number = "1",
    pages = "015008",
    year = "2022"
}

@article{Neuweiler:2022eum,
    author = "Neuweiler, Anna and Dietrich, Tim and Bulla, Mattia and Chaurasia, Swami Vivekanandji and Rosswog, Stephan and Ujevic, Maximiliano",
    title = "{Long-term simulations of dynamical ejecta: Homologous expansion and kilonova properties}",
    eprint = "2208.13460",
    archivePrefix = "arXiv",
    primaryClass = "astro-ph.HE",
    doi = "10.1103/PhysRevD.107.023016",
    journal = "Phys. Rev. D",
    volume = "107",
    number = "2",
    pages = "023016",
    year = "2023"
}

@article{Neuweiler:2025klw,
    author = "Neuweiler, Anna and others",
    title = "{General-relativistic radiation magnetohydrodynamics simulations of binary neutron star mergers: The influence of spin on the multimessenger picture}",
    eprint = "2510.14850",
    archivePrefix = "arXiv",
    primaryClass = "astro-ph.HE",
    doi = "10.1103/mxlf-8sbm",
    journal = "Phys. Rev. D",
    volume = "113",
    number = "4",
    pages = "043038",
    year = "2026"
}

@article{Nedora:2024vrv,
    author = "Nedora, Vsevolod and Menegazzi, Ludovica Crosato and Peretti, Enrico and Dietrich, Tim and Shibata, Masaru",
    title = "{Multiphysics framework for fast modelling of gamma-ray burst afterglows}",
    eprint = "2409.16852",
    archivePrefix = "arXiv",
    primaryClass = "astro-ph.HE",
    doi = "10.1093/mnras/staf302",
    journal = "Mon. Not. Roy. Astron. Soc.",
    volume = "538",
    number = "3",
    pages = "2089--2115",
    year = "2025"
}

@online{ngVLA_limits,
    title = "{ngVLA Performance Estimates (December 2021)}",
    author = {NRAO},
    year = 2021,
    url = {https://ngvla.nrao.edu/page/performance},
    urldate = {2026-10-30}
}

@article{Nicholl:2021rcr,
    author = "Nicholl, Matt and Margalit, Ben and Schmidt, Patricia and Smith, Graham P. and Ridley, Evan J. and Nuttall, James",
    title = "{Tight multimessenger constraints on the neutron star equation of state from GW170817 and a forward model for kilonova light-curve synthesis}",
    eprint = "2102.02229",
    archivePrefix = "arXiv",
    primaryClass = "astro-ph.HE",
    doi = "10.1093/mnras/stab1523",
    journal = "Mon. Not. Roy. Astron. Soc.",
    volume = "505",
    number = "2",
    pages = "3016--3032",
    year = "2021"
}

@article{Pandey:2024mlo,
    author = "Pandey, Shiksha and Gupta, Ish and Chandra, Koustav and Sathyaprakash, Bangalore S.",
    title = "{The Critical Role of LIGO-India in the Era of Next-generation Observatories}",
    eprint = "2411.10349",
    archivePrefix = "arXiv",
    primaryClass = "gr-qc",
    doi = "10.3847/2041-8213/add15f",
    journal = "Astrophys. J. Lett.",
    volume = "985",
    number = "1",
    pages = "L17",
    year = "2025"
}

@article{Pang:2022rzc,
    author = "Pang, Peter T. H. and others",
    title = "{An updated nuclear-physics and multi-messenger astrophysics framework for binary neutron star mergers}",
    eprint = "2205.08513",
    archivePrefix = "arXiv",
    primaryClass = "astro-ph.HE",
    reportNumber = "LA-UR-22-23872, LIGO-P2200150",
    doi = "10.1038/s41467-023-43932-6",
    journal = "Nature Commun.",
    volume = "14",
    number = "1",
    pages = "8352",
    year = "2023"
}

@article{Pei:1992ub,
    author = "Pei, Yichuan C.",
    title = "{Interstellar dust from the Milky Way to the Magellanic Clouds}",
    doi = "10.1086/171637",
    journal = "Astrophys. J.",
    volume = "395",
    pages = "130--139",
    year = "1992"
}

@article{Perego:2014fma,
    author = {Perego, A. and Rosswog, Stephan and Cabez{\'o}n, Ruben M. and Korobkin, Oleg and K{\"a}ppeli, Roger and Arcones, Almudena and Liebend{\"o}rfer, Matthias},
    title = "{Neutrino-driven winds from neutron star merger remnants}",
    eprint = "1405.6730",
    archivePrefix = "arXiv",
    primaryClass = "astro-ph.HE",
    doi = "10.1093/mnras/stu1352",
    journal = "Mon. Not. Roy. Astron. Soc.",
    volume = "443",
    number = "4",
    pages = "3134--3156",
    year = "2014"
}

@article{Piro:2021oaa,
    author = "Piro, Luigi and others",
    title = "{Athena synergies in the multi-messenger and transient universe}",
    eprint = "2110.15677",
    archivePrefix = "arXiv",
    primaryClass = "astro-ph.HE",
    doi = "10.1007/s10686-022-09865-6",
    journal = "Exper. Astron.",
    volume = "54",
    number = "1",
    pages = "23--117",
    year = "2022"
}

@article{Planck:2018vyg,
    author = "Aghanim, N. and others",
    collaboration = "Planck",
    title = "{Planck 2018 results. VI. Cosmological parameters}",
    eprint = "1807.06209",
    archivePrefix = "arXiv",
    primaryClass = "astro-ph.CO",
    doi = "10.1051/0004-6361/201833910",
    journal = "Astron. Astrophys.",
    volume = "641",
    pages = "A6",
    year = "2020",
    note = "[Erratum: Astron.Astrophys. 652, C4 (2021)]"
}

@article{Pratten:2020fqn,
    author = "Pratten, Geraint and Husa, Sascha and Garcia-Quiros, Cecilio and Colleoni, Marta and Ramos-Buades, Antoni and Estelles, Hector and Jaume, Rafel",
    title = "{Setting the cornerstone for a family of models for gravitational waves from compact binaries: The dominant harmonic for nonprecessing quasicircular black holes}",
    eprint = "2001.11412",
    archivePrefix = "arXiv",
    primaryClass = "gr-qc",
    reportNumber = "LIGO-P2000018",
    doi = "10.1103/PhysRevD.102.064001",
    journal = "Phys. Rev. D",
    volume = "102",
    number = "6",
    pages = "064001",
    year = "2020"
}

@article{Podsiadlowski:2003py,
    author = "Podsiadlowski, Philipp and Langer, N. and Poelarends, A. J. T. and Rappaport, S. and Heger, A. and Pfahl, E.",
    title = "{The effects of binary evolution on the dynamics of core collapse and neutron - star kicks}",
    eprint = "astro-ph/0309588",
    archivePrefix = "arXiv",
    doi = "10.1086/421713",
    journal = "Astrophys. J.",
    volume = "612",
    pages = "1044--1051",
    year = "2004"
}

@article{Pol:2020tfz,
    author = "Pol, Nihan and McLaughlin, Maura and Lorimer, Duncan",
    title = "{An Updated Galactic Double Neutron Star Merger Rate Based on Radio Pulsar Populations}",
    eprint = "2002.10225",
    archivePrefix = "arXiv",
    primaryClass = "astro-ph.HE",
    doi = "10.3847/2515-5172/ab7307",
    journal = "Res. Notes AAS",
    volume = "4",
    number = "2",
    pages = "22",
    year = "2020"
}

@article{Puecher:2024dhl,
    author = "Puecher, Anna and Dietrich, Tim",
    title = "{Machine-learning classifier for the postmerger remnant of binary neutron stars}",
    eprint = "2408.10678",
    archivePrefix = "arXiv",
    primaryClass = "astro-ph.HE",
    doi = "10.1103/PhysRevD.110.123038",
    journal = "Phys. Rev. D",
    volume = "110",
    number = "12",
    pages = "123038",
    year = "2024"
}

@article{Punturo:2010zz,
    author = "Punturo, M. and others",
    editor = "Ricci, Fulvio",
    title = "{The Einstein Telescope: A third-generation gravitational wave observatory}",
    doi = "10.1088/0264-9381/27/19/194002",
    journal = "Class. Quant. Grav.",
    volume = "27",
    pages = "194002",
    year = "2010"
}

@article{Raaijmakers:2021slr,
    author = "Raaijmakers, Geert and others",
    title = "{The Challenges Ahead for Multimessenger Analyses of Gravitational Waves and Kilonova: A Case Study on GW190425}",
    eprint = "2102.11569",
    archivePrefix = "arXiv",
    primaryClass = "astro-ph.HE",
    doi = "10.3847/1538-4357/ac222d",
    journal = "Astrophys. J.",
    volume = "922",
    number = "2",
    pages = "269",
    year = "2021"
}

@article{Raaijmakers:2021uju,
    author = "Raaijmakers, G. and Greif, S. K. and Hebeler, K. and Hinderer, T. and Nissanke, S. and Schwenk, A. and Riley, T. E. and Watts, A. L. and Lattimer, J. M. and Ho, W. C. G.",
    title = "{Constraints on the Dense Matter Equation of State and Neutron Star Properties from NICER{\textquoteright}s Mass{\textendash}Radius Estimate of PSR J0740+6620 and Multimessenger Observations}",
    eprint = "2105.06981",
    archivePrefix = "arXiv",
    primaryClass = "astro-ph.HE",
    doi = "10.3847/2041-8213/ac089a",
    journal = "Astrophys. J. Lett.",
    volume = "918",
    number = "2",
    pages = "L29",
    year = "2021"
}

@article{Radice:2017lry,
    author = "Radice, David and Perego, Albino and Zappa, Francesco and Bernuzzi, Sebastiano",
    title = "{GW170817: Joint Constraint on the Neutron Star Equation of State from Multimessenger Observations}",
    eprint = "1711.03647",
    archivePrefix = "arXiv",
    primaryClass = "astro-ph.HE",
    reportNumber = "LIGO-P1700421, VIR-0894A-17",
    doi = "10.3847/2041-8213/aaa402",
    journal = "Astrophys. J. Lett.",
    volume = "852",
    number = "2",
    pages = "L29",
    year = "2018"
}

@article{Radice:2018pdn,
    author = "Radice, David and Perego, Albino and Hotokezaka, Kenta and Fromm, Steven A. and Bernuzzi, Sebastiano and Roberts, Luke F.",
    title = "{Binary Neutron Star Mergers: Mass Ejection, Electromagnetic Counterparts and Nucleosynthesis}",
    eprint = "1809.11161",
    archivePrefix = "arXiv",
    primaryClass = "astro-ph.HE",
    doi = "10.3847/1538-4357/aaf054",
    journal = "Astrophys. J.",
    volume = "869",
    number = "2",
    pages = "130",
    year = "2018"
}

@article{Raithel:2022aee,
    author = "Raithel, Carolyn A. and Most, Elias R.",
    title = {{Tidal deformability doppelg{\"a}nger: Implications of a low-density phase transition in the neutron star equation of state}},
    eprint = "2208.04295",
    archivePrefix = "arXiv",
    primaryClass = "astro-ph.HE",
    doi = "10.1103/PhysRevD.108.023010",
    journal = "Phys. Rev. D",
    volume = "108",
    number = "2",
    pages = "023010",
    year = "2023"
}

@article{Regimbau:2012ir,
    author = "Regimbau, Tania and others",
    title = "{A Mock Data Challenge for the Einstein Gravitational-Wave Telescope}",
    eprint = "1201.3563",
    archivePrefix = "arXiv",
    primaryClass = "gr-qc",
    doi = "10.1103/PhysRevD.86.122001",
    journal = "Phys. Rev. D",
    volume = "86",
    pages = "122001",
    year = "2012"
}

@article{Reitze:2019iox,
    author = "Reitze, David and others",
    title = "{Cosmic Explorer: The U.S. Contribution to Gravitational-Wave Astronomy beyond LIGO}",
    eprint = "1907.04833",
    archivePrefix = "arXiv",
    primaryClass = "astro-ph.IM",
    reportNumber = "LIGO-P1900316",
    journal = "Bull. Am. Astron. Soc.",
    volume = "51",
    number = "7",
    pages = "035",
    year = "2019"
}

@article{Reynolds:2023vvf,
    author = "Reynolds, Christopher S. and others",
    title = "{Overview of the advanced x-ray imaging satellite (AXIS)}",
    eprint = "2311.00780",
    archivePrefix = "arXiv",
    primaryClass = "astro-ph.IM",
    doi = "10.1117/12.2677468",
    journal = "Proc. SPIE Int. Soc. Opt. Eng.",
    volume = "12678",
    pages = "126781E",
    year = "2023"
}

@online{Roman-technical-details,
        title="{Roman Space Telescope - Scientific Observatory Details}",
        author = {NASA},
        year = 2026,
        url = {https://science.nasa.gov/mission/roman-space-telescope/observatory-technical/#Field-of-Regard-and-Optical-Field-Layout},
        ulrdate = {2026-10-30}
}

@article{Ronchini:2022gwk,
    author = "Ronchini, Samuele and Branchesi, Marica and Oganesyan, Gor and Banerjee, Biswajit and Dupletsa, Ulyana and Ghirlanda, Giancarlo and Harms, Jan and Mapelli, Michela and Santoliquido, Filippo",
    title = "{Perspectives for multimessenger astronomy with the next generation of gravitational-wave detectors and high-energy satellites}",
    eprint = "2204.01746",
    archivePrefix = "arXiv",
    primaryClass = "astro-ph.HE",
    doi = "10.1051/0004-6361/202243705",
    journal = "Astron. Astrophys.",
    volume = "665",
    pages = "A97",
    year = "2022"
}

@article{Rose:2021ykg,
    author = "Rose, Jonah C. and Torrey, Paul and Lee, K. H. and Bartos, I.",
    title = "{Where binary neutron stars merge: predictions from IllustrisTNG}",
    eprint = "2101.10343",
    archivePrefix = "arXiv",
    primaryClass = "astro-ph.HE",
    doi = "10.3847/1538-4357/abe405",
    journal = "Astrophys. J.",
    volume = "909",
    number = "2",
    pages = "207",
    year = "2021"
}

@article{Rosswog:2023rqa,
    author = "Rosswog, S. and Diener, P. and Torsello, F. and Tauris, T. M. and Sarin, N.",
    title = "{Mergers of double NSs with one high-spin component: brighter kilonovae and fallback accretion, weaker gravitational waves}",
    eprint = "2310.15920",
    archivePrefix = "arXiv",
    primaryClass = "astro-ph.HE",
    doi = "10.1093/mnras/stae454",
    journal = "Mon. Not. Roy. Astron. Soc.",
    volume = "530",
    number = "2",
    pages = "2336--2354",
    year = "2024"
}

@article{Ruhe:2022ddi,
    author = "Ruhe, David and Wong, Kaze and Cranmer, Miles and Forr{\'e}, Patrick",
    title = "{Normalizing Flows for Hierarchical Bayesian Analysis: A Gravitational Wave Population Study}",
    eprint = "2211.09008",
    archivePrefix = "arXiv",
    primaryClass = "astro-ph.IM",
    month = "11",
    year = "2022"
}

@article{Rudolph:2023auv,
    author = "Rudolph, Annika and Tamborra, Irene and Gottlieb, Ore",
    title = "{Subphotospheric Emission from Short Gamma-Ray Bursts: Protons Mold the Multimessenger Signals}",
    eprint = "2309.08667",
    archivePrefix = "arXiv",
    primaryClass = "astro-ph.HE",
    doi = "10.3847/2041-8213/ad1525",
    journal = "Astrophys. J. Lett.",
    volume = "961",
    number = "1",
    pages = "L7",
    year = "2024"
}

@article{Ruiz:2021gsv,
    author = "Ruiz, Milton and Shapiro, Stuart L. and Tsokaros, Antonios",
    title = "{Multimessenger Binary Mergers Containing Neutron Stars: Gravitational Waves, Jets, and $\gamma$-Ray Bursts}",
    eprint = "2102.03366",
    archivePrefix = "arXiv",
    primaryClass = "astro-ph.HE",
    doi = "10.3389/fspas.2021.656907",
    journal = "Front. Astron. Space Sci.",
    volume = "8",
    pages = "39",
    year = "2021"
}

@article{Ryan:2019fhz,
    author = "Ryan, Geoffrey and van Eerten, Hendrik and Piro, Luigi and Troja, Eleonora",
    title = "{Gamma-Ray Burst Afterglows in the Multimessenger Era: Numerical Models and Closure Relations}",
    eprint = "1909.11691",
    archivePrefix = "arXiv",
    primaryClass = "astro-ph.HE",
    doi = "10.3847/1538-4357/ab93cf",
    journal = "Astrophys. J.",
    volume = "896",
    number = "2",
    pages = "166",
    year = "2020"
}

@article{Sachdev:2020lfd,
    author = "Sachdev, Surabhi and others",
    title = "{An Early-warning System for Electromagnetic Follow-up of Gravitational-wave Events}",
    eprint = "2008.04288",
    archivePrefix = "arXiv",
    primaryClass = "astro-ph.HE",
    doi = "10.3847/2041-8213/abc753",
    journal = "Astrophys. J. Lett.",
    volume = "905",
    number = "2",
    pages = "L25",
    year = "2020"
}

@article{sadeh_non-thermal_2023,
	title = {Non-thermal emission from mildly relativistic dynamical ejecta of neutron star mergers},
	volume = {518},
	issn = {0035-8711},
	url = {https://doi.org/10.1093/mnras/stac3260},
	doi = {10.1093/mnras/stac3260},
	number = {2},
	urldate = {2023-09-11},
	journal = {Monthly Notices of the Royal Astronomical Society},
	author = {Sadeh, Gilad and Guttman, Or and Waxman, Eli},
	month = jan,
	year = {2023},
	pages = {2102--2112},
}

@article{sadeh_non-thermal_2024,
	title = {Non-thermal emission from mildly relativistic dynamical ejecta of neutron star mergers: spectrum and sky image},
	volume = {531},
	issn = {0035-8711},
	shorttitle = {Non-thermal emission from mildly relativistic dynamical ejecta of neutron star mergers},
	url = {https://ui.adsabs.harvard.edu/abs/2024MNRAS.531.3279S},
	doi = {10.1093/mnras/stae1286},
	urldate = {2025-02-25},
	journal = {Monthly Notices of the Royal Astronomical Society},
	author = {Sadeh, Gilad and Linder, Noya and Waxman, Eli},
	month = jul,
	year = {2024},
	note = {Publisher: OUP
ADS Bibcode: 2024MNRAS.531.3279S},
	pages = {3279--3286},
}

@article{sadeh_late-time_2024,
	title = {Late-time non-thermal emission from mildly relativistic tidal ejecta of compact objects merger},
	volume = {535},
	issn = {0035-8711},
	url = {https://doi.org/10.1093/mnras/stae2561},
	doi = {10.1093/mnras/stae2561},
	number = {4},
	urldate = {2024-12-08},
	journal = {Monthly Notices of the Royal Astronomical Society},
	author = {Sadeh, Gilad},
	month = dec,
	year = {2024},
	pages = {3252--3261},
}

@article{sadeh_synchrotron_2024,
	title = {Synchrotron {Break} {Frequencies} of {Mildly} to {Highly} {Relativistic} {Outflows} {Observed} {Off}-axis},
	volume = {977},
	issn = {0004-637X},
	url = {https://dx.doi.org/10.3847/1538-4357/ad9684},
	doi = {10.3847/1538-4357/ad9684},
	language = {en},
	number = {2},
	urldate = {2024-12-12},
	journal = {The Astrophysical Journal},
	publisher = {The American Astronomical Society},
	author = {Sadeh, Gilad},
	month = dec,
	year = {2024},
	pages = {181},
}

@article{sadeh_nonthermal_2025,
	title = {The {Nonthermal} {Emission} {Following} {GW170817} is {Consistent} with a {Conical} {Radially} {Stratified} {Outflow} with {Initial} {Lorentz} {Factor} ≲10},
	volume = {987},
	issn = {0004-637X},
	url = {https://dx.doi.org/10.3847/1538-4357/ade150},
	doi = {10.3847/1538-4357/ade150},
	language = {en},
	number = {2},
	urldate = {2025-07-09},
	journal = {The Astrophysical Journal},
	publisher = {The American Astronomical Society},
	author = {Sadeh, Gilad and Waxman, Eli},
	month = jul,
	year = {2025},
	pages = {178},
}

@article{Safarzadeh:2019znp,
    author = "Safarzadeh, Mohammadtaher and Berger, Edo",
    title = "{Measuring the delay time distribution of binary neutron stars. I. Through Scaling Relations of the Host Galaxies of Gravitational Wave Events}",
    eprint = "1904.08436",
    archivePrefix = "arXiv",
    primaryClass = "astro-ph.GA",
    doi = "10.3847/2041-8213/ab24df",
    journal = "Astrophys. J. Lett.",
    volume = "878",
    number = "1",
    pages = "L12",
    year = "2019"
}

@article{Salafia:2015vla,
    author = "Salafia, O. S. and Ghisellini, G. and Pescalli, A. and Ghirlanda, G. and Nappo, F.",
    title = "{Structure of Gamma-Ray Burst jets: intrinsic versus apparent properties}",
    eprint = "1502.06608",
    archivePrefix = "arXiv",
    primaryClass = "astro-ph.HE",
    doi = "10.1093/mnras/stv766",
    journal = "Mon. Not. Roy. Astron. Soc.",
    volume = "450",
    number = "4",
    pages = "3549--3558",
    year = "2015"
}

@article{Salafia:2017ebv,
    author = "Salafia, Om Sharan and Colpi, Monica and Branchesi, Marica and Chassande-Mottin, Eric and Ghirlanda, Giancarlo and Ghisellini, Gabriele and Vergani, Susanna",
    title = "{Where and When: Optimal Scheduling of the Electromagnetic Follow-up of Gravitational-wave Events Based on Counterpart Light-curve Models}",
    eprint = "1704.05851",
    archivePrefix = "arXiv",
    primaryClass = "astro-ph.HE",
    doi = "10.3847/1538-4357/aa850e",
    journal = "Astrophys. J.",
    volume = "846",
    number = "1",
    pages = "62",
    year = "2017"
}

@article{Salafia:2019off,
    author = "Salafia, O. S. and Ghirlanda, G. and Ascenzi, S. and Ghisellini, G.",
    title = "{On-axis view of GRB 170817A}",
    eprint = "1905.01190",
    archivePrefix = "arXiv",
    primaryClass = "astro-ph.HE",
    doi = "10.1051/0004-6361/201935831",
    journal = "Astron. Astrophys.",
    volume = "628",
    pages = "A18",
    year = "2019"
}

@article{Salafia:2020jro,
    author = "Salafia, Om S. and Giacomazzo, Bruno",
    title = "{Accretion-to-jet energy conversion efficiency in GW170817 (Corrigendum)}",
    eprint = "2006.07376",
    archivePrefix = "arXiv",
    primaryClass = "astro-ph.HE",
    doi = "10.1051/0004-6361/202038590",
    journal = "Astron. Astrophys.",
    volume = "645",
    pages = "A93",
    year = "2021",
    note = "[Erratum: Astron.Astrophys. 660, C1 (2022)]"
}

@article{Salafia:2022xjd,
    author = "Salafia, O. S. and Colombo, A. and Gabrielli, F. and Mandel, I.",
    title = "{Constraints on the merging binary neutron star mass distribution and equation of state based on the incidence of jets in the population}",
    eprint = "2202.01656",
    archivePrefix = "arXiv",
    primaryClass = "astro-ph.HE",
    doi = "10.1051/0004-6361/202243260",
    journal = "Astron. Astrophys.",
    volume = "666",
    pages = "A174",
    year = "2022"
}

@article{Salvarese:2024jpq,
    author = "Salvarese, Alberto and Chen, Hsin-Yu",
    title = "{Mitigating the Binary Viewing Angle Bias for Standard Sirens}",
    eprint = "2406.11126",
    archivePrefix = "arXiv",
    primaryClass = "astro-ph.CO",
    doi = "10.3847/2041-8213/ad7bbc",
    journal = "Astrophys. J. Lett.",
    volume = "974",
    number = "1",
    pages = "L16",
    year = "2024"
}

@article{Samajdar:2018dcx,
    author = "Samajdar, Anuradha and Dietrich, Tim",
    title = "{Waveform systematics for binary neutron star gravitational wave signals: effects of the point-particle baseline and tidal descriptions}",
    eprint = "1810.03936",
    archivePrefix = "arXiv",
    primaryClass = "gr-qc",
    doi = "10.1103/PhysRevD.98.124030",
    journal = "Phys. Rev. D",
    volume = "98",
    number = "12",
    pages = "124030",
    year = "2018"
}

@article{Samajdar:2019ulq,
    author = "Samajdar, Anuradha and Dietrich, Tim",
    title = "{Waveform systematics for binary neutron star gravitational wave signals: Effects of spin, precession, and the observation of electromagnetic counterparts}",
    eprint = "1905.03118",
    archivePrefix = "arXiv",
    primaryClass = "gr-qc",
    doi = "10.1103/PhysRevD.100.024046",
    journal = "Phys. Rev. D",
    volume = "100",
    number = "2",
    pages = "024046",
    year = "2019"
}

@article{Sarin:2020gxb,
    author = "Sarin, Nikhil and Lasky, Paul D.",
    title = "{The evolution of binary neutron star post-merger remnants: a review}",
    eprint = "2012.08172",
    archivePrefix = "arXiv",
    primaryClass = "astro-ph.HE",
    doi = "10.1007/s10714-021-02831-1",
    journal = "Gen. Rel. Grav.",
    volume = "53",
    number = "6",
    pages = "59",
    year = "2021"
}

@article{Sarin:2024tja,
    author = "Sarin, Nikhil and Rosswog, Stephan",
    title = "{Cautionary Tales on Heating-rate Prescriptions in Kilonovae}",
    eprint = "2404.07271",
    archivePrefix = "arXiv",
    primaryClass = "astro-ph.HE",
    doi = "10.3847/2041-8213/ad739d",
    journal = "Astrophys. J. Lett.",
    volume = "973",
    number = "1",
    pages = "L24",
    year = "2024"
}

@article{Savchenko:2017ffs,
    author = "Savchenko, V. and others",
    title = "{INTEGRAL Detection of the First Prompt Gamma-Ray Signal Coincident with the Gravitational-wave Event GW170817}",
    eprint = "1710.05449",
    archivePrefix = "arXiv",
    primaryClass = "astro-ph.HE",
    doi = "10.3847/2041-8213/aa8f94",
    journal = "ApJL",
    volume = "848",
    number = "2",
    pages = "L15",
    year = "2017"
}

@article{Schianchi:2024vvi,
    author = "Schianchi, Federico and Ujevic, Maximiliano and Neuweiler, Anna and Gieg, Henrique and Markin, Ivan and Dietrich, Tim",
    title = "{Black-hole formation in binary neutron star mergers: The impact of spin on the prompt-collapse scenario}",
    eprint = "2402.16626",
    archivePrefix = "arXiv",
    primaryClass = "astro-ph.HE",
    doi = "10.1103/PhysRevD.109.123011",
    journal = "Phys. Rev. D",
    volume = "109",
    number = "12",
    pages = "123011",
    year = "2024"
}

@article{Shvartzvald:2023ofi,
    author = "Shvartzvald, Y. and others",
    title = "{ULTRASAT: A Wide-field Time-domain UV Space Telescope}",
    eprint = "2304.14482",
    archivePrefix = "arXiv",
    primaryClass = "astro-ph.IM",
    doi = "10.3847/1538-4357/ad2704",
    journal = "Astrophys. J.",
    volume = "964",
    number = "1",
    pages = "74",
    year = "2024"
}

@article{Siegel:2017nub,
    author = "Siegel, Daniel M. and Metzger, Brian D.",
    title = "{Three-Dimensional General-Relativistic Magnetohydrodynamic Simulations of Remnant Accretion Disks from Neutron Star Mergers: Outflows and $r$-Process Nucleosynthesis}",
    eprint = "1705.05473",
    archivePrefix = "arXiv",
    primaryClass = "astro-ph.HE",
    doi = "10.1103/PhysRevLett.119.231102",
    journal = "Phys. Rev. Lett.",
    volume = "119",
    number = "23",
    pages = "231102",
    year = "2017"
}

@article{Singh:2021zah,
    author = "Singh, Neha and Bulik, Tomasz and Belczynski, Krzysztof and Askar, Abbas",
    title = "{Exploring compact binary populations with the Einstein Telescope}",
    eprint = "2112.04058",
    archivePrefix = "arXiv",
    primaryClass = "astro-ph.HE",
    reportNumber = "Virgo document number VIR-0785A-21 and ET document number
  ET-0010A-22, Virgo document number VIR-0785A-21",
    doi = "10.1051/0004-6361/202142856",
    journal = "Astron. Astrophys.",
    volume = "667",
    pages = "A2",
    year = "2022"
}

@online{SKA2,
    title = "{SKA Telescope Specifications }",
    author = {SKAO},
    year = 2025,
    url = {https://www.skao.int/en/science-users/118/ska-telescope-specifications#__otpm6},
    urldate = {2026-10-30}
}

@article{Somasundaram:2021clp,
    author = "Somasundaram, Rahul and Tews, Ingo and Margueron, J{\'e}r{\^o}me",
    title = "{Investigating signatures of phase transitions in neutron-star cores}",
    eprint = "2112.08157",
    archivePrefix = "arXiv",
    primaryClass = "nucl-th",
    reportNumber = "LA-UR-21-22340",
    doi = "10.1103/PhysRevC.107.025801",
    journal = "Phys. Rev. C",
    volume = "107",
    number = "2",
    pages = "025801",
    year = "2023"
}

@article{Speagle:2019ivv,
    author = "Speagle, Joshua S.",
    title = "{dynesty: a dynamic nested sampling package for estimating Bayesian posteriors and evidences}",
    eprint = "1904.02180",
    archivePrefix = "arXiv",
    primaryClass = "astro-ph.IM",
    doi = "10.1093/mnras/staa278",
    journal = "Mon. Not. Roy. Astron. Soc.",
    volume = "493",
    number = "3",
    pages = "3132--3158",
    year = "2020"
}

@article{Spergel:2015sza,
    author = "Spergel, D. and others",
    title = "{Wide-Field InfrarRed Survey Telescope-Astrophysics Focused Telescope Assets WFIRST-AFTA 2015 Report}",
    eprint = "1503.03757",
    archivePrefix = "arXiv",
    primaryClass = "astro-ph.IM",
    month = "3",
    year = "2015"
}

@article{Stairs:2004ye,
    author = "Stairs, Ingrid H. and Thorsett, S. E. and Arzoumanian, Z.",
    title = "{Measurement of gravitational spin-orbit coupling in a binary pulsar system}",
    eprint = "astro-ph/0408457",
    archivePrefix = "arXiv",
    doi = "10.1103/PhysRevLett.93.141101",
    journal = "Phys. Rev. Lett.",
    volume = "93",
    pages = "141101",
    year = "2004"
}

@article{Steinle:2025xae,
    author = "Steinle, Nathan and Safi-Harb, Samar and Nicholl, Matt and Worssam, Isabelle and Gompertz, Benjamin P.",
    title = "{Binary Neutron Star Mergers: Multi-Messenger Systematics and Prospects with Next-Generation Facilities}",
    eprint = "2507.20258",
    archivePrefix = "arXiv",
    primaryClass = "astro-ph.HE",
    month = "7",
    year = "2025"
}

@article{Stevance:2023byv,
    author = "Stevance, Heloise F. and Eldridge, Jan J. and Stanway, Elizabeth R. and Lyman, Joe and McLeod, Anna F. and Levan, Andrew J.",
    title = "{End-to-end study of the host galaxy and genealogy of the first binary neutron star merger}",
    eprint = "2301.05236",
    archivePrefix = "arXiv",
    primaryClass = "astro-ph.HE",
    doi = "10.1038/s41550-022-01873-y",
    journal = "Nature Astron.",
    volume = "7",
    number = "4",
    pages = "444--450",
    year = "2023"
}

@article{Stevenson:2025fqt,
    author = {Stevenson, Simon and M{\"o}ller, Anais and Powell, Jade},
    title = "{Strategy for Identifying Vera C. Rubin Observatory Kilonova Candidates for Targeted Gravitational-wave Searches}",
    eprint = "2510.12932",
    archivePrefix = "arXiv",
    primaryClass = "astro-ph.HE",
    doi = "10.3847/1538-4357/ae244e",
    journal = "Astrophys. J.",
    volume = "998",
    number = "1",
    pages = "8",
    year = "2026"
}

@article{STROBE-XScienceWorkingGroup:2019cyd,
    author = "Ray, Paul S. and others",
    collaboration = "STROBE-X Science Working Group",
    title = "{STROBE-X: X-ray Timing and Spectroscopy on Dynamical Timescales from Microseconds to Years}",
    eprint = "1903.03035",
    archivePrefix = "arXiv",
    primaryClass = "astro-ph.IM",
    month = "3",
    year = "2019"
}

@article{Suleiman:2024ztn,
    author = "Suleiman, Lami and Read, Jocelyn",
    title = "{Quasiuniversal relations in the context of future neutron star detections}",
    eprint = "2402.01948",
    archivePrefix = "arXiv",
    primaryClass = "astro-ph.HE",
    doi = "10.1103/PhysRevD.109.103029",
    journal = "Phys. Rev. D",
    volume = "109",
    number = "10",
    pages = "103029",
    year = "2024"
}

@article{Tanvir:2017pws,
    author = "Tanvir, N. R. and others",
    title = "{The Emergence of a Lanthanide-Rich Kilonova Following the Merger of Two Neutron Stars}",
    eprint = "1710.05455",
    archivePrefix = "arXiv",
    primaryClass = "astro-ph.HE",
    doi = "10.3847/2041-8213/aa90b6",
    journal = "Astrophys. J. Lett.",
    volume = "848",
    number = "2",
    pages = "L27",
    year = "2017"
}

@article{Tchekhovskoy:2009ba,
    author = "Tchekhovskoy, Alexander and Narayan, Ramesh and McKinney, Jonathan C.",
    title = "{Black Hole Spin and the Radio Loud/Quiet Dichotomy of Active Galactic Nuclei}",
    eprint = "0911.2228",
    archivePrefix = "arXiv",
    primaryClass = "astro-ph.HE",
    doi = "10.1088/0004-637X/711/1/50",
    journal = "Astrophys. J.",
    volume = "711",
    pages = "50--63",
    year = "2010"
}

@article{Tews:2018iwm,
    author = "Tews, I. and Margueron, J. and Reddy, S.",
    title = "{Critical examination of constraints on the equation of state of dense matter obtained from GW170817}",
    eprint = "1804.02783",
    archivePrefix = "arXiv",
    primaryClass = "nucl-th",
    reportNumber = "INT-PUB-18-014",
    doi = "10.1103/PhysRevC.98.045804",
    journal = "Phys. Rev. C",
    volume = "98",
    number = "4",
    pages = "045804",
    year = "2018"
}

@article{Troja:2017nqp,
    author = "Troja, E. and others",
    title = "{The X-ray counterpart to the gravitational wave event GW 170817}",
    eprint = "1710.05433",
    archivePrefix = "arXiv",
    primaryClass = "astro-ph.HE",
    doi = "10.1038/nature24290",
    journal = "Nature",
    volume = "551",
    pages = "71--74",
    year = "2017"
}

@article{Valenti:2017ngx,
    author = "Valenti, Stefano and Sand, David J. and Yang, Sheng and Cappellaro, Enrico and Tartaglia, Leonardo and Corsi, Alessandra and Jha, Saurabh W. and Reichart, Daniel E. and Haislip, Joshua and Kouprianov, Vladimir",
    title = "{The discovery of the electromagnetic counterpart of GW170817: kilonova AT 2017gfo/DLT17ck}",
    eprint = "1710.05854",
    archivePrefix = "arXiv",
    primaryClass = "astro-ph.HE",
    doi = "10.3847/2041-8213/aa8edf",
    journal = "Astrophys. J. Lett.",
    volume = "848",
    number = "2",
    pages = "L24",
    year = "2017"
}

@article{vanPutten:2022xyx,
    author = "van Putten, Maurice H. P. M. and Della Valle, Massimo",
    title = "{Central engine of GRB170817A: Neutron star versus Kerr black hole based on multimessenger calorimetry and event timing}",
    eprint = "2212.03295",
    archivePrefix = "arXiv",
    primaryClass = "astro-ph.HE",
    doi = "10.1051/0004-6361/202142974",
    journal = "Astron. Astrophys.",
    volume = "669",
    pages = "A36",
    year = "2023"
}

@report{Vikhlinin:2022sde,
  author       = {Vikhlinin, Alexey and
                  {\"O}zel, Feryal and
                  Gaskin, Jessica},
  title        = {{LYNX X-ray Observatory Concept Study Report}},
  year         = {2022},
  institution  = {{LYNX} Team},
  url          = {https://wwwastro.msfc.nasa.gov/lynx/docs/LynxConceptStudy.pdf},
}

@article{Virtuoso:2024kyp,
    author = "Virtuoso, Andrea and Milotti, Edoardo",
    title = "{Beyond the long wavelength approximation: Next-generation gravitational-wave detectors and frequency-dependent antenna patterns}",
    eprint = "2412.01693",
    archivePrefix = "arXiv",
    primaryClass = "gr-qc",
    doi = "10.1103/PhysRevD.111.064058",
    journal = "Phys. Rev. D",
    volume = "111",
    number = "6",
    pages = "064058",
    year = "2025"
}

@article{Vitale:2018yhm,
    author = "Vitale, Salvatore and Farr, Will M. and Ng, Ken and Rodriguez, Carl L.",
    title = "{Measuring the star formation rate with gravitational waves from binary black holes}",
    eprint = "1808.00901",
    archivePrefix = "arXiv",
    primaryClass = "astro-ph.HE",
    reportNumber = "LIGO Document P1800219",
    doi = "10.3847/2041-8213/ab50c0",
    journal = "Astrophys. J. Lett.",
    volume = "886",
    number = "1",
    pages = "L1",
    year = "2019"
}

@article{vonKienlin:2020xvz,
    author = "von Kienlin, A. and others",
    title = "{The Fourth Fermi-GBM Gamma-Ray Burst Catalog: A Decade of Data}",
    eprint = "2002.11460",
    archivePrefix = "arXiv",
    primaryClass = "astro-ph.HE",
    doi = "10.3847/1538-4357/ab7a18",
    journal = "Astrophys. J.",
    volume = "893",
    pages = "46",
    year = "2020"
}

@article{Wade:2014vqa,
    author = "Wade, Leslie and Creighton, Jolien D. E. and Ochsner, Evan and Lackey, Benjamin D. and Farr, Benjamin F. and Littenberg, Tyson B. and Raymond, Vivien",
    title = "{Systematic and statistical errors in a bayesian approach to the estimation of the neutron-star equation of state using advanced gravitational wave detectors}",
    eprint = "1402.5156",
    archivePrefix = "arXiv",
    primaryClass = "gr-qc",
    doi = "10.1103/PhysRevD.89.103012",
    journal = "Phys. Rev. D",
    volume = "89",
    number = "10",
    pages = "103012",
    year = "2014"
}

@article{Walker:2024loo,
    author = "Walker, Kris and Smith, Rory and Thrane, Eric and Reardon, Daniel J.",
    title = "{Precision constraints on the neutron star equation of state with third-generation gravitational-wave observatories}",
    eprint = "2401.02604",
    archivePrefix = "arXiv",
    primaryClass = "astro-ph.HE",
    reportNumber = "LIGO-P2300446",
    doi = "10.1103/PhysRevD.110.043013",
    journal = "Phys. Rev. D",
    volume = "110",
    number = "4",
    pages = "043013",
    year = "2024"
}

@article{Wanderman:2014eza,
    author = "Wanderman, David and Piran, Tsvi",
    title = "{The rate, luminosity function and time delay of non-Collapsar short GRBs}",
    eprint = "1405.5878",
    archivePrefix = "arXiv",
    primaryClass = "astro-ph.HE",
    doi = "10.1093/mnras/stv123",
    journal = "Mon. Not. Roy. Astron. Soc.",
    volume = "448",
    number = "4",
    pages = "3026--3037",
    year = "2015"
}

@article{Wang:2015vpa,
    author = "Wang, Xiang-Gao and others",
    title = "{How bad or Good are the External Forward Shock Afterglow Models of Gamma-ray Bursts?}",
    eprint = "1503.03193",
    archivePrefix = "arXiv",
    primaryClass = "astro-ph.HE",
    doi = "10.1088/0067-0049/219/1/9",
    journal = "Astrophys. J. Suppl.",
    volume = "219",
    number = "1",
    pages = "9",
    year = "2015"
}

@article{Wang:2020vgr,
    author = "Wang, Hao and Giannios, Dimitrios",
    title = "{Multimessenger parameter estimation of GW170817: from jet structure to the Hubble constant}",
    eprint = "2009.04427",
    archivePrefix = "arXiv",
    primaryClass = "astro-ph.HE",
    doi = "10.3847/1538-4357/abd39c",
    journal = "Astrophys. J.",
    volume = "908",
    number = "2",
    pages = "200",
    year = "2021"
}

@software{ward2023flowjax,
  title = {FlowJAX: Distributions and Normalizing Flows in Jax},
  author = {Daniel Ward},
  url = {https://github.com/danielward27/flowjax},
  version = {17.2.1},
  year = {2026},
  doi = {10.5281/zenodo.10402073},
}

@inproceedings{Wei:2016eox,
    author = "Wei, J. and Cordier, B.",
    title = "{The Deep and Transient Universe in the SVOM Era: New Challenges and Opportunities - Scientific prospects of the SVOM mission}",
    eprint = "1610.06892",
    archivePrefix = "arXiv",
    primaryClass = "astro-ph.IM",
    month = "10",
    year = "2016"
}

@article{Williams:2026jqv,
    author = "Williams, Natalie and Puecher, Anna and Grams, Guilherme and Flores, C{\'e}sar V. and Dietrich, Tim",
    title = "{Investigating the impact of quasiuniversal relations on neutron star constraints in third-generation detectors}",
    eprint = "2602.14659",
    archivePrefix = "arXiv",
    primaryClass = "gr-qc",
    doi = "10.1103/c862-2dvz",
    journal = "Phys. Rev. D",
    volume = "114",
    number = "2",
    pages = "024063",
    year = "2026"
}

@article{Wong:2022xvh,
    author = "Wong, Kaze W. k. and Gabri{\'e}, Marylou and Foreman-Mackey, Daniel",
    title = "{flowMC: Normalizing flow enhanced sampling package for probabilistic inference in JAX}",
    eprint = "2211.06397",
    archivePrefix = "arXiv",
    primaryClass = "astro-ph.IM",
    doi = "10.21105/joss.05021",
    journal = "J. Open Source Softw.",
    volume = "8",
    number = "83",
    pages = "5021",
    year = "2023"
}

@article{Wouters:2025zju,
    author = "Wouters, Thibeau and Pang, Peter T. H. and Koehn, Hauke and Rose, Henrik and Somasundaram, Rahul and Tews, Ingo and Dietrich, Tim and Van Den Broeck, Chris",
    title = "{Leveraging differentiable programming in the inverse problem of neutron stars}",
    eprint = "2504.15893",
    archivePrefix = "arXiv",
    primaryClass = "astro-ph.HE",
    reportNumber = "LA-UR-25-23486",
    doi = "10.1103/v2y8-kxvx",
    journal = "Phys. Rev. D",
    volume = "112",
    number = "4",
    pages = "043037",
    year = "2025"
}

@article{Wysocki:2020myz,
    author = "Wysocki, Daniel and O'Shaughnessy, Richard and Wade, Leslie and Lange, Jacob",
    title = "{Inferring the neutron star equation of state simultaneously with the population of merging neutron stars}",
    eprint = "2001.01747",
    archivePrefix = "arXiv",
    primaryClass = "gr-qc",
    reportNumber = "LIGO-P1900359",
    month = "1",
    year = "2020"
}

@article{Yagi:2016bkt,
    author = "Yagi, Kent and Yunes, Nicol{\'a}s",
    title = "{Approximate Universal Relations for Neutron Stars and Quark Stars}",
    eprint = "1608.02582",
    archivePrefix = "arXiv",
    primaryClass = "gr-qc",
    doi = "10.1016/j.physrep.2017.03.002",
    journal = "Phys. Rept.",
    volume = "681",
    pages = "1--72",
    year = "2017"
}

@article{Yi:2021wqf,
    author = "Yi, Shu-Xu and Nelemans, Gijs and Brinkerink, Christiaan and Kostrzewa-Rutkowska, Zuzanna and Timmer, Sjoerd T. and Stoppa, Fiorenzo and Rossi, Elena M. and Portegies Zwart, Simon F.",
    title = "{The Gravitational Wave Universe Toolbox - A software package to simulate observations of the gravitational wave universe with different detectors,}",
    eprint = "2106.13662",
    archivePrefix = "arXiv",
    primaryClass = "astro-ph.HE",
    doi = "10.1051/0004-6361/202141634",
    journal = "Astron. Astrophys.",
    volume = "663",
    pages = "A155",
    year = "2022"
}

@article{Yuan:2025cbh,
    author = "Yuan, Weimin and others",
    title = "{Science objectives of the Einstein Probe mission}",
    eprint = "2501.07362",
    archivePrefix = "arXiv",
    primaryClass = "astro-ph.HE",
    doi = "10.1007/s11433-024-2600-3",
    journal = "Sci. China Phys. Mech. Astron.",
    volume = "68",
    number = "3",
    pages = "239501",
    year = "2025"
}

@article{Zhao:2017imr,
    author = "Zhao, Wen and Santos, Larissa",
    title = "{Model-independent measurement of the absolute magnitude of Type Ia Supernovae with gravitational-wave sources}",
    eprint = "1710.10055",
    archivePrefix = "arXiv",
    primaryClass = "astro-ph.HE",
    doi = "10.1088/1475-7516/2019/11/009",
    journal = "JCAP",
    volume = "11",
    pages = "009",
    year = "2019"
}

@article{Zhang:2020axa,
    author = "Zhang, Sixuan and Cao, Shuo and Zhang, Jia and Liu, Tonghua and Liu, Yuting and Geng, Shuaibo and Lian, Yujie",
    title = "{A model-independent constraint on the Hubble constant with gravitational waves from the Einstein Telescope}",
    eprint = "2009.04204",
    archivePrefix = "arXiv",
    primaryClass = "astro-ph.CO",
    month = "9",
    year = "2020"
}

@article{Zhu:2017znf,
    author = "Zhu, Xingjiang and Thrane, Eric and Oslowski, Stefan and Levin, Yuri and Lasky, Paul D.",
    title = "{Inferring the population properties of binary neutron stars with gravitational-wave measurements of spin}",
    eprint = "1711.09226",
    archivePrefix = "arXiv",
    primaryClass = "astro-ph.HE",
    reportNumber = "LIGO-P1700400",
    doi = "10.1103/PhysRevD.98.043002",
    journal = "Phys. Rev. D",
    volume = "98",
    pages = "043002",
    year = "2018"
}

@article{Zhu:2020zij,
    author = "Zhu, Xing-Jiang and Ashton, Gregory",
    title = "{Characterizing Astrophysical Binary Neutron Stars with Gravitational Waves}",
    eprint = "2007.08198",
    archivePrefix = "arXiv",
    primaryClass = "astro-ph.HE",
    doi = "10.3847/2041-8213/abb6ea",
    journal = "Astrophys. J. Lett.",
    volume = "902",
    number = "1",
    pages = "L12",
    month = "7",
    year = "2020"
}

@article{Iacovelli:2023nbv,
    author = "Iacovelli, Francesco and Mancarella, Michele and Mondal, Chiranjib and Puecher, Anna and Dietrich, Tim and Gulminelli, Francesca and Maggiore, Michele and Oertel, Micaela",
    title = "{Nuclear physics constraints from binary neutron star mergers in the Einstein Telescope era}",
    eprint = "2308.12378",
    archivePrefix = "arXiv",
    primaryClass = "gr-qc",
    doi = "10.1103/PhysRevD.108.122006",
    journal = "Phys. Rev. D",
    volume = "108",
    number = "12",
    pages = "122006",
    year = "2023"
}

@inbook{MCMC_handbook,
    author = {Geyer, Charles J},
    title = {Handbook of Markov chain Monte Carlo},
    publisher = {Chapman and Hall / CRC},
    year = "2011",
    edition = 1,
    chapter = 1,
    doi = "10.1201/b10905"
}

@article{Skilling:2004pqw,
    author = "Skilling, John",
    title = "{Nested Sampling}",
    doi = "10.1063/1.1835238",
    journal = "AIP Conf. Proc.",
    volume = "735",
    number = "1",
    pages = "395",
    year = "2004"
}

@article{Skilling:2006gxv,
    author = "Skilling, John",
    title = "{Nested sampling for general Bayesian computation}",
    doi = "10.1214/06-BA127",
    journal = "Bayesian Analysis",
    volume = "1",
    number = "4",
    pages = "833--859",
    year = "2006"
}

@article{Kobyzev:2019ydm,
    author = "Kobyzev, Ivan and Prince, Simon J. D. and Brubaker, Marcus A.",
    title = "{Normalizing Flows: An Introduction and Review of Current Methods}",
    eprint = "1908.09257",
    archivePrefix = "arXiv",
    primaryClass = "stat.ML",
    doi = "10.1109/tpami.2020.2992934",
    journal = "IEEE Trans. Pattern Anal. Machine Intell.",
    volume = "43",
    number = "11",
    pages = "3964--3979",
    year = "2021"
}

@article{Papamakarios:2019fms,
    author = "Papamakarios, George and Nalisnick, Eric and Rezende, Danilo Jimenez and Mohamed, Shakir and Lakshminarayanan, Balaji",
    title = "{Normalizing Flows for Probabilistic Modeling and Inference}",
    eprint = "1912.02762",
    archivePrefix = "arXiv",
    primaryClass = "stat.ML",
    doi = "10.5555/3546258.3546315",
    journal = "J. Machine Learning Res.",
    volume = "22",
    number = "1",
    pages = "2617--2680",
    year = "2021"
}

@article{Douchin:2001sv,
    author = "Douchin, F. and Haensel, P.",
    title = "{A unified equation of state of dense matter and neutron star structure}",
    eprint = "astro-ph/0111092",
    archivePrefix = "arXiv",
    doi = "10.1051/0004-6361:20011402",
    journal = "Astron. Astrophys.",
    volume = "380",
    pages = "151",
    year = "2001"
}

@article{Somasundaram:2020chb,
    author = "Somasundaram, R. and Drischler, C. and Tews, I. and Margueron, J.",
    title = "{Constraints on the nuclear symmetry energy from asymmetric-matter calculations with chiral $NN$ and $3N$ interactions}",
    eprint = "2009.04737",
    archivePrefix = "arXiv",
    primaryClass = "nucl-th",
    reportNumber = "LA-UR-20-26894",
    doi = "10.1103/PhysRevC.103.045803",
    journal = "Phys. Rev. C",
    volume = "103",
    number = "4",
    pages = "045803",
    year = "2021"
}

@article{Grams:2021lzx,
    author = "Grams, G. and Somasundaram, R. and Margueron, J. and Reddy, S.",
    title = "{Properties of the neutron star crust: Quantifying and correlating uncertainties with improved nuclear physics}",
    eprint = "2110.00441",
    archivePrefix = "arXiv",
    primaryClass = "nucl-th",
    doi = "10.1103/PhysRevC.105.035806",
    journal = "Phys. Rev. C",
    volume = "105",
    number = "3",
    pages = "035806",
    year = "2022"
}

@article{del2006sequential,
  title={Sequential monte carlo samplers},
  author={Del Moral, Pierre and Doucet, Arnaud and Jasra, Ajay},
  journal={Journal of the Royal Statistical Society Series B: Statistical Methodology},
  volume={68},
  number={3},
  pages={411--436},
  year={2006},
  publisher={Oxford University Press}
}

@book{chopin2020introduction,
  title={An introduction to sequential Monte Carlo},
  author={Chopin, Nicolas and Papaspiliopoulos, Omiros and others},
  volume={4},
  year={2020},
  publisher={Springer},
  series={}
}

@article{Kumar:2026ckp,
    author = "Kumar, Sumit and Melching, Max and Ohme, Frank and Narola, Harsh and Dooney, Tom and Van Den Broeck, Chris",
    title = "{Mitigating Systematic Errors in Parameter Estimation of Binary Black Hole Mergers in O1-O3 LIGO-Virgo Data}",
    eprint = "2604.21859",
    archivePrefix = "arXiv",
    primaryClass = "astro-ph.HE",
    month = "4",
    year = "2026"
}

@article{Goncharov:2022dgl,
    author = "Goncharov, Boris and Nitz, Alexander H. and Harms, Jan",
    title = "{Utilizing the null stream of the Einstein Telescope}",
    eprint = "2204.08533",
    archivePrefix = "arXiv",
    primaryClass = "gr-qc",
    doi = "10.1103/PhysRevD.105.122007",
    journal = "Phys. Rev. D",
    volume = "105",
    number = "12",
    pages = "122007",
    year = "2022"
}

@article{Wong:2025iaf,
    author = "Wong, Isaac C. F. and Cireddu, Francesco and Wils, Milan and Colemont, Tom and Narola, Harsh and Van Den Broeck, Chris and Li, Tjonnie G. F.",
    title = "{Bayesian Calibration of Gravitational-Wave Detectors Using Null Streams Without Waveform Assumptions}",
    eprint = "2510.06327",
    archivePrefix = "arXiv",
    primaryClass = "gr-qc",
    month = "10",
    year = "2025"
}

@article{Vallisneri:2007ev,
    author = "Vallisneri, Michele",
    title = "{Use and abuse of the Fisher information matrix in the assessment of gravitational-wave parameter-estimation prospects}",
    eprint = "gr-qc/0703086",
    archivePrefix = "arXiv",
    reportNumber = "LIGO-P070009-00-Z",
    doi = "10.1103/PhysRevD.77.042001",
    journal = "Phys. Rev. D",
    volume = "77",
    pages = "042001",
    year = "2008"
}

@article{Rodriguez:2013mla,
    author = "Rodriguez, Carl L. and Farr, Benjamin and Farr, Will M. and Mandel, Ilya",
    title = "{Inadequacies of the Fisher Information Matrix in gravitational-wave parameter estimation}",
    eprint = "1308.1397",
    archivePrefix = "arXiv",
    primaryClass = "astro-ph.IM",
    doi = "10.1103/PhysRevD.88.084013",
    journal = "Phys. Rev. D",
    volume = "88",
    number = "8",
    pages = "084013",
    year = "2013"
}

@article{Dhani:2024jja,
    author = {Dhani, Arnab and V{\"o}lkel, Sebastian H. and Buonanno, Alessandra and Estelles, Hector and Gair, Jonathan and Pfeiffer, Harald P. and Pompili, Lorenzo and Toubiana, Alexandre},
    title = "{Systematic Biases in Estimating the Properties of Black Holes Due to Inaccurate Gravitational-Wave Models}",
    eprint = "2404.05811",
    archivePrefix = "arXiv",
    primaryClass = "gr-qc",
    doi = "10.1103/5pks-qz6b",
    journal = "Phys. Rev. X",
    volume = "15",
    number = "3",
    pages = "031036",
    year = "2025"
}

@article{Janssens:2022xmo,
    author = "Janssens, Kamiel and Boileau, Guillaume and Christensen, Nelson and Badaracco, Francesca and van Remortel, Nick",
    title = "{Impact of correlated seismic and correlated Newtonian noise on the Einstein Telescope}",
    eprint = "2206.06809",
    archivePrefix = "arXiv",
    primaryClass = "astro-ph.IM",
    doi = "10.1103/PhysRevD.106.042008",
    journal = "Phys. Rev. D",
    volume = "106",
    number = "4",
    pages = "042008",
    year = "2022"
}

@article{Leyde:2026hvm,
    author = "Leyde, Konstantin and Green, Stephen R. and Dax, Maximilian and Mould, Matthew and Fabbri, Cecilia Maria and Gair, Jonathan",
    title = "{End-to-End Population Inference from Gravitational-Wave Strain using Transformers}",
    eprint = "2605.11274",
    archivePrefix = "arXiv",
    primaryClass = "gr-qc",
    month = "5",
    year = "2026"
}

@article{Tagliazucchi:2026dpr,
    author = "Tagliazucchi, Matteo and Moresco, Michele and Agapito, Alessandro and Mancarella, Michele and Ferraiuolo, Sarah and Mastrogiovanni, Simone and Borghi, Nicola and Pannarale, Francesco and Bonacorsi, Daniele",
    title = "{Pushing spectral siren cosmology into the third-generation era: A blinded mock data challenge}",
    eprint = "2602.17756",
    archivePrefix = "arXiv",
    primaryClass = "astro-ph.CO",
    doi = "10.1051/0004-6361/202659517",
    journal = "Astron. Astrophys.",
    volume = "711",
    pages = "A120",
    year = "2026"
}

@article{Borghi:2023opd,
    author = "Borghi, Nicola and Mancarella, Michele and Moresco, Michele and Tagliazucchi, Matteo and Iacovelli, Francesco and Cimatti, Andrea and Maggiore, Michele",
    title = "{Cosmology and Astrophysics with Standard Sirens and Galaxy Catalogs in View of Future Gravitational Wave Observations}",
    eprint = "2312.05302",
    archivePrefix = "arXiv",
    primaryClass = "astro-ph.CO",
    doi = "10.3847/1538-4357/ad20eb",
    journal = "Astrophys. J.",
    volume = "964",
    number = "2",
    pages = "191",
    year = "2024"
}

@article{Tagliazucchi:2025ofb,
    author = "Tagliazucchi, Matteo and Moresco, Michele and Borghi, Nicola and Fiebig, Manfred",
    title = "{Accelerating the standard siren method: Improved constraints on modified gravitational-wave propagation with future data}",
    eprint = "2504.02034",
    archivePrefix = "arXiv",
    primaryClass = "astro-ph.CO",
    doi = "10.1051/0004-6361/202554827",
    journal = "Astron. Astrophys.",
    volume = "702",
    pages = "A244",
    year = "2025"
}

@article{Mastrogiovanni:2023zbw,
    author = "Mastrogiovanni, Simone and Pierra, Gr{\'e}goire and Perri{\`e}s, St{\'e}phane and Laghi, Danny and Caneva Santoro, Giada and Ghosh, Archisman and Gray, Rachel and Karathanasis, Christos and Leyde, Konstantin",
    title = "{ICAROGW: A python package for inference of astrophysical population properties of noisy, heterogeneous, and incomplete observations}",
    eprint = "2305.17973",
    archivePrefix = "arXiv",
    primaryClass = "astro-ph.CO",
    doi = "10.1051/0004-6361/202347007",
    journal = "Astron. Astrophys.",
    volume = "682",
    pages = "A167",
    year = "2024"
}

@article{Mould:2025dts,
    author = "Mould, Matthew and Wolfe, Noah E. and Vitale, Salvatore",
    title = "{Rapid inference and comparison of gravitational-wave population models with neural variational posteriors}",
    eprint = "2504.07197",
    archivePrefix = "arXiv",
    primaryClass = "astro-ph.IM",
    doi = "10.1103/xk1z-fxnm",
    journal = "Phys. Rev. D",
    volume = "111",
    number = "12",
    pages = "123049",
    year = "2025"
}

@article{Papadopoulos:2026puy,
    author = "Papadopoulos, Alexander and Chapman-Bird, Christian E. A. and Gray, Rachel and Messenger, Christopher and Bertheas, Tom",
    title = "{Scalable dark siren cosmology with gwcosmo: GPU acceleration, validation, and systematics}",
    eprint = "2605.23538",
    archivePrefix = "arXiv",
    primaryClass = "astro-ph.CO",
    doi = "10.1103/53c5-mtk5",
    journal = "Phys. Rev. D",
    volume = "114",
    number = "6",
    pages = "063528",
    year = "2026"
}

@article{Talbot:2024yqw,
    author = "Talbot, Colm and Farah, Amanda and Galaudage, Shanika and Golomb, Jacob and Tong, Hui",
    title = "{GWPopulation: Hardware agnostic population inference for compact binaries and beyond}",
    eprint = "2409.14143",
    archivePrefix = "arXiv",
    primaryClass = "astro-ph.IM",
    doi = "10.21105/joss.07753",
    journal = "J. Open Source Softw.",
    volume = "10",
    number = "109",
    pages = "7753",
    year = "2025"
}

@article{Mondal:2021vzt,
    author = "Mondal, Chiranjib and Gulminelli, Francesca",
    title = "{Can we decipher the composition of the core of a neutron star?}",
    eprint = "2111.04520",
    archivePrefix = "arXiv",
    primaryClass = "nucl-th",
    doi = "10.1103/PhysRevD.105.083016",
    journal = "Phys. Rev. D",
    volume = "105",
    number = "8",
    pages = "083016",
    year = "2022"
}

\appendix
\section{\textsc{possis} geometry}
\label{app:possis_geometry}
\textsc{possis} is a 3D Monte-Carlo radiation-transfer code to determine kilonova emission~\citep{Bulla:2019muo, Bulla:2022mwo}. 
It assumes homologous expansion for the ejecta and determines the emitted flux by propagating photon packets according to the local heating rates, thermalisation efficiencies and interactions through wavelength-dependent opacities~\citep{Bulla:2022mwo}.

It is possible to use the output of \ac{NR} simulations directly as input geometry~\citep{Neuweiler:2022eum}, but for our general \ac{KN} model that is supposed span a wide range of masses and velocities, we rely on the idealized geometry from \citet{Kawaguchi:2019nju}.
This geometry proposes with two components, one being the dynamical ejecta
\begin{align}
    \rho_{\text{dyn}}(v, \theta) &= \rho_{0,\text{d}} \sin^2(\theta) \begin{cases}
        v^{-\alpha_{1}} \quad \quad \quad \quad \ v_{\text{min,d}} \leq v \leq v_{\text{mid}}\\
        v_{\text{mid}}^{-\alpha_1 + \alpha_2} v^{-\alpha_2}\ \  v_{\text{mid}} \leq v \leq v_{\text{max,d}} \\
        0 \qquad \qquad \qquad \text{else}
    \end{cases}
\end{align}
and the other the wind ejecta
\begin{align}
    \rho_{\text{wind}}(v) &= \rho_{0,\text{w}}\begin{cases}
        v^{-\alpha_w} \quad v_\text{min,w} \leq v \leq v_{\text{max,w}}\\
        0 \quad \quad \quad \text{else}
    \end{cases}\ .
\end{align}
The power law slopes, normalization, and velocity boundaries are determined by the input parameters of our \ac{KN} model.
The six input parameters are $m_{\text{ej, dyn}}$, $m_{\text{ej, wind}}$, $\bar{v}_{\text{ej, dyn}}$, $\bar{v}_{\text{ej,wind}}$, $\bar{Y}_{e,\text{dyn}}$, and $Y_{e,\text{wind}}$.
The bars indicate mass averaged quantities and hence $v_{\text{max}}$, $v_{\text{min}}$, and $\alpha_1$ are chosen such that 
\begin{align}
    \frac{\int v^3 \rho(v, \theta)\ dv d\Omega}{\int v^2 \rho(v, \theta)\ dv d\Omega}  = \bar{v}\ .
    \label{eq:average_velocity}
\end{align}
Specifically, for the wind ejecta, we fix $\alpha_1 = 3$ and $v_{\text{min}}=0.02c$ and adjust $v_{\text{max}}$ so that the requested $\bar{v}_{\text{ej,wind}}$ is met via Eq.~\eqref{eq:average_velocity}. 
The overall normalization $\rho_{0,\text{w}}$ is then determined by the requested mass $m{_\text{ej, wind}}$.
We proceed similarly for the dynamical ejecta, where we fix $\alpha_1=4$, $\alpha_2=8$, $v_{\text{min}}=0.1c$, $v_{\text{mid}}=0.4c$, but if the requested $\bar{v}_{\text{ej, dyn}}$ exceeds 0.204$c$, the resulting maximum velocity would have to be bigger than 0.9$c$.
Once this happens, we instead keep $v_{\text{max,dyn}}=0.9c$ and start to alter $\alpha_1$ and $v_{\text{min, dyn}}$.
Specifically, we increase $v_{\text{min}}$ to match the average velocity, while simultaneously decreasing $\alpha_1$. 
During this process, the two parameters are linked through $\alpha_1 = 6- 2\frac{v_{\text{min,dyn}}}{0.1c}$. 
Both the dynamical and wind component are then combined to yield the total density 
\begin{align}
    \rho(v, \theta) &= \begin{cases}
        C_{\text{w}}\ \rho_{\text{wind}}(v) \ \ \ \ \text{if condition 1} \\
        C_{\text{d}}\ \rho_{\text{dyn}}(v, \theta) \ \ \text{elif condition 2} \\
        0 \qquad \qquad \ \ \ \ \text{ else}
    \end{cases}\ .
\end{align}
Condition 1 is shorthand for $v_{\text{min,wind}}\leq v\leq v_{\text{max,wind}}\ \land (v\leq v_{\text{min,dyn}} \lor \theta <\pi/4 \lor \theta>3\pi/4)$, condition 2 is simply $v_{\text{min,dyn}}\leq v\leq v_{\text{max,dyn}}$~\citep{Kawaguchi:2019nju}.
The normalizations $C_{\text{w}}$ and $C_{\text{d}}$ are chosen such that the wind and dynamical ejecta mass matches the requested $m_{\text{ej, dyn}}$ and $m_{\text{ej, wind}}$.

Based on 3660 \textsc{possis} radiative-transfer simulations conducted with this setup (each simulation provides 11 light curves at different inclinations), we use \textsc{fiesta}~\citep{Koehn:2025zzb} to train the machine-learning surrogate that produces the \ac{KN} light curves for this work.
The number of photon packets for each simulation was set to $10^7$.

\section{\ac{EM} counterparts with reduced wind ejecta}
\label{app:reduced_wind_ejecta}
\begin{table}
\caption{
    Modified detection numbers for \ac{EM} counterparts when the wind ejecta masses are reduced by half from the values in the main text. 
    The columns display the total number of short-term UVOIR counterparts and long-term afterglows. 
    The last three italicized columns reveal the actual transient types according to the post hoc criterion laid out in Sec.~\ref{subsec:detection_prospects}.
    The corresponding values from Table~\ref{tab:detections} are shown in brackets.
}
\begin{tblr}{
    colspec = {Q[c, 0.3cm] Q[c, 1cm]  Q[c, 1.3cm]  Q[c, 1.3cm] Q[c, 0.6cm] Q[c, 0.6cm] Q[c, 0.6cm]},
    rowsep = 3pt
}
\toprule 
\toprule 
& & Short-term UVOIR & Long-term afterglow & \textit{KN} & \textit{GRB afg.} & \textit{KN afg.} \\ 
\midrule 
\SetCell[r=8]{c} \rotatebox{90}{Narrow} & \SetCell[r=2]{c}\ett  & 52 & 63 & \textit{36} & \textit{49} & \textit{33}  \\[-3pt] 
 &  & (59) & (67) & \textit{(46)} & \textit{(53)} & \textit{(35)}  \\ 
 & \SetCell[r=2]{c}ETL  & 75 & 81 & \textit{60} & \textit{61} & \textit{39}  \\[-3pt] 
 &  & (91) & (89) & \textit{(78)} & \textit{(66)} & \textit{(44)}  \\ 
 & \SetCell[r=2]{c}\ett+CE  & 200 & 189 & \textit{191} & \textit{121} & \textit{103}  \\[-3pt] 
 &  & (242) & (212) & \textit{(236)} & \textit{(131)} & \textit{(118)}  \\ 
 & \SetCell[r=2]{c}ETL+CE  & 370 & 297 & \textit{354} & \textit{176} & \textit{165}  \\[-3pt] 
 &  & (497) & (349) & \textit{(489)} & \textit{(199)} & \textit{(198)}  \\ 
\midrule 
 \SetCell[r=8]{c} \rotatebox{90}{Wide} & \SetCell[r=2]{c}\ett  & 35 & 62 & \textit{19} & \textit{60} & \textit{11}  \\[-3pt] 
 &  & (43) & (67) & \textit{(28)} & \textit{(64)} & \textit{(13)}  \\ 
 & \SetCell[r=2]{c}ETL  & 51 & 84 & \textit{28} & \textit{79} & \textit{18}  \\[-3pt] 
 &  & (61) & (88) & \textit{(39)} & \textit{(83)} & \textit{(21)}  \\ 
 & \SetCell[r=2]{c}\ett+CE  & 88 & 148 & \textit{80} & \textit{128} & \textit{44}  \\[-3pt] 
 &  & (107) & (159) & \textit{(98)} & \textit{(136)} & \textit{(49)}  \\ 
 & \SetCell[r=2]{c}ETL+CE  & 139 & 200 & \textit{127} & \textit{174} & \textit{51}  \\[-3pt] 
 &  & (195) & (231) & \textit{(181)} & \textit{(200)} & \textit{(59)}  \\ 
\bottomrule 

\end{tblr}
\label{tab:detections_reduced_wind}
\end{table}

Whether a particular \ac{KN} light curve is bright enough to be detected depends mostly on its wind ejecta mass.
At the moment, it is not well understood how much of the post-merger disc is blown away by its wind and therefore the \ac{KN} detection estimates in Sec.~\ref{subsec:detection_prospects} carry an additional uncertainty.
In this Appendix, we quantify how the \ac{KN} counts change when all wind ejecta masses are reduced by 50\% compared to the values taken in the main Sec.~\ref{subsec:bns_catalogues}.
We run the same detection algorithm as outlined in Sec.~\ref{subsec:EM counterparts detection} and Appendices~\ref{app:observation_strategy_UVOIR} and \ref{app:observation_strategy_afterglow}, with the same mass distribution catalogues and \ac{GW} detector configurations.
Table~\ref{tab:detections_reduced_wind} compares the detection numbers from the case with the reduced wind ejecta to the numbers given in the main Table~\ref{tab:detections}.
Across catalogues, the number of short-term UVOIR counterparts decreases by about 20\%.
The largest relative reduction is seen in the ETL+CE catalogues, since originally these had the highest share of \acp{KN} in their UVOIR counterparts.
With lower UVOIR counterparts being detected, the number of subsequent afterglow detections also decreases, because many of them are found through deep observations once the sky localization is known exactly.

\section{Observation algorithm for UVOIR counterpart detection}
\label{app:observation_strategy_UVOIR}
\begin{table}
\caption{Telescope times needed for the UVOIR counterpart searches from the different catalogues, assuming different \ac{GW} detectors.
The entries are expressed as percentage of the total telescope time available in one year.
}
\begin{tblr}{
    colspec = {Q[c, 0.3cm] Q[c, 1cm]  Q[c, 1cm]  Q[c, 1cm] Q[c, 0.8cm] Q[c, 0.8cm] Q[c, 0.8cm]}
}
\toprule 
\toprule 
& & Rubin & PAN\-STARRS & ZTF & \textit{Roman} & 
\textit{ULTRA}\-\textit{SAT} \\ 
\midrule 
\SetCell[r=4]{c} \rotatebox{90}{Narrow} & \ett  & 6.7 & 1.2 & 0.2 & 0.6 & 0.1  \\ 
 & ETL  & 7.1 & 2.8 & 0.3 & 0.7 & 0.2  \\ 
 & \ett+CE  & 10.0 & 4.5 & 0.3 & 0.1 & 0.2  \\ 
 & ETL+CE  & 26.4 & 4.2 & 0.3 & 0.1 & 0.2  \\ 
\midrule 
 \SetCell[r=4]{c} \rotatebox{90}{Wide} & \ett  & 8.0 & 1.4 & 0.1 & 0.8 & 0.1  \\ 
 & ETL  & 9.1 & 3.1 & 0.3 & 0.9 & 0.2  \\ 
 & \ett+CE  & 14.1 & 4.8 & 0.3 & 0.1 & 0.2  \\ 
 & ETL+CE  & 35.7 & 4.4 & 0.2 & 0.2 & 0.2  \\ 
\bottomrule 

\end{tblr}
\label{tab:telescope_time}
\end{table}
To keep telescope time within reasonable bounds, observers will need to prioritize which \ac{GW} signal they want to follow up on.
This will depend on the sensitivity of their telescope and the overall number of \acp{BNS} with decent sky localization.
Hence, we set different cutoffs for $d_L$ and $\Delta\Omega$ a \ac{GW} signal must not exceed in order to receive a follow-up campaign by a particular telescope. 
These cutoffs also depend on which \ac{GW} detectors are employed, since this impacts how many \acp{BNS} have good localization.
Generally, the brightest \acp{KN} are expected to peak at $-16$ absolute magnitude, hence, with a limiting magnitude of 22, ZTF would only start observations for events within $d_L\leq400$~Mpc.
Similarly, we require $d_L\leq4000$~Mpc for Rubin, $d_L\leq1000$~Mpc for PAN-STARRs, $d_L\leq2000$~Mpc for \textit{Roman}, and $d_L\leq500$~Mpc for \textit{ULTRASAT}.
If at the same time there is an associated \ac{GRB}, we increase these cutoffs to $d_L\leq1000$~Mpc for ZTF and \textit{ULTRASAT}, $d_L\leq7000$~Mpc for Rubin, and $d_L\leq3000$~Mpc for PAN-STARRs and \textit{Roman}.
This is because observers would attempt to find the early afterglow, which at low inclination can be significantly brighter than a \ac{KN}.

Moreover, ZTF only starts a campaign for sources with $\Delta\Omega \leq 200$~deg$^2$.
If CE is in the detector setup, we even require $\Delta\Omega \leq 100$~deg$^2$.
Likewise, the threshold for Rubin is $\Delta\Omega \leq 100$~deg$^2$ if there is only \ett, 50~deg$^2$ with only ETL, and 10~deg$^2$ with CE.
In the latter case, farther events with $0.2 \leq z \leq 1$ additionally need to have $\Delta\Omega \leq 7$~deg$^2$, otherwise the total required telescope time becomes too large.
For a Pan-STARRS survey, we require $\Delta\Omega\leq50$ (30)~deg$^2$ without (with) CE.
\textit{Roman} observes if $\Delta\Omega\leq10$ (1)~deg$^2$ without (with) CE.
\textit{ULTRASAT} targets triggers with $\Delta\Omega \leq 400$~deg$^2$ regardless of the \ac{GW} detector network. 
With these cutoffs, the telescope times stay within reasonable limits.
We provide the corresponding values after one year in Table~\ref{tab:telescope_time}.

Ground based telescopes attempt to find the transient in up to three observing periods.
Observations only take place as long as the transient is visible during astronomical night, i.e., the sun at telescope location is more than $18^\circ$ below the horizon and the transient is higher than $20^\circ$ altitude.
If a period is still ongoing when the transient ceases to be visible, it will be resumed the next night if the above conditions can be fulfilled again.
During each observation period, the sky area $\Delta\Omega$ is divided into 
\begin{align}
    n_{\text{tiles}} = \biggl\lceil \frac{\Delta\Omega}{\text{FOV}} \biggr\rceil + 1
    \label{eq:tiling}
\end{align}
tiles, although we assume that transients with $\Delta\Omega \leq 0.8\times\text{FOV}$ can be covered through a single tile.
We assume that exactly one random tile contains the transient, though in reality tiling sky localization areas is more complex~\citep{Almualla:2020hbs, Gupte:2020mfp} and instead of starting in a random order, one would begin with the posterior likelihood peak.
Each tile is observed with the exposure time and limiting magnitude listed in Table~\ref{tab:kn_instruments}.
A detection flag is set if the brightness of the joint \ac{KN} and \ac{GRB} afterglow light curve exceeds the limiting magnitude at the time of observing the correct tile.
Once every tile has been observed in $g$, the scan starts again in the $i$ filter.
One observation period thus consists of two complete scans of the tiling.
The first observation starts 5 hours after the trigger alert or as soon as the transient becomes night-visible. 
The second period starts 13 hours after the trigger, also subject to the night visibility condition.
A third period is scheduled at 17 hours after trigger, if the total telescope time spent on this transient is still below 6 hours and so far only one detection flag in one filter at one of the previous periods has been reported.
Our setup ignores weather and cloud statistics for the ground-based sites, which in reality could cause sub-optimal observing conditions in certain nights.

Our algorithm for the space-based telescopes \textit{Roman} and \textit{ULTRASAT} follows a similar approach, with minor adaptions.
At any given time, \textit{ULTRASAT}'s instantaneous sky coverage is 51\%, thus, we assume observations start with a probability of 51\% 15~min after trigger.
If this fails, there is another attempt at 3~h with a probability of 75\%, corresponding to the cumulative sky fraction available by then~\citep{Shvartzvald:2023ofi}.
\textit{ULTRASAT} only observes one filter in one observation period and resumes normal operations no later than 3~h after trigger.
For \textit{Roman}, a target is within the field of regard as long as the angle to the Sun is more than 54$^\circ$ but less than 126$^\circ$~\citep{Roman-technical-details}.
We assume two observing periods, the first 12~h after merger, the second 3~days after the first.

\section{Observation algorithm for wide-field afterglow searches}
\label{app:observation_strategy_afterglow}

If no UVOIR counterpart is found in the immediate aftermath of the merger, we suppose that Rubin and the telescopes listed in Table~\ref{tab:afterglow_instruments} consider a wide-field follow-up campaign for the afterglow, provided certain conditions are met.
Concretely, we require $d_L\leq 11$~Gpc and $\Delta\Omega\leq100$~deg$^2$ if there is only \ett, $\Delta\Omega\leq50$~deg$^2$ if there is only ETL, and $\Delta\Omega\leq10$~deg$^2$ if there is CE.
The difficulty in performing these follow-ups lies in the fact that the time at which the afterglow peaks depends sensitively on several factors such as inclination, opening angle, interstellar medium density, kinetic energy, and eventually redshift.
The inclination might be estimated from the \ac{GW} signal, although even for high \ac{SNR} the uncertainty remains at $\gtrsim 0.1$~rad.
Thus, although analytical estimates for the peak time of a \ac{GRB}~\citep{Ryan:2019fhz} or \ac{KN} afterglow~\citep{Kathirgamaraju:2019xwu} exist, their use for determining optimal afterglow observation times after a \ac{GW} event is limited.
For this reason, we simply spread ten observation periods log-uniformly between 10 days to 10 years.
The last observation period serves specifically to explore the possibility for a \ac{KN} afterglow detection.
For radio telescopes, observations take place once the transient region is above 20$^\circ$ altitude and with the tiling according to Eq.~\eqref{eq:tiling}.
This is the same for Rubin, which here only observes in the $g$ band, except that additionally the transient needs to be visible at astronomical night.
If the sky location is inaccessible, we also allow later observation periods to be postponed by half a year and check whether the visibility conditions can be met then.
Similarly, the space-based \textit{Einstein Probe} has access to the entire night-side of the sky.
If the transient is on the day side, we simply postpone the observation for half a year until the transient becomes accessible.
An afterglow detection is noticed once the flux at one observing period exceeds the limits listed in Table~\ref{tab:afterglow_instruments}.

\section{GW injection and recovery}
\label{app:gw_injections}
To obtain individual posteriors from the \ac{GW} events, we inject the signal into \ett or ETL noise.
Parameter estimation is performed using \textsc{bilby} and the \textsc{dynesty} sampler~\citep{Speagle:2019ivv} with a minimum frequency of 5~Hz.
To accelerate the inference we use the multi-banding approximation~\citep{Morisaki:2021ngj} and marginalize over phase and distance.
Priors are chosen to be uniform or isotropic.
For the masses, the prior is uniform in the detector frame component masses $m_1^{\text{det}}$ and $m_2^{\text{det}}$. 
In order to save computational costs, we additionally restrict the detector frame chirp mass to values of $\pm 0.01~\msun$ around the injected value and the mass ratios to $m_2/m_1 \geq 0.4$.
Priors on $\Lambda_1$, $\Lambda_2$ are uniform between 0 and 5000.
We use aligned spin priors with spin magnitude from 0 to 0.2.
The luminosity distance prior is uniform in $(1+z)^{-1} V_c$ between 40 and 8000~Mpc.
The coalescence time prior is uniform in the interval $\pm 0.1~$s around the injected value.
Right ascension and sky declination are fixed.

Since we do not sample directly in the source frame, the prior on the source frame component masses is only implicit
\begin{align}
\begin{split}
    \pi(m_1, m_2) &= \int \pi(m_1, m_2, z)\ dz\\
    &= \int \pi(m_1^{\text{det}}, m_2^{\text{det}})\ \pi(z) \left|\frac{\partial (m_1^{\text{det}}, m_2^{\text{det}}, z)}{\partial(m_1, m_2, z)}\right|\ dz \\
    &\propto \int_{z_{\text{min}}(m_1, m_2)}^{z_{\text{max}}(m_1, m_2)} \pi(z)\ (1+z)^2 dz\ \ ,
\end{split}
\end{align}
where the limits of the integral result from the restriction on the detector frame chirp mass that we impose during sampling
\begin{align}
    \begin{split}
        z_{\text{min}}(m_1, m_2) &= \frac{\mathcal{M}^{\text{det}}_{\text{min}}}{\mathcal{M}^{\text{source}}(m_1,m_2)} -1 \ ,\\        
        z_{\text{max}}(m_1, m_2) &= \frac{\mathcal{M}^{\text{det}}_{\text{max}}}{\mathcal{M}^{\text{source}}(m_1,m_2)} -1\ .
    \end{split}
\end{align}
It is this prior we divide by in Eqs.~\eqref{eq:gw_eos_likelihood} and \eqref{eq:mm_eos_likelihood}.
For the inference where we only use \ac{GW} data, the prior on $\pi(z)$ is derived from the luminosity distance prior, whereas in the multimessenger case, we assume the redshift of the host galaxy leads to $\pi(z)$ being a normal distribution with 1\% uncertainty.

\section{EM injection and recovery}
\label{app:em_injections}

\begin{table}
\caption{Priors used for the EM inference. The first section lists the \ac{KN} parameters, the second one the parameters for the GRB afterglow.
Uniform priors are denoted by $\mathcal{U}$, log-uniform priors by Log.
}
\begin{tblr}{
    colspec = {Q[c, 3.5 cm] Q[c, 1.5cm] Q[c, 1.5cm]}
}
\toprule 
\toprule 
Parameter & Symbol & Prior \\ 
\midrule 

Log dynamical ejecta mass~[M$_\odot$] & $\log_{10}(m_{\text{ej, dyn}})$ & $\mathcal{U}(-4, -1.3)$ \\
Average dynamical ejecta velocity~[$c$] & $\bar{v}_{\text{ej, dyn}}$ & $\mathcal{U}(0.12, 0.35)$ \\
Average dynamical ejecta electron fraction & $\bar{Y}_{\text{e, dyn}}$ & $\mathcal{U}(0.15, 0.35)$ \\
Log wind ejecta mass [M$_\odot$] & $\log_{10}(m_{\text{ej, wind}})$ & $\mathcal{U}(-4, -0.55)$ \\
Average wind ejecta velocity [$c$] & $\bar{v}_{\text{ej, wind}}$ & $\mathcal{U}(0.05, 0.15)$ \\
Wind ejecta electron fraction & $Y_{\text{e, wind}}$ & $\mathcal{U}(0.2, 0.4)$ \\
\midrule 
Log isotropic kinetic energy~[erg] & $\log_{10}(E_{\text{kin, iso}})$ & $\mathcal{U}(47, 57)$ \\
Jet core angle~[rad] & $\theta_{\text{c}}$ & Log$(0.01, \pi/5)$ \\
Wing factor & $\alpha_{\text{w}}$ & $\mathcal{U}(0.2, 3.5)$   \\
Log interstellar medium density~[cm$^{-3}$] & $\log_{10}(n_{\text{ism}})$ & $\mathcal{U}(-6, 2)$ \\
Electron power index & $p$ & $\mathcal{U}(2, 3)$ \\
Log electron energy fraction & $\log_{10}(\epsilon_e)$ & $\mathcal{U}(-4, 0)$ \\
Log magnetic energy fraction & $\log_{10}(\epsilon_B)$ & $\mathcal{U}(-8, 0)$ &\\
Initial Lorentz factor & $\Gamma_0$ & $\mathcal{U}(100, 1000)$ \\
\bottomrule

\end{tblr}
\label{tab:em_priors}
\end{table}
To obtain individual posteriors for the UVOIR counterparts, we create mock light curve measurements.
Like for the detection algorithm in Sec.~\ref{subsec:EM counterparts detection}, light curves are calculated from the \textsc{pyblastafterglow} and \textsc{possis} surrogates.
We assume that for each event 30 observations are taken, where the data points are spread randomly across the $J$, $i$, $v$, and $u$ bands.
If a GRB afterglow is present, we increase the number of observations to 60 and also include the 1.4~GHz and 1~keV bands.
Occasionally, we use even more magnitude measurements to bolster the convergence of the posterior to the injected values.
The time points are randomly log-uniformly distributed between 0.5 and 10~days.
If the light curve at a particular data time point falls below the detection limit, the data point becomes an upper limit.
The detection limits are set to 3.63$\times 10^{-6}$~mJy in the radio, 29.5~mag in the $J$, $i$, and $v$ bands, 26~mag in $u$, and 9.12 $\times 10^{-39}$~erg\,cm$^{-2}$\,Hz$^{-1}$ in the X-ray.

The posteriors from these mock light curve measurements are sampled with the \textsc{flowMC} sampler~\citep{Wong:2022xvh}.
The recovery always uses both the \ac{KN} and GRB afterglow model, i.e., we assume it is not a-priori known which transient types are present.
Table~\ref{tab:em_priors} summarizes the priors used to sample on the model's parameters.
The inclination is isotropic,  the luminosity distance is uniform in $(1+z)^{-1} V_c$, and the redshift prior is a normal distribution around the true value with a 1\% uncertainty.
Since in reality, \acp{KN} might not be perfectly matched by the model used for the  inference, we introduce a systematic uncertainty parameter $\sigma_{\text{sys}}$ to broaden the posterior. 
This parameter is applied to all data points and during sampling is drawn from a uniform prior between 0.3 and 1~mag, see e.g. \citet{Koehn:2025zzb} for more details.

\section{Wind ejecta in the multimessenger likelihood}
\label{app:wind_ejecta_likelihood}
\begin{figure}
    \centering
    \includegraphics[width=\linewidth]{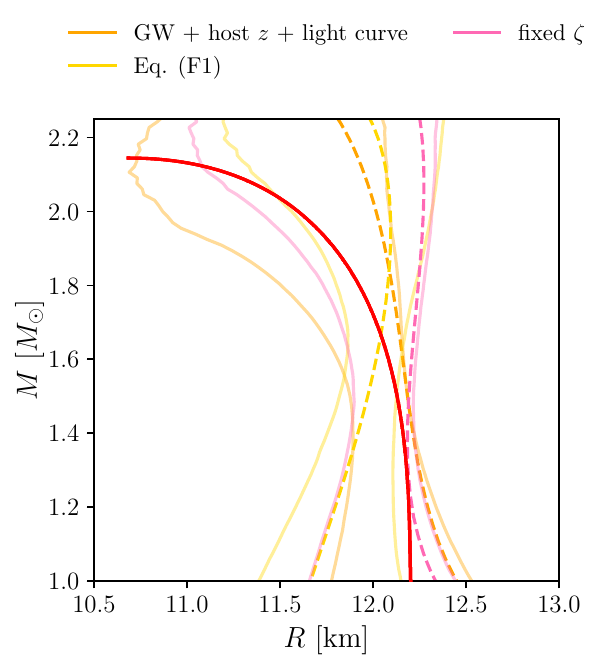}
    \caption{Joint inference of the \ac{EOS} and mass distribution from the \ac{KN} events in narrow+ETL, using different methods to incorporate the wind ejecta.
    We show the posteriors using the regular multimessenger likelihood Eq.~\eqref{eq:mm_eos_likelihood} (orange, the same as in Fig.~\ref{fig:eos_inference_narrow_ETL} but without selection effects), using the likelihood from Eq.~\eqref{eq:mm_eos_likelihood_wind} (yellow), and where we assume $\zeta$ from Eq.~\eqref{eq:disc_to_wind_mass} is known exactly for each \ac{BNS} (pink).
    The 95\% credible limits are drawn as solid lines and the posteriors are not reweighted by the detection probabilities.}
    \label{fig:eos_inference_different_mm}
\end{figure}
Although the wind and subsequently disc mass can be inferred from the \ac{KN} light curves, Eq.~\eqref{eq:mm_eos_likelihood} ignores this information, despite the potential for additional constraints on the \ac{EOS}.
Including this element, however, is not straight-forward, since in our setup only the mapping from the EOS to the disc mass is deterministic, whereas the wind ejecta are then obtained from the randomly drawn ratio $\zeta$ in Eq.~\eqref{eq:disc_to_wind_mass}.
In theory, one therefore would need to sample $\zeta$ separately for each event, but this inflates the dimensionality of the problem and makes the computation infeasible.

One approach would be to sample $\zeta$ from the same truncated normal distribution used to generate the population, but for a real application it is implausible that the correct distribution for $\zeta$ is known beforehand.
Because the disc is always heavier than the wind mass, an alternative would be to use the cumulative probability, evaluated through a conditional \ac{NF}.
In this case, Eq.~\eqref{eq:mm_eos_likelihood} becomes
\begin{align}
\begin{split}
&\mathcal{L}(\text{EOS}, \lambda_{\text{pop}} \mid \text{GW}_j, \text{EM}_j)
= \\
&\int  \biggl[ \frac{p(m_1, m_2, \Lambda_1^{\text{EOS}}(m_1), \Lambda_2^{\text{EOS}}(m_2) ,d_L, \iota \mid \text{GW}_j)}{\pi(m_1, m_2) \pi(d_L) \pi(\iota)}\ \times \\
& \qquad \ \int_{-\infty}^{\log_{10}(m_{\text{disc}}^{\text{fit}})} \frac{p(\log_{10}m_{\text{ej, dyn}}^{\text{fit}}, u, d_L, \iota \mid \text{EM}_j)}{\pi(d_L) \pi(\iota)}\  du\ \times \\
& \qquad \  p(m_1, m_2 \mid \lambda_{\text{pop}})  \biggr]\ dm_1\,dm_2\ .
\label{eq:mm_eos_likelihood_wind}
\end{split}
\end{align}
Here, the variable $u$ stands for the integral over the $\log_{10}$ wind ejecta mass, the fit superscript for the disc and dynamical ejecta masses means that these are evaluated by using the fitting relations in Eqs.~(\ref{eq:pc_fittings})--(\ref{eq:nopc_disc}).

In Fig.~\ref{fig:eos_inference_different_mm}, we show the resulting $M$-$R$ posteriors from the narrow+ETL catalogue.
We show the posteriors when either the usual Eq.~\eqref{eq:mm_eos_likelihood} (orange) or Eq.~\eqref{eq:mm_eos_likelihood_wind} is used (yellow).
For reference, we also show the posterior obtained by assuming that $\zeta$ from Eq.~\eqref{eq:disc_to_wind_mass} is known exactly for each \ac{BNS} (pink).
To allow for a direct comparison of the likelihood formulae, no selection effects are applied here.
We see that the default posterior from the main section and the version where $\zeta$ is fixed recover very similar radii.
The posterior using Eq.~\eqref{eq:mm_eos_likelihood_wind} infers notably softer \acp{EOS}, because in this version of the likelihood it is beneficial to overestimate the disc mass, which per Eqs.~\eqref{eq:nopc_disc} and \eqref{eq:pc_fittings} is achieved by \acp{EOS} with higher $\mtov$ but smaller $R_{1.6}$.
We thus conclude that the inclusion of the wind ejecta via Eq.~\eqref{eq:mm_eos_likelihood_wind} can even lead to flawed results, while fixing $\zeta$ seems to have no further benefit for the events considered here. 
This is expected, given that the overall impact of the light-curve data even with just the dynamical ejecta on the \ac{EOS} appears negligible.
A potential method to effectively incorporate the wind ejecta in hierarchical multimessenger inference for fewer events with lesser \ac{SNR} would be a sequential approach, where the hierarchical posterior is updated event for event and $\zeta$ is then added as one additional sampling parameter.
The utility of incorporating wind ejecta might be explored further in future work though, to see if better constraints for other event sets can be achieved.

\section{Estimating detection probabilities}
\label{app:detection_probability}

\begin{table}
\centering
\caption{
    Posterior credibility intervals for the inference results from Sec.~\ref{subsec:cosmo_inference} without selection effects.
    We show the corresponding intervals from Table~\ref{tab:cosmo_inference_results} (i.e. with selection effects applied) in brackets. 
    The credibility limit is set to 95\%.
}
\label{tab:no_selection_effects}
\begin{tblr}{
    colspec = {Q[c, 0.8cm] Q[c, 1.2cm]  Q[c, 1.2cm] Q[c, 1.2cm]  Q[c, 1.2cm] Q[c, 0.6cm]},
}
\toprule 
\toprule
 & Narrow+ETL & Wide+ETL & Narrow+\ett & Wide+\ett  & Truth\\
\midrule 
\SetCell[r=2]{{c}} $\Lambda_{1.4}$ & ${372}^{+12}_{-11}$ & ${397}^{+19}_{-20}$ & ${379}^{+16}_{-14}$ & ${382}^{+27}_{-26}$ &\SetCell[r=2]{c} 383 \\ 
 & $({371}^{+12}_{-11})$ & $({395}^{+18}_{-22})$ & $({380}^{+16}_{-14})$ & $({379}^{+20}_{-23})$ & \\[0.5em] 
\SetCell[r=2]{{c}} $R_{1.4}$ \newline [km] & ${12.13}^{+0.08}_{-0.11}$ & ${12.06}^{+0.14}_{-0.22}$ & ${11.95}^{+0.25}_{-0.20}$ & ${12.00}^{+0.19}_{-0.23}$ &\SetCell[r=2]{c} 12.16 \\ 
 & $({12.12}^{+0.08}_{-0.11})$ & $({12.05}^{+0.14}_{-0.17})$ & $({11.95}^{+0.25}_{-0.20})$ & $({11.99}^{+0.15}_{-0.21})$ & \\[0.5em] 
\SetCell[r=2]{{c}} $M_{\rm{TOV}}$ \newline $[\msun]$ & ${2.11}^{+0.14}_{-0.13}$ & ${2.22}^{+0.15}_{-0.25}$ & ${2.09}^{+0.26}_{-0.14}$ & ${2.13}^{+0.23}_{-0.17}$ &2.14 \\ 
 & $({2.11}^{+0.15}_{-0.13})$ & $({2.21}^{+0.14}_{-0.24})$ & $({2.08}^{+0.25}_{-0.13})$ & $({2.13}^{+0.21}_{-0.18})$ & \\[0.5em] 
\SetCell[r=2]{{c}} $p(3\nsat)$ [MeV~fm$^{{-3}}$] & ${71}^{+18}_{-7}$ & ${90}^{+27}_{-16}$ & ${73}^{+28}_{-10}$ & ${75}^{+22}_{-10}$ &75 \\[0.5em] 
 & $({70}^{+14}_{-6})$ & $({86}^{+24}_{-15})$ & $({73}^{+27}_{-9})$ & $({74}^{+18}_{-11})$ & \\ 
\midrule 
\SetCell[r=2]{c} $\mu_1~[\msun]$ & ${1.32}_{-0.01}^{+0.03}$ & - & ${1.34}_{-0.05}^{+0.06}$ & - & \SetCell[r=2]{c} 1.34 \\ 
 & $({1.32}_{-0.01}^{+0.02})$ & - & $({1.33}_{-0.04}^{+0.05})$ & - &  \\[0.5em] 
\SetCell[r=2]{c} $\sigma_1~[\msun]$ & ${0.03}_{-0.01}^{+0.02}$ & - & ${0.06}_{-0.04}^{+0.04}$ & - & \SetCell[r=2]{c} 0.02 \\ 
 & $({0.03}_{-0.01}^{+0.02})$ & - & $({0.05}_{-0.04}^{+0.04})$ & - &  \\[0.5em] 
\SetCell[r=2]{c} $\mu_2~[\msun]$ & ${1.71}_{-0.22}^{+0.18}$ & - & ${1.68}_{-0.28}^{+0.18}$ & - & \SetCell[r=2]{c} 1.47 \\ 
 & $({1.69}_{-0.19}^{+0.20})$ & - & $({1.69}_{-0.23}^{+0.15})$ & - &  \\[0.5em] 
\SetCell[r=2]{c} $\sigma_2~[\msun]$ & ${0.16}_{-0.07}^{+0.04}$ & - & ${0.16}_{-0.11}^{+0.04}$ & - & \SetCell[r=2]{c} 0.15 \\ 
 & $({0.15}_{-0.06}^{+0.05})$ & - & $({0.15}_{-0.11}^{+0.05})$ & - &  \\[0.5em] 
\SetCell[r=2]{c} $w$ & ${0.94}_{-0.11}^{+0.05}$ & - & ${0.86}_{-0.41}^{+0.10}$ & - & \SetCell[r=2]{c} 0.68 \\ 
 & $({0.94}_{-0.13}^{+0.04})$ & - & $({0.87}_{-0.35}^{+0.08})$ & - &  \\[0.5em] 
\SetCell[r=2]{c} $m^l~[\msun]$ & ${1.16}_{-0.03}^{+0.03}$ & - & ${1.17}_{-0.07}^{+0.07}$ & - & \SetCell[r=2]{c} 1.16 \\ 
 & $({1.16}_{-0.03}^{+0.03})$ & - & $({1.18}_{-0.07}^{+0.07})$ & - &  \\[0.5em] 
\SetCell[r=2]{c} $m^u~[\msun]$ & ${1.48}_{-0.03}^{+0.02}$ & - & ${1.43}_{-0.09}^{+0.06}$ & - & \SetCell[r=2]{c} 1.42 \\ 
 & $({1.48}_{-0.03}^{+0.01})$ & - & $({1.46}_{-0.10}^{+0.04})$ & - &  \\[0.5em] 
\midrule 
\SetCell[r=2]{c} $m_{\rm min}$ $[\msun]$ & - & ${1.11}_{-0.05}^{+0.04}$ & - & ${1.11}_{-0.07}^{+0.05}$ & \SetCell[r=2]{c} 1.1 \\ 
 & - & $({1.13}_{-0.04}^{+0.03})$ & - & $({1.14}_{-0.05}^{+0.03})$ &  \\[0.5em] 
\SetCell[r=2]{c} $m_{\rm max}$ $[\msun]$ & - & ${1.84}_{-0.05}^{+0.09}$ & - & ${1.88}_{-0.08}^{+0.12}$ & \SetCell[r=2]{c} 2.0 \\ 
 & - & $({1.90}_{-0.08}^{+0.08})$ & - & $({1.94}_{-0.09}^{+0.10})$ &  \\[0.5em] 
\SetCell[r=2]{c} $\alpha$ & - & ${0.93}_{-0.89}^{+2.21}$ & - & ${0.75}_{-0.72}^{+2.46}$ & \SetCell[r=2]{c} 2 \\ 
 & - & $({2.43}_{-2.02}^{+1.48})$ & - & $({2.29}_{-2.11}^{+1.54})$ &  \\[0.5em] 
\midrule 
\SetCell[r=2]{{c}} $H_0$ [km~s$^{-1}$ Mpc$^{-1}$] & ${67.8}_{-0.9}^{+0.9}$ & ${67.7}_{-1.5}^{+1.7}$ & ${68.1}_{-1.4}^{+1.4}$ & ${67.8}_{-3.0}^{+2.6}$ &\SetCell[r=2]{c} 67.7 \\ 
 & $({68.2}_{-0.9}^{+0.9})$ & $({67.5}_{-1.7}^{+1.4})$ & $({68.1}_{-1.2}^{+1.4})$ & $({68.1}_{-3.2}^{+2.3})$ & \\[0.5em] 
\SetCell[r=2]{{c}} $\Omega_0$ & ${0.42}_{-0.13}^{+0.17}$ & ${0.34}_{-0.11}^{+0.11}$ & ${0.34}_{-0.11}^{+0.14}$ & ${0.41}_{-0.18}^{+0.26}$ &\SetCell[r=2]{c} 0.31 \\ 
 & $({0.32}_{-0.07}^{+0.11})$ & $({0.34}_{-0.13}^{+0.15})$ & $({0.30}_{-0.11}^{+0.11})$ & $({0.34}_{-0.15}^{+0.22})$ & \\[0.5em] 
\bottomrule 

\end{tblr}
\end{table}

To estimate the multimessenger selection effects for hierarchical posteriors, we essentially repeat the steps from Secs.~\ref{sec:events} and \ref{sec:observations}, but with a few simplifications to efficiently evaluate the detection probability for the thousands of hypersamples.
For a given hierarchical posterior, we first determine the GW detection probability for each hyperparameter sample $\lambda$.
This is done by sampling 2000 component masses from $p(m_1, m_2 \mid \lambda_{\text{pop}})$ and calculating the \ac{GW} \ac{SNR} for the corresponding detector network.
Exterior parameters are drawn isotropically or uniformly; the redshifts are drawn from Eq.~\eqref{eq:BNS_observation_rate} within $z\leq1.2$ using the same rate parameters as in Sec.~\ref{subsec:bns_catalogues}.
If the hierarchical posterior also contains samples for $H_0$ and $\Omega_0$, the conversion to luminosity distance is performed using those, otherwise this happens with the Planck18 values.
To make the \ac{SNR} calculation efficient, we follow the approach of \citet{Gerosa:2020pgy} and train a neural network to learn the relationship between \ac{SNR}, $m_1$, $m_2$ and exterior parameters on $10^6$ training data points.
The \ac{GW} detection probability is then proportional to the number of binaries where the \ac{SNR} exceeds 12.

In the next step, we also compute the \ac{EM} detection probability for each hyperparameter sample.
Using the \ac{EOS} from the hyperparameter sample and the prompt collapse criterion in Eq.~\eqref{eq:pc_simple_criterion}, we determine the \ac{KN} light curve for each of the 2000 \ac{BNS} mass pairs through the same procedure as in Sec.~\ref{subsec:kilonova}.
Since the detailed observation algorithm developed in Sec.~\ref{sec:observations} is too costly to run for each hyperparameter sample point, we simply assume that the \ac{EM} detection probability is proportional to the number of \ac{KN} light curves where the brightness exceeds the 25.6 mag in the Rubin $i$ band.

Typically, the detection probabilities for our hyperposterior samples are very similar, though a few samples ($\sim$10 out of normally 7000 hierarchical posterior samples) have a detection probability that is a factor of 2--3 lower than the typical detection probability.
Since we reweight the hyperposterior according to $p_{\text{det}}(\lambda)^{-n}$, where $n\gtrsim20$ is the number of events, this sometimes leads to situations where a single sample with low detection probability would make up 20\% or more of the reweighted posterior.
Such extreme distortions do not occur when we neglect the \ac{EM} selection effect.
Hence, this is a reflection of our simplified treatment for the counterpart observations.
In particular, \acp{EOS} with $\mtov\lesssim2.0~\msun$ or mass distributions with a heavy tail at large masses are disproportionally disfavoured by our detection probability algorithm and become dominant after reweighting.
For this reason, we clip their detection probability values to the 99\% quantile of the overall set of values for $p_{\text{det}}(\lambda)$.
The choice for the clipping quantile is based on the fact that $\lesssim 50$ samples out of typically 7000 run into this issue.
The specific level chosen does not impact final posterior estimates in a strong way.
If we would clip instead at the 95\%, 90\%, or 80\% quantile, we always find $R_{1.4} = {12.12}^{+0.09}_{-0.11}$~km for the representative narrow+ETL multimessenger inference from Sec.~\ref{subsec:cosmo_inference}.
The estimates for $H_0$ change only slightly to ${68.2}_{-0.8}^{+0.9},\ {68.1}_{-0.8}^{+0.9},\ {68.0}_{-0.9}^{+0.9}$~km~s$^{-1}$~Mpc$^{-1}$.

Generally, posterior estimates for the \ac{EOS} stay relatively unchanged after applying the selection effect, while mass distribution and cosmological parameters can display sensitive changes.
In Table~\ref{tab:no_selection_effects}, we compare the posterior credibility intervals before and after the posteriors are reweighted with the detection probability.
The bracketed values correspond to the reweighted posteriors from the main section and confirm that the largest changes occur for $\Omega_0$ and $\alpha$, while \ac{EOS} quantities remain largely unaffected.

\section{Hierarchical recovery with a stiffer \ac{EOS}}
\label{app:wide_ETT_stiffer}

\begin{figure}
    \centering
    \includegraphics[width=1\linewidth]{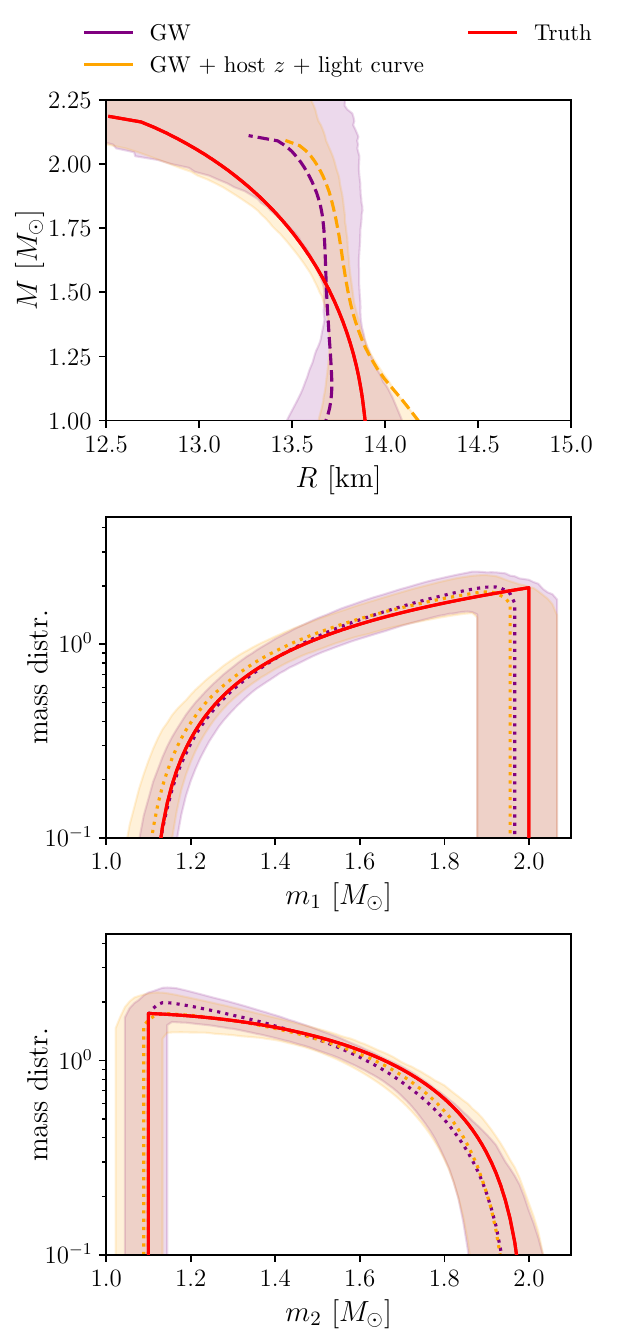}
    \caption{
    \ac{EOS} and mass distribution inference from the \ac{KN} events in the wide+\ett catalogue and a stiffer \ac{EOS}.
    The layout is equivalent to Fig.~\ref{fig:eos_inference_wide_ETL}.
    }
    \label{fig:eos_inference_wide_ETT_stiffer}
\end{figure}

To assess the robustness of our hierarchical inference method, we repeat the hierarchical inference for wide+\ett using a stiffer injection \ac{EOS}.
Specifically, we use an \ac{EOS} that is parametrized through the meta-model with \ac{CSE}, i.e., the same \ac{EOS} parametrization used in the recovery.
Our injected \ac{EOS} leads to a canonical \ac{NS} with $R_{1.4} = 13.76$~km, $\Lambda_{1.4} = 891$, and a mass limit of $\mtov = 2.18~\msun$.
We do not regenerate the full multimessenger population from scratch, using our detection algorithm, but simply take the events of the wide+\ett catalogue from the main text and replace the tidal deformabilities and ejecta masses with new values according to the stiffer \ac{EOS}.
We then follow the same approach as in Sec.~\ref{subsec:eos_inference}, by first obtaining individual \ac{GW} and \ac{EM} posteriors and combining them in the hierarchical likelihoods of Eqs.~\eqref{eq:gw_eos_likelihood} and \eqref{eq:mm_eos_likelihood}.

The resulting posteriors are shown in Fig.~\ref{fig:eos_inference_wide_ETT_stiffer}.
The radius from the \ac{GW}-only (full multimessenger) posterior is ${13.78}^{+0.08}_{-0.11}\ ({13.78}^{+0.07}_{-0.09})$~km,  likewise $\Lambda_{1.4}={926}^{+44}_{-51}\ ({897}^{+36}_{-40})$.
The pressure is now best constrained at densities between $1$ and $2~\nsat$, with $p_{1.5\nsat} = {14.9}_{-1.8}^{+1.2}\ ({14.4}^{+1.0}_{-1.2})$~MeV~fm$^{-3}$, where the injected stiff \ac{EOS} has a value of $p_{1.5\nsat}=15.1$~MeV~fm$^{-3}$.
Using the stiffer meta-model-\ac{CSE} \ac{EOS} appears to have two main effects. 
First, there is no offset in $R_{1.4}$ any more, because both injection and recovery now use the same \ac{EOS} model and are therefore congruent.
Secondly, the uncertainties are tighter compared with the wide+\ett results from the main section.
This effect is partly related to the congruence between injection and recovery, but also to the fact that stiffer \acp{EOS} produce more pronounced tidal signatures overall.
We also note that the \ac{GW}-only posterior slightly overestimates the tidal deformabilities, especially at higher \ac{NS} masses.
The reason for this is not entirely clear, but could be related to degeneracy with mass ratio and $\Lambda_1$ that is present in many \ac{GW} posteriors for events with small mass ratios.

Lastly, in regards to the mass distribution, we find that $m_{\text{min}} = {1.10}_{-0.06}^{+0.04}\ ({1.08}_{-0.06}^{+0.05})$~$\msun$ and $m_{\text{max}} = {1.97}_{-0.09}^{+0.10}\ ({1.96}_{-0.07}^{+0.11})$~$\msun$ are well recovered, whereas for $\alpha = {0.82}_{-0.77}^{+1.88}\ ({1.93}_{-1.76}^{+1.87})$, the \ac{GW}-only posterior significantly underestimates the power-law index $\alpha$.

\vfill
\bsp	
\label{lastpage}
\end{document}